\documentclass{aa}  

\usepackage{graphicx}
\usepackage{txfonts}
\usepackage{mathrsfs}
\usepackage{graphicx}   
\usepackage{amsmath}    
\usepackage{todonotes}
\usepackage{mathtools}
\usepackage{bm}
\usepackage{multicol}
\usepackage{graphicx}  
\usepackage{soul}
\usepackage[utf8]{inputenc}
\usepackage{lmodern}
\usepackage{diagbox}
\usepackage{listings}
\usepackage{ulem}
\usepackage{lscape}

\usepackage{natbib}
\usepackage{longtable}
\usepackage{multirow}
\usepackage{amssymb}
\usepackage{nicematrix}
\usepackage{placeins}
\usepackage{rotating}

\usepackage[hidelinks,citecolor=blue,linkcolor=blue]{hyperref}
\hypersetup{
    colorlinks=true,
    linkcolor=blue,
    urlcolor=blue,
    citecolor=blue
} 

\begin{document} 
\definecolor{forestgreen}{RGB}{24, 129, 24}
\newcommand\luba[1]{\emph{{\color{forestgreen}#1}}}
\newcommand\sm[1]{\emph{{\color{red}#1}}}
\newcommand\libu[1]{\emph{{\color{blue}#1}}}
\newcommand{\Msol}{M_\odot}
\newcommand\MdB{\cite{Mathis2011Low-frequencyField}}
\newcommand\Malk{\hl{Malk66}}
\newcommand\TT{\hl{TT22}}
\newenvironment{bluetext}{\color{purple}}{}
\newcommand\orc[1]{\href{https://orcid.org/#1}{\includegraphics[width=3mm]{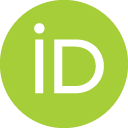}}}
\newenvironment{redtext}{\color{purple}}{}
\newenvironment{greentext}{\color{forestgreen}}{}

   \title{Exploring the probing power of $\gamma$-Dor's inertial dip \\ for core magnetism: case of a toroidal field}

    \titlerunning{The probing power of $\gamma$-Dor's inertial dip to core magnetism}

   \author{L. Barrault\inst{1} \orc{0009-0007-7748-900X}
          \and
          L. Bugnet\inst{1}\orc{0000-0003-0142-4000}
          \and
          S. Mathis\inst{2}\orc{0000-0001-9491-8012}
          \and
          J.S.G. Mombarg\inst{2}\orc{0000-0002-9901-3113}
          }

   \institute{Institute of Science and Technology Austria (ISTA), Am Campus 1, 3400 Klosterneuburg, Austria \\
            \email{lucas.barrault@ist.ac.at}
         \and
             Université Paris-Saclay, Université Paris Cité, CEA, CNRS, AIM, 91191, Gif-sur-Yvette, France\\
             }

   \date{Received XXX}

 
  \abstract
   {$\gamma$-Dor stars are ideal targets for studies of stellar innermost dynamical properties due to their rich asteroseismic spectrum of gravity modes. Integrating internal magnetism to the picture appears as the next milestone of detailed asteroseismic studies, for its prime importance on stellar evolution. The inertial dip in prograde dipole modes period-spacing pattern of $\gamma$-Dors stands out as a unique window on the convective core structure and dynamics. Recent studies have highlighted the dependence of the dip structure on core density stratification, contrast of the near-core Brunt-Väisälä frequency and rotation rate, as well as the core-to-near-core differential rotation. In the meantime, the effect of envelope magnetism has been derived on low-frequency magneto-gravito-inertial waves.}
   {We aim to revisit the inertial dip formation including core and envelope magnetism, and explore the probing power of this feature on dynamo-generated core fields.}
   {We consider as a first step a toroidal magnetic field with a bi-layer (core and envelope) Alfvén frequency. This configuration allows us to revisit the coupling problem using our knowledge on both core magneto-inertial modes and envelope magneto-gravito-inertial modes. Using this configuration, we can stay in an analytical framework to exhibit the magnetic effects on the inertial dip shape and location. This configuration allows to set up a laboratory towards the comprehension of magnetic effects on the dip structure.}
   {We show a shift of the inertial dip towards lower spin parameter values and a thinner dip with increasing core magnetic field's strength, quite similar to the signature of differential rotation. The magnetic effects become sizeable when the ratio between the magnetic and the Coriolis effects is large enough. We explore the potential degeneracy of the magnetic effects with differential rotation. We study the detectability of core magnetism, considering both observational constraints on the periods of the modes and potential gravito-inertial mode suppression.}
   {}


   \maketitle
%
\defcitealias{Barrault2025ConstrainingDips}{BMB25}
\defcitealias{Tokuno2022AsteroseismologyOscillations}{TT22}

\section{Introduction}

Magnetic fields can be considered as an ubiquitous stellar property, playing a key role in stellar dynamics at each evolutionnary stage \citep{Mestel1984TheoryStars,Donati2009MagneticStars,Brun2017MagnetismConnection}. Yet, getting a coherent picture of magnetic stellar evolution is of an utmost difficulty, as it is in its essence a multi-dimensional process, comprising a wide range of lengths,  time scales, and generation mechanisms \citep{Maeder2009PhysicsStars,Mathis2013TransportInteriors,Aerts2019AngularInteriors}. 
Getting a deeper understanding of the evolution of internal magnetic fields and evaluating the relative weights of each formation and survival scenario is a major task of modern stellar physics, and could bring invaluable inputs on a wide number of topics: internal angular momentum transport \citep[e.g.][]{Eggenberger2005StellarFields,Eggenberger2008TheCode,Cantiello2014ANGULARSTARS,Eggenberger2019RotationInteriors,Takahashi2021ModelingEvolution,Moyano2023AngularPulsators,Moyano2024AngularStars} and chemicals distribution \citep{Eggenberger2010EffectsStars,Eggenberger2022TheAbundances}, stellar age estimation \citep{Keszthelyi2019TheBraking,Keszthelyi2020TheStars}, compact objects formation \citep{Heger2005PresupernovaFields,Suijs2008WhiteModels,Lebreton2014AsteroseismologyWeighing,Petit2017MagneticHoles}, gyrochronology \citep{Barnes2003OnGyrochronology,Meynet2011MassiveBraking,Reville2016AGESTUDY} to name a few. \\
\indent Asteroseismology stands out as a unique way to probe internal magnetic fields throughout stellar evolution. Alteration of the frequencies \citep{Gomes2020CoreStars,Mathis2021ProbingMagneto-asteroseismology,Bugnet2021MagneticFrequencies,Bugnet2022MagneticFrequencies,Li2022MagneticStars,Dhouib2022DetectingField,Lignieres2024PerturbativeModes,Rui2024AsteroseismicStars} or the amplitudes \citep{Fuller2015AsteroseismologyStars,Lecoanet2017ConversionWaves,Rui2023GravityFields} by the action of the Lorentz force as an additional restoring force for the oscillations, allows to retrieve pieces of information about the magnetic field intensity, geometry and topology. These theoretical breakthroughs have led to measurements at the red giant branch (RGB) stage of magnetic field strenghts from mixed-mode asymmetries in an intermediate field amplitude regime \citep{Li2022MagneticStars,Li2023InternalAsteroseismology,Deheuvels2023StrongSpacings,Hatt2024AsteroseismicGiants}, where the asteroseismic probes are sensitive to the H-burning shell region \citep{Li2022MagneticStars,Bhattacharya2024DetectabilityStars,Das2024UnveilingStars}. In a strong field regime, lower-limits for the internal field strength have been measured from mode suppression \citep{Garcia2014StudyModes,Stello2016SuppressioniKepler/i,Stello2016AStars}. 
From this magnetic revolution on the RGB, it is now up to theorists to develop seismic probes sensitive to internal magnetism, at various evolutionary stages and for the different types of stellar regions, to get a dynamic view of the evolution of stellar magnetism across the Hertzprung-Russel diagram. \\
\indent For these analyses to be pursued, intermediate-mass main sequence (MS) stars showing gravity-mode pulsations (hereafter $g$-mode pulsators) stand out as unique targets. The asteroseismic frequency spectrum of these MS stars is very rich and probes the inner regions of the star. In addition, they are the progenitors of the RGB stars for which an internal magnetic measurement is now available \citep{Li2022MagneticStars,Deheuvels2023StrongSpacings,Li2023InternalAsteroseismology,Hatt2024AsteroseismicGiants}. For $g$-mode pulsators, a unique observable is the period-spacing pattern (hereafter PSP), the period spacing between modes of consecutive radial orders but same horizontal structure, varying with the period of the modes. In the classical, non-rotating asymptotic theory, the period-spacing is known to be constant \citep{Tassoul1980AsymptoticPulsations}, resulting in a flat horizontal line for the PSP. The analysis of the PSP has already provided unprecedented results for the measurement of rotation in the near-core region of the radiative envelope of $\gamma$-Dor \citep{VanReeth2015DetectingStudies,Ouazzani2017AStars,Christophe2018DecipheringStars} and SPB stars \citep{Papics2017SignaturesStars, Pedersen2021InternalModes}. Indeed, the PSPs of gravito-inertial (hereafter $g$-$i$) modes, which are $g$-modes modified by the Coriolis acceleration, show a slope whose value is linked to the rotation rate in this precise region in which the modes reach their highest sensitivity in a differentially rotating radiative envelope \citep{VanReeth2018SensitivityStars}. Combining these results to a measurement of the rotation at the surface layers, by means of an analysis of the $p$-modes in mixed $\delta$-Scuti-$\gamma$-Dor pulsators \citep{Kurtz2014Asteroseismic11145123,Saio2015Asteroseismic9244992}, or rotational spot modulation \citep{VanReeth2018SensitivityStars}, one can now access near-core-to-surface differential rotation in the radiative zone. Differential rotation was proven to be limited, with a surface to near-core differential rotation ranging from 0.97 to 1.02 in the latter study. The most complete state-of-the-art sample of $\gamma$-Dor stars' PSPs was provided by \citet{Li2020Gravity-modeKepler}, comprising 611 stars analysed from the \textsl{Kepler} mission \citep{Borucki2010KeplerResults}. It contains a wide range of rotation rates and $g$-$i$ modes series of different angular degree $l$ and azimuthal order $m$, prograde Kelvin modes PSPs representing the most numerous part of them, complemented by the retrograde $r$-modes, and marginaly other types of modes, or modes for which the classification was inconclusive. \\
\indent As PSPs of $g$-$i$ modes corrected from the effect of the Coriolis acceleration in the frame co-rotating with the near-core region would result in a flat baseline in the standard asymptotic theory, any deviation from this pattern can hint towards a peculiar supplementary process, such as mode trapping, or mode coupling. On the former case, modulations in the PSPs were found to be a signature of strong thermal or chemical stratification gradients in the radiative zone (see e.g. \citeauthor{Miglio2008ProbingStars} \citeyear{Miglio2008ProbingStars}; \citeauthor{Cunha2019AnalyticalDiagram} \citeyear{Cunha2019AnalyticalDiagram}, \citeyear{Cunha2024BuoyancyRevisited} in a non-rotating case and \citeauthor{Bouabid2013EffectsStars} \citeyear{Bouabid2013EffectsStars} in the rotating one), and can be used as a probe of the transition region between the convective core and the envelope \citep{Michielsen2019ProbingAsteroseismology,Pedersen2021InternalModes}. On the latter, since the seminal work of \citet{Ouazzani2020FirstRevealed}, a dip structure in the PSP has been proven to result from the interaction of the envelope $g$-$i$ modes with core pure inertial modes restored only by the Coriolis acceleration in fast-rotating pulsators.\\
\indent This inertial dip has gained significant interest over the last few years, driven by its unprecedented probing power of the convective core of intermediate-mass MS stars. Its shape and location were proven to depend on (1) the core density stratification, (2) the near-core stratification profile, (3) the rotation rate of the near-core region, each of those parameters influenced by the age and mass of the pulsator \citep[see][for numerical computations]{Ouazzani2020FirstRevealed,Saio2021RotationModes,Galoy2024PropertiesStars}. \citeauthor{Tokuno2022AsteroseismologyOscillations} (\citeyear{Tokuno2022AsteroseismologyOscillations}, hereafter \citetalias{Tokuno2022AsteroseismologyOscillations}) provided a first analytical understanding of the interaction, later extended in Appendix D of \citet{Galoy2024PropertiesStars} to account for multi-mode interactions from both sides of the convective-radiative boundary. These works derived a Lorentzian shape of the dip in the PSP, characteristic of this coupling mechanism compared to periodic modulations created by strong gradients of thermal or chemical stratification \citep{Kurtz2014Asteroseismic11145123,Saio2015Asteroseismic9244992,Schmid2016Asteroseismic10080943B,Murphy2016Near-uniform7661054,Pedersen2018TheSpacings,Michielsen2019ProbingAsteroseismology,Li2019PeriodStars,Wu2020AsteroseismicCore}. As analytical works remained in the framework of solid-body rotation, and the numerical work of \citet{Saio2021RotationModes} showed a sensitivity of the dip location in the PSP to core rotation, we aimed in \citeauthor{Barrault2025ConstrainingDips} (\citeyear{Barrault2025ConstrainingDips}, hereafter \citetalias{Barrault2025ConstrainingDips}) to extend \cite{Tokuno2022AsteroseismologyOscillations}'s model to include a convective core to radiative envelope differential rotation, and finely investigate the variation of the dip structure, alongside with its location in the PSP. We investigated as well the potentiality of measuring this differential rotation from an inversion of the dip structure using our model in realistic \textsl{Kepler} data \citepalias{Barrault2025ConstrainingDips}. Results showed that for most of the values of near-core stratification inferred from a sample of 37 $\gamma$-Dor stars by \citet{Aerts2023ModeStars}, the convective core rotation would be retrieved in the regime where the convective core rotates faster than the radiative envelope, as the inertial dip is displaced towards low periods, in a region of the PSP less affected by the observational noise.\\
\indent Given the unprecedented sensitivity of the inertial dip to the convective core structure and dynamics, and the importance that a convective core magnetism measurement would bear on constraining the different scenarios of magnetic field generation, we investigate in this work the sensitivity of the inertial dip to magnetism, both in the convective core and in the near-core region. In the radiative zone, \citet{Dhouib2022DetectingField}, \citet{Lignieres2024PerturbativeModes} and \citet{Rui2024AsteroseismicStars} have investigated the effect of a magnetic field on the PSP of the $g$-$i$ modes, from now on referred to as magneto-gravito-inertial ($m$-$g$-$i$) modes because of their modification by the Lorentz force, with different magnetic topologies. All of those studies point towards an additional curvature in the PSP compared to the sole impact of rotation. We place ourselves in the framework of a toroidal field corresponding to a bi-layer Alfv\'en frequency, with two uniform values in the core and in the envelope. This framework, even if simplified compared to the complex magnetic configurations potentially present from both sides of the boundary \citep{Brun2005SimulationsAction, Featherstone2009EffectsStars, Augustson2016TheStars}, can be seen as a laboratory towards the fine comprehension of the effect of a magnetic field on the interaction between convective core magneto-inertial and radiative envelope $m$-$g$-$i$ modes. Building on the previous analytical hydrodynamical results of \citetalias{Barrault2025ConstrainingDips}, our framework allows us to exhibit the region to which each magnetic probe, i.e. the inertial dip and the curvature of PSP, is sensitive.\\
\indent The outline of this paper is as follows: in Section 2, we expose the generation mechanisms and the expected characteristics of magnetic fields in both the core and the envelope, and describe the models used in this work. In Section 3, we present the hypotheses and approximations required by our model, and recall the structure of the envelope and core oscillation modes in this context. In Section 4, we rewrite the coupling problem exposed in \citetalias{Tokuno2022AsteroseismologyOscillations} in our MHD framework and take profit of the analytical development of \citetalias{Barrault2025ConstrainingDips}, solving both numerically and analytically the coupling equation. We discuss our results in Section 5, focusing on a comparison between the purely hydrodynamical regime and the regime of field intensities potentially accessible by an analysis of the inertial dip. We then conclude in Section 6 on this new window on core magnetism, keeping in mind the simplifications and the hypotheses made in our model, and further giving leads for future studies on the inertial dips in the case of magnetic stars.

\section{Magnetic framework and modes description}

In this section, we first summarize theoretical elements on the theory of magnetic field generation and relaxation, both in the radiative envelope and in the convective core, focusing on the structure of the magnetic field awaited in different scenarios. We then describe the magnetic model and the assumptions we choose to adopt. We focus on our particular magnetic framework and related hypotheses in \ref{subsec:hyp_mag}, while we reserve considerations common to the hydrodynamical study \citepalias{Barrault2025ConstrainingDips} to section \ref{sec:modes}.

\subsection{Scenarios for the presence of magnetic fields in the radiative envelope of intermediate-mass MS stars}
\label{subsec:scenar}
Two main scenarios are generally considered for a generation of magnetic fields in the radiative zone of an intermediate-mass MS star evolving as a single star: an in-situ dynamo triggered by the Tayler-Spruit instability, or a fossil field originating from a past convective episode.
The Tayler-Spruit dynamo mechanism, originating from the seminal papers \citet{Tayler1973TheFIELDS}, \citet{Spruit1999DifferentialInteriors} and \citet{Spruit2002DynamoInterior}, originates from an interplay between differential rotation in a stellar radiative zone and magnetic fields. When not frozen by the Lorentz force, differential rotation generates a strong toroidal magnetic field, which becomes unstable towards the Tayler instability. This instability generates 3D motions of material, inducing an electromotive force which can sustain a dynamo mechanism in the stratified layer. This type of scenario has been extensively discussed in the literature since then, with different saturation hypotheses \citep{Braithwaite2006AStar,Zahn2007OnZones,Gellert2008HelicityRotation,Gellert2011HelicityFields,Fuller2019SlowingCores}. Tayler-Spruit-like mechanisms very efficiently transport angular momentum from the core to the envelope thanks to magnetic torques, partially explaining the spinning down of the cores of evolved stars, depending on the precise implementation \citep{Cantiello2014ANGULARSTARS,Fuller2019SlowingCores}. Recent 3D simulations (\citeauthor{Petitdemange2023Spin-downLayers} \citeyear{Petitdemange2023Spin-downLayers, Petitdemange2024TaylerSpruitLayers} for MS stars, \citeauthor{Barrere2023NumericalProto-magnetars} \citeyear{Barrere2023NumericalProto-magnetars} for proto-magnetars)  show a remarkable versatility of the dynamo settlement from MHD instabilities among different regimes of diffusion and stratification. They point towards the generation of a large toroidal magnetic field localized in the deep radiative zone, while reaching an intensity at the surface compatible with the low surface magnetic fields intensities found in 90 \% of early-type stars \citep{Petit2010TheVega,Blazere2016DetectionLeonis}.\\
\indent The fossil field scenario is the second main candidate for the settlement of a large scale magnetic field in stellar radiative zones. Due to the small magnetic diffusivity in stellar interiors, a field generated by a dynamo mechanism in a past convective layer could then relax into a stable, large-scale field in the newly radiative region \citep[e.g.][]{Arlt2011AmplificationStars,Emeriau-Viard2017OriginChanges}, and contribute to the efficient angular momentum transport in the radiative zone. Pure toroidal of poloidal configurations have been demonstrated to be unstable \citep{Tayler1973TheFIELDS,Markey1973TheFIELDS,Braithwaite2006AStar,Braithwaite2007TheStars}. A number of works have computed stable magnetic configurations either analytically \citep{Broderick2007MagneticFields,Lyutikov2010StructureStars,Duez2010ONSTARS,Akgun2013StabilityTreatment} or numerically \citep{Braithwaite2004ADwarfs,Braithwaite2006StableInteriors,Kaufman2022TheFields,Becerra2022StabilityStars}.\\
\indent A generation of a dynamo-originated stochastic field resulting in a fossil field can occur at various stages of stellar evolution: during the pre main sequence (PMS) for all stars, and in the core of intermediate-mass MS stars for stars with a mass of $\mathrm{M} \gtrsim 1.1 \mathrm{M_{\odot}}$. In the fossil field scenario, the radial magnetic field strengths now measured in RGB stars (we refer the reader to \citeauthor{Li2022MagneticStars} \citeyear{Li2022MagneticStars}; \citeauthor{Deheuvels2023StrongSpacings} \citeyear{Deheuvels2023StrongSpacings} and \citeauthor{Hatt2024AsteroseismicGiants} \citeyear{Hatt2024AsteroseismicGiants} for measurements of field of intermediate amplitude and to \citeauthor{Stello2016AStars} \citeyear{Stello2016AStars} for lower limits of a strong field) would be the result of these past convective episodes. \\
\indent The two scenarios described here can compete with each other, leading to a magnetic dichotomy: differential rotation triggering the TS instability and a dynamo-generated field in stellar radiation zones would be allowed by the presence of a low-amplitude field, whereas a strong pre-existing field would flatten the radial rotation gradient and favour a relaxation in a stable fossil field \citep{Spruit1999DifferentialInteriors}. In this regard, strong differential rotation and strong fossil magnetic fields are antagonists \citep[see][for works tackling this magnetic dichotomy]{Moss1982MagneticEnvelope,Auriere2007WeakStars,Gaurat2015EvolutionZone,Jouve2020InterplayZone}.

\subsection{Characteristics of core dynamo and further evolution in radiative zones}

The characteristics of dynamos are accessible through 3D MHD simulations of core convection. \citet{Lecoanet2023MultidimensionalConvection} listed the different codes currently available and their own specificities. Core dynamos departs from dynamo of convective envelopes in lower-mass stars by their different regimes: the magnetic Prandtl number ($\mathrm{Pm} = \nu/\eta$, with $\nu$ the kinematic viscosity and $\eta$ the magnetic diffusivity) is high for core dynamo and low for envelope one, which changes the prevalence of lengthscales, with more energy for large-scale structures in the case of the envelope dynamo \citep[see Fig.1 in][]{Augustson2019RossbyDynamos}. Furthermore, the kinetic energy is higher for a convective core than for a convective envelope, due to the increased density. Additionnaly, a convective core is almost adiabatic, whereas a convective envelope displays a superadiabatic gradient.\\
\indent Rotation has a strong influence on core convection, hence dynamo. Convective motions that would mainly be dipolar in the non-rotating case are organized in large columnar structures with rotation, with lengthscales perpendicular to the axis of rotation much smaller than parallel ones \citep{Davidson2013TurbulenceFluids}. This can be seen at first with an argument based on the Taylor-Proudmann theorem, and it has been observed in realistic simulations (\citeauthor{Brun2005SimulationsAction} \citeyear{Brun2005SimulationsAction}; \citeauthor{Featherstone2009EffectsStars} \citeyear{Featherstone2009EffectsStars} for a $2.0 \, \rm M_{\odot}$ star, \citeauthor{Augustson2016TheStars} \citeyear{Augustson2016TheStars} for a $10 \, \rm M_{\odot}$ star). Importantly, the level of equipartition of the magnetic field energy compared to the kinetic energy has been proven to depend on the Rossby number ($\mathrm{Ro} = V_{\rm conv}/2\Omega L_{\rm conv}$, $V_{\rm conv}$ and $L_{\rm conv}$ being respectively a characteristic velocity and a lengthscale of convection, and $\Omega$ the rotation rate). At low rotation, hence high Rossby number, an equipartition is found, whereas a magnetostrophic regime with a superequipartition state is observed for high rotation rates, or low Rossby numbers. \citet{Augustson2019RossbyDynamos} compiled several results from MHD simulations with a range of Rossby numbers and confirmed this enhancement of magnetic energy compared to kinetic energy at a low Rossby number regime (we refer the reader to their Figure 4). \\
\indent Another key question is if this core dynamo could establish a large scale structure for the magnetic field as in the case of the solar dynamo, and what would be the geometry of such a field. \citet{Featherstone2009EffectsStars} and \citet{Augustson2016TheStars} found increased magnetic energy along large-scale columnar structures of the velocity field caused by strong rotation. Interestingly, \citet{Augustson2016TheStars} found a mean magnetic energy of the toroidal component approximately fifty times higher compared to the one of the poloidal component (see their Fig. 9).\\
\indent This dynamo-generated field, advected by rising material that mixes the core boundary, extends to the neighbouring radiative zone, and stratification alters its characteristics. \citet{Brun2005SimulationsAction} and \citet{Featherstone2009EffectsStars} agree on a large ribbon of toroidal field at this location. This was further seen in the recent work of \citet{Ratnasingam2024OnStars}, keeping in mind the higher mass of the modelled star ($\rm M= 7 \, \rm M_{\odot}$). The zone at which the Brunt-Väisälä profile peaks is a shear layer that produces strong toroidal fields by an $\Omega$-effect. \\
\indent The transition from these dynamo fields displaying a broad distribution of length scales to large-scale magnetic configuration on the RGB has been tackled in \citet{Braithwaite2006StableInteriors}, \citet{Cantiello2016ASTEROSEISMICFIELDS}, \citet{Bugnet2021MagneticFrequencies} and \citet{Becerra2022EvolutionStars} and their long-term stability questioned \citep{Kaufman2022TheFields}. Interestingly, fields coming from different origins or stellar stages can interact in a highly non-linear way. As investigated in \citet{Featherstone2009EffectsStars}, a case in which an input fossil field is present around the core of an A-type star shows a state of super-equipartition of the magnetic-to-kinetic energy for its core dynamo. The magnetic field reaches a strength of several hundreds of kG.\\

\subsection{Modelled star and modes considered from both sides of the boundary}
\begin{figure}
    \centering
    \includegraphics[width=\linewidth]{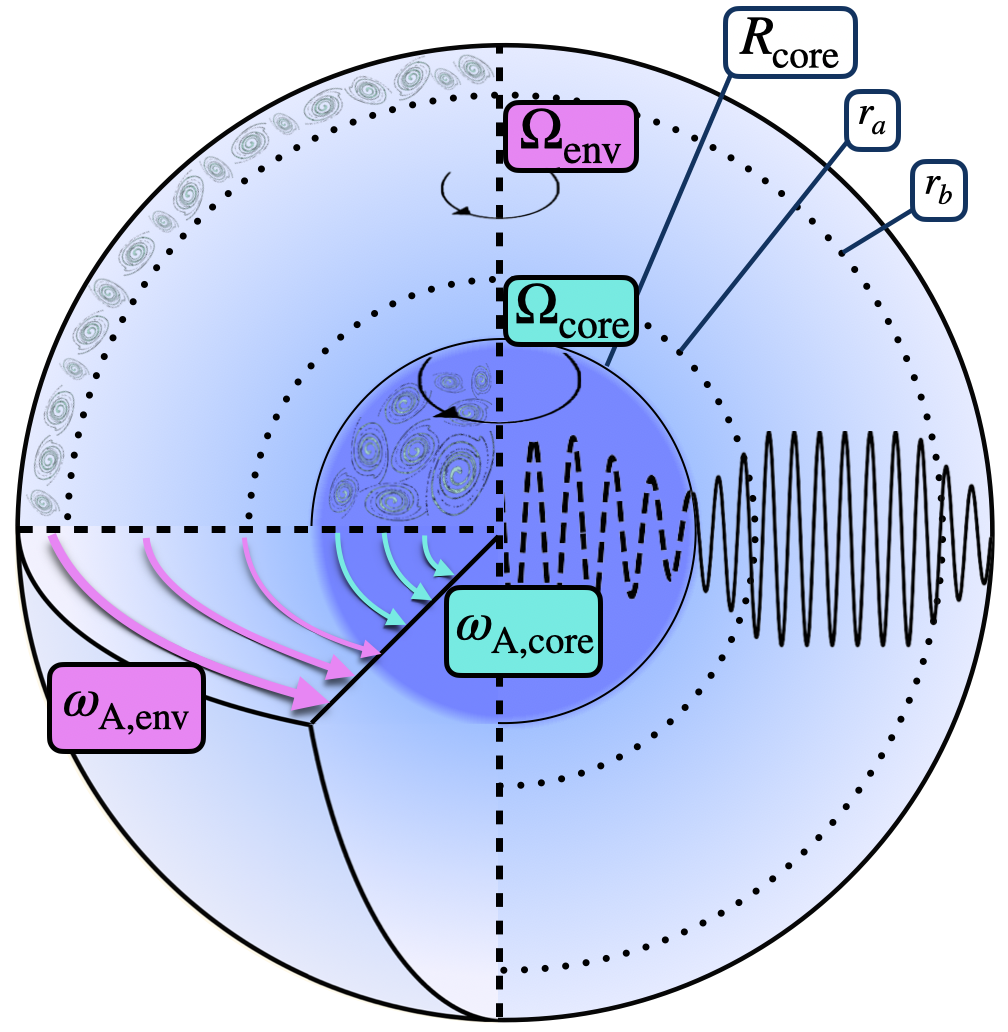}
    \caption{Magnetic star with a bi-layer Alfvén frequency, $\omega_{\rm A, core}$ in the core, $\omega_{\rm A, env}$ in the envelope, and a bi-layer rotation rate, $\Omega_{\rm core}$ in the core, $\Omega_{\rm env}$ in the envelope. The cavity for $m$-$g$-$i$ modes lies between $r_{\rm a}$ and $r_{\rm b}$ in the radiative zone. They get an evanescent character in the region $[R_{\rm core};r_{\rm a}]$ when applying the TARM. Magneto-inertial modes propagate in the convective core below the location $R_{\rm core}$.}
    \label{fig:sketch_mag}
\end{figure}

The present study, as well as previous ones concerning the inertial dip in the PSP, would apply to any intermediate to fast-rotating star presenting a structure with a convective core surrounded by a radiative zone in which $g$-$i$ modes can propagate (see Fig.\ref{fig:sketch_mag}). From an observational point of view, the spectrum of $g$-$i$ modes must also contain many modes, so that the period at which the interaction studied would occur is comprised in the PSP, and the inertial dip can be analysed \citep{Saio2021RotationModes}.\\
\indent This is classically the case in two classes of pulsators: $\gamma$-Dor and SPB stars. 
In this study, we focus on the case of $\gamma$-Dor stars, as (1) their PSPs comprise a wider extent of radial orders compared to SPBs and have been already used to infer radiative zone properties with a great precision, allowing for more in-depth studies comprising the inertial dip and (2) the absence of an extended convective envelope inhibits magnetic braking, thus $\gamma$-Dor stars are in general intermediate to fast rotators \citep{Aerts2024AsteroseismicAstrophysics}, rotating faster than SPBs \citep{Aerts2023ModeStars}.
This class of pulsators classically comprises zero-age main sequence (ZAMS) stars of 1.5 $\rm M_{\odot}$ to terminal-age main sequence (TAMS) stars of 2.0 $\rm M_{\odot}$. However, the recent analyses of thousands of targets from \textsl{Gaia} \citep{Prusti2016TheMission} data suggests that g-mode pulsators and especially $\gamma$-Dor type ones span accross a much more extended region of the Hertzprung-Russel diagram \citep[See Fig.5 of][]{DeRidder2023iGaia/i3}.\\
\indent We consider for our study 3 models computed with MESA \citep[version 23.05.1,][]{Paxton2011MODULESMESA,Paxton2013MODULESSTARS,Paxton2018ModulesExplosions,Paxton2019ModulesConservation,Jermyn2023ModulesInfrastructure} in \citet{Mombarg2024ProbabilityStars}\footnote{The inlists used can be found on \url{https://zenodo.org/records/10629035}}, with different rotation rates and age. We retain an intermediate value for the overshooting parameter of $f_{\rm ov} = 0.02$ in an exponentially-diffusive prescription \citep{Freytag1996HydrodynamicalSun} and a solar metallicity $Z = 0.014$ \citep{Asplund2009TheSun}. \\
\indent To choose the models, we first consider that the frequencies in the dip region of the PSP must not suffer from high uncertainties due to the finite observing time. Second, the rotation rate at the surface $\Omega_{\rm surf}$ must not be too high compared to the surface critical rotation rate $\Omega_{\rm crit}$ for TAR calculations to hold. \citet{Mathis2019TheStars} found this limit to be 40\% of the surface critical rotation rate, while \citet{Dhouib2021ThePlanets,Dhouib2021ThePlanetsb} considered a more conservative limit of 20\%. The latter limit appears as the most reliable, since it relies on non-perturbative calculations based on 2D stellar models computed with ESTER \citep{EspinosaLara2013Self-consistentStars}. However, the impact of the centrifugal force is small on the structure of $g$-$i$ modes near the core, and the deviation in fast-rotating stars from the frequencies obtained with the TAR was proven to be small: as argued in \citet{Dhouib2021ThePlanets}, current uncertainties on e.g. rotational mixing and atomic diffusion would mask the effect of centrifugal deformation. 
We thus choose to consider stars rotating up to 40 \% of their critical surface rotation rate, keeping in mind the potential improvement of this model to the TAR in deformed stars. \\
\indent We retain two ZAMS models (central H fraction $X_{\rm H} = 0.70$) of $1.5 \, \rm M_{\odot}$ stars, with rotation rates $\Omega/2\pi = 2.29 \, \mathrm{c.d^{-1}}$ (hereafter $fz$ model) and $1.22 \, \mathrm{c.d^{-1}}$ (hereafter $iz$ model), corresponding to 40 and 20\% of their critical rotation rate, respectively. These rotation rates correspond approximatively to the minimum and maximum rotation rates in the sample of $\gamma$-Dor harbouring inertial dips analyzed by \citet{Saio2021RotationModes}.
As for older and higher-mass stars, the critical rotation rate decreasing with evolution, and the star breaking during the MS, we retain one model of a 1.8 $\rm M_{\odot}$ with $X_{\rm H} = 0.30$, rotating at $1.14 \rm \, c.d^{-1}$ corresponding to 42 \% of the critical rotation rate (hereafter $im$ model). A list of the relevant physical quantities used in the present study for each model can be found in Table~\ref{tab:models}.\\
\begin{table}[]
    \centering
    \begin{tabular}{|p{2.8cm}|p{1.5cm}|p{1.5cm}|p{1.5cm}|}
    \hline
         model name & $fz$ &  $iz$ & $im$\\
         \hline
        $X_{\rm H}$ &  0.70&  0.70& 0.30\\
        $\Omega_{\rm nc}/2\pi$ ($\rm c.d^{-1}$)&  2.25 & 1.22 & 1.14 \\
        $\Omega_{\rm surf}/\Omega_{\rm crit} \, (\%)$ &  40 &  20& 42\\
        $\Omega_{\rm nc}/\Omega_{\rm crit}\, (\%)$ &  3.6 &  2.0& 2.4\\
        $\Pi_{0} \, (s)$ & 4609 & 4623 & 4670\\
        $N_{\rm max}/2\pi \, (\mu \rm Hz)$ & 544 & 496 & 1027\\
        $\bar{N}/2\pi \, (\mu \rm Hz)$ &  293 & 296 & 246\\
        $R_{\rm core} \, (R_{\odot})$ & 0.141 & 0.142& 0.165\\
        $\bar{\rho}|_{R_{\rm core}} \, (\rm g.cm^{-3})$ & 64.7& 64.5& 53.9\\
        $\bar{\rho}|_{R_{\rm core}/2} \, (\rm g.cm^{-3})$ & 79.9 & 79.8 & 75.4\\
        $\mathrm{B_{ms}}|_{R_{\rm core}/2} \, (\rm MG)$ &  1.40& 1.03 & 1.63\\
        $\mathrm{B_{equi}}|_{R_{\rm core}/2} \, (\rm kG)$ &  75& 76 & 151\\
        $\mathrm{Ro}|_{R_{\mathrm{core}}/2}$&  $2\times 10^{-4}$& $3.7 \times 10^{-4}$ & $9\times 10^{-4}$ \\
        \hline
    \end{tabular}
    \caption{Relevant quantities used throughout the study for the three considered considered. $f$ stands for fast rotation, $i$ for intermediate rotation, $z$ for ZAMS and $m$ for mid-MS. We refer the reader to Appendix~\ref{Appendix:models} for their definition.}
    \label{tab:models}
\end{table}
\indent $g$-$i$ modes are of four different types: Poincar\'e, r-modes, Yanai and Kelvin. Descriptions of those modes are given in \citet{Townsend2003AsymptoticStars} and \citet{Mathis2008AngularWaves}. In the magnetic context, we refer to subsection 4.2 of \citet{Mathis2011Low-frequencyField}. In the context of a differentially-rotating envelope, each type of $g$-$i$ mode reach their highest sensitivity to a different depth of the star \citep{VanReeth2018SensitivityStars}. \\

\indent Kelvin modes are of particular interest because of their high occurrence rate in the most up-to-date $\gamma$-Dor sample observed by \textsl{Kepler} \citep{Li2020Gravity-modeKepler}, due to their high visibility. They possess no equatorial node, their angular degree $l$ equating the azimuthal number $m$. They hence benefit from low surface cancellation. They exist in both the sub-inertial and super-inertial regimes. We focus on Kelvin modes in our study, as they propagate in the sub-inertial regime in the convective core \citep{Prat2018AsymptoticRotation} and become magneto-inertial ($m\text{-}i$) modes, as buoyancy is no longer a restoring force.
The geometry of the stellar layer in which $m\text{-}i$ modes propagate holds a great influence on their properties. In a non-differentially rotating full sphere, in an inviscid and incompressible framework, the configuration that we are interested in, the spectrum is dense in the interval $[-2\Omega,2\Omega]$, $\Omega$ being the rotation rate of the considered zone.

\subsection{Magnetic configuration and the Traditional Approximation of Rotation and Magnetism}\label{subsec:hyp_mag}
\begin{figure}
    \centering
    \includegraphics[width=\linewidth]{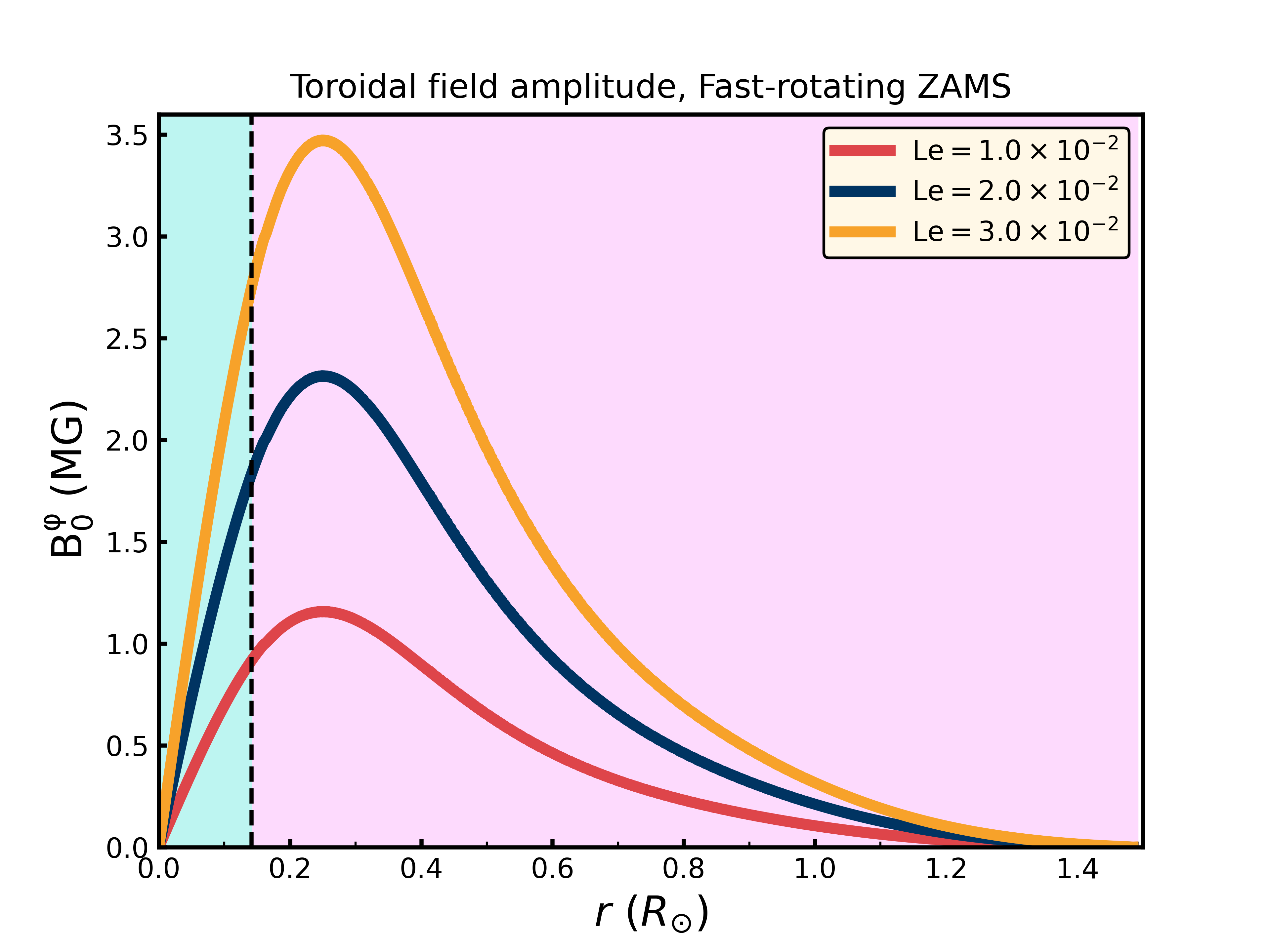}
    \caption{Magnetic field profiles of uniform Alfv\'en frequency, taking the 1.5 $\rm M_{\odot}$ $fz$ model, at the equator. As an example, a magnetic field profile with a bi-layer Alfv\'en frequency corresponding to $\rm Le_{core} = 1.0\times 10^{-2}$ and $\rm Le_{env} = 2.0\times 10^{-2}$ would follow the blue profile in the core (turquoise zone) and a red profile in the envelope (pink zone).}
    \label{fig:field-amp}
\end{figure}

\begin{figure*}
    \centering
    \includegraphics[width=1\linewidth]{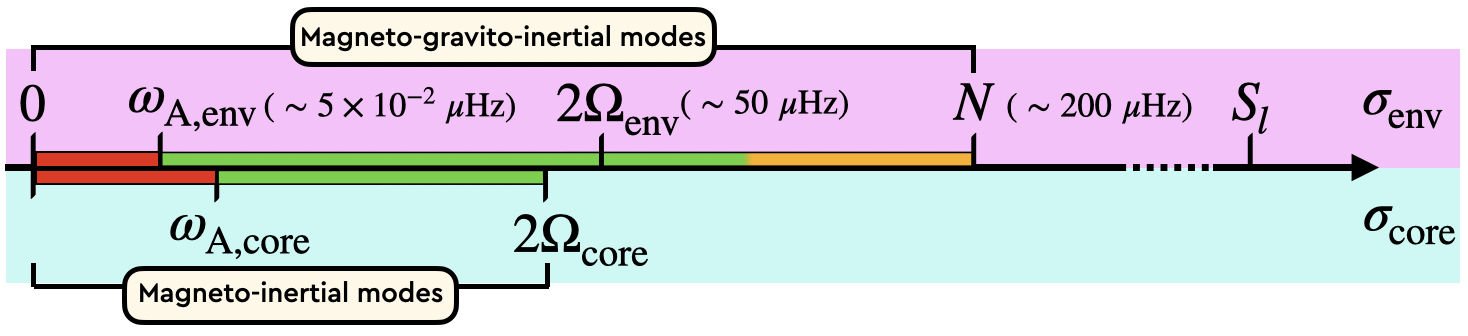}
    \caption{Hierarchy of the characteristic angular frequencies hypothesized in this work. Quantities related to the envelope are specified in the top panel, while quantities related to the core are noted in the bottom panel. Values for the linear frequencies for the $fz$ model are denoted in parentheses. An Alfvén frequency of $\omega_{\rm A,env}\approx5\times10^{-2}\mu\rm Hz$ corresponds to an equatorial field strength of $\rm B_{0}^{\varphi}\approx 9\times 10^{4}\rm G$ at the bottom of the radiative zone for the three considered models.
    }
    \label{fig:hierarchy}
\end{figure*}

We choose an azimuthal axi-symmetric magnetic field (see Fig.~\ref{fig:sketch_mag}) with a bi-layer Alfv\'en angular frequency $\omega_{\rm A} = \mathrm{B}_{0}^{\varphi}/(\sqrt{\mu_{0}\bar{\rho}}r\sin \theta)$, such that $\omega_{\rm A} = \omega_{\rm A, core}$ in the convective core and $\omega_{\rm A} = \omega_{\rm A, env}$  in the radiative envelope, both assumed to be uniform in their respective regions (see Fig.\ref{fig:field-amp}), with $\mathrm{B}_{0}^{\varphi}$ the toroidal background field and $\bar{\rho}$ the background hydrostatic density. We use $\mu_{0}$ the magnetic permeability of the vacuum, corresponding to the one of the plasma in stars. This framework allows us to understand the respective effects of rotation or magnetism on the dip formation picture, building on the previous analytical works of \citet{Malkus1967HydromagneticWaves} for the convective core and \citet{Mathis2011Low-frequencyField} for the radiative envelope. Even though being simplified compared to the topology and geometry of magnetic fields hypothesized in both regions, this configuration retains a toroidal component present in both the scenarios of relaxed fossil field, and dynamo-generated field near the core by a Tayler-Spruit like mechanism (see section~\ref{subsec:scenar}). Additionally, we allow for different magnetic fields amplitudes from both sides of the boundary, as different magnetic fields formation mechanisms are at play in the two zones and can lead to significantly dissimilar amplitudes \citep{Featherstone2009EffectsStars}. We will point out the key importance of adopting this field profile to maintain the analytical description of core modes in subsection~\ref{subsec:core_modes}, while describing ways to consider more realistic profiles in subsection~\ref{subsec:limit}.\\
We adopt the Traditional Approximation of Rotation and Magnetism in the radiative envelope: this approximation was introduced in \citet{Mathis2011Low-frequencyField} and later used in \citet{Dhouib2022DetectingField} and in \citet{Rui2024AsteroseismicStars}. It consists, in a highly stratified medium where $[\omega_{\rm A},2\Omega] \ll N$, with $N$ the Brunt-Väisälä (angular) frequency such that $N^{2} = -\bar{g}(\mathrm{d}\ln\bar{\rho}/\mathrm{d}r-1/\Gamma_{1}\cdot\mathrm{d}\ln\bar{P}/\mathrm{d}r)$ with $\bar{g}$ the background self-gravity, $\bar{P}$ the background gaseous pressure and $\Gamma_{1}$ the first adiabatic exponent, to only retain the transverse component of both the Coriolis acceleration and the Lorentz force, as transverse displacement are prominent compared to radial ones in such regimes. As in \citetalias{Barrault2025ConstrainingDips}, we choose a bi-layer rotation profile, $\Omega_{\rm core}$ in the convective core, $\Omega_{\rm env}$ in the radiative envelope. We provide a graphical summary of the hierarchy of relevant frequencies for both the radiative envelope (upper part) and the convective core (lower part) in Fig.\ref{fig:hierarchy}. In each part, a green color bar indicates the considered local angular wave frequencies interval.\\
 \indent We verify that this regime of strong stratification is attained even for the fast rotators considered here, showing the typical values of Brunt-Väisälä, rotation and Alfv\'en frequencies for the $fz$ model in parenthesis on the upper panel of Fig.~\ref{fig:hierarchy}. We point out in this figure an orange zone in the frequency domain for which an envelope $g$-$i$ mode of this local wave frequency would deviate appreciably from the TARM, and we refer to \citet{Dhouib2022DetectingField} for a discussion of this limit value of the rotation rate compared to $N$ (see their table 1).\\
\indent We emphasize that we consider a regime of fields of intermediate strengths, high enough for a perturbative treatment of the Lorentz force not to be applicable \citep{Rui2024AsteroseismicStars}, but lower than a regime of magnetic mode suppression ($|m|\omega_{\rm A, zone} > \sigma_{\rm zone}$). We will discuss this further in Section~\ref{sec:discuss}.\\

\section{Expressions of mode structures from both sides of the boundary}\label{sec:modes}

In this section, we first set up in \ref{subsec:eq} the MHD framework that we are using for both the convective core and the radiative envelope. We further recall the results of \citet{Mathis2011Low-frequencyField} on envelope $m$-$g$-$i$ modes in \ref{subsec:mdb}, then derive the structure of these modes at the location of the convective core - radiative envelope boundary by means of a JWKB analysis. As in \citetalias{Tokuno2022AsteroseismologyOscillations}, calculations differ if we consider a continuity or discontinuity of the $N$ profile at the core-to-envelope boundary. We further give the structure of core $m\text{-}i$ modes in \ref{subsec:core_modes}, exploiting the results of \citet{Malkus1967HydromagneticWaves}. Relevant parameters and notations are gathered and defined in Appendix ~\ref{App:notations}.\\

\subsection{System of equations in the ideal MHD framework}\label{subsec:eq}

As in \citetalias{Barrault2025ConstrainingDips}, we place ourselves in a sub-inertial regime for which the local angular wave frequencies in the two zones are inferior to the inertial frequencies $(2\Omega_{\rm core},2\Omega_{\rm env})$. $m\text{-}g\text{-}i$ modes from the envelope can propagate in the core as $m\text{-}i$ modes in this framework.\\
We adopt an hypothesis of anelasticity in both the convective core and the radiative envelope. This filters out the high-frequency acoustic waves. As we are interested in low-frequency waves, the anelasticity is valid in the quasi-entire part of the propagating cavity, and especially near the boundary where the modified Lamb frequency $\tilde{S}$ is much higher than the frequency of the mode, where $\tilde{S}^{2} = \Lambda_{k}^{m}c_{S}^{2}/r^{2}$, with $c_{S}$ the sound speed and $\Lambda_{k}^{m}$ the eigenvalue of the Laplace Tidal Equation, defined further in Section~\ref{subsec:eq}. This can be questionable near the outer boundary of the cavity $r_{\rm b}$, as $\tilde{S}$ can be inferior to $N$ there. For the low-frequency modes we are interested in, we check that we still have $N<\tilde{S}$ at $r_{b}$ (see Fig.~\ref{fig:prop-diag}). This would not change fundamentally the results near the convective core, and would leave an imprint as a curvature of the PSP in the co-rotating frame, which is caused by a receding outer boundary $r_{b}$ of the $m\text{-}g\text{-}i$ modes propagation cavity with increased mode frequency \citep[See Appendix A of][]{Tokuno2022AsteroseismologyOscillations}. We use as well the Cowling approximation \citep{Cowling1941TheStars}, as the series of Kelvin modes found in $\gamma$-Dors are of a high radial order \citep{Li2020Gravity-modeKepler}. Additionally, these modes have a strongly oscillating character near the core \citep[see Fig.6 of][]{Galoy2024PropertiesStars}. \\
We assume adiabaticity in the whole region of propagation of both types of modes. We thus neglect heat, Ohmic and viscous diffusions. \\
We neglect the centrifugal force, as the rotation rate of the most internal layers of the radiative zone is negligible compared to the critical rotation rate (it is inferior to $4 \%$ of this critical rotation in the $fz$ model; see Appendix~\ref{Appendix:models}). All the more, we neglect the indirect effect of the magnetic field on the hydrostatic equilibrium, as $\omega_{\rm A, zone} <\Omega_{\rm zone}$ (see Appendix~\ref{App:hierarchy}). The gaseous pressure is more significant than the magnetic pressure in such internal layers: in our $fz$ model, the magnetic pressure is $\sim10^{6}$ times lower than the gaseous pressure for a magnetic field of $1\, \mathrm{MG}$.\\
\indent With these hypotheses, we have to consider the same system of equations as \citet{Mathis2011Low-frequencyField}:\\
First, the ideal induction equation:
\begin{equation}
    \partial_{t} \mathbf{B} = \nabla \times (\mathbf{V} \times \mathbf{B}) \, ,
\end{equation}
Second, the inviscid momentum equation:
\begin{equation}
  D_{t} \mathbf{V} = -\frac{1}{\rho}\boldsymbol{\nabla}{P} - \boldsymbol{\nabla}{\Phi} + \frac{1}{\rho}\left[\frac{1}{\mu_{0}}(\boldsymbol{\nabla} \times \mathbf{B}) \times \mathbf{B}\right]  \, ,
\end{equation}
Then, the continuity equation:
\begin{equation}
    D_{t}\rho+\rho\boldsymbol{\nabla}\cdot\mathbf{V} = 0 \, ,
\end{equation}
the heat transport equation in the adiabatic limit:
\begin{equation}
    \frac{1}{\Gamma_{1}}D_{t}\ln{P}-D_{t}\ln{\rho} = 0 \, .
\end{equation}
and the Poisson equation:
\begin{equation}
    \nabla^{2}\Phi = 4\pi \mathcal{G} \rho \, .
\end{equation}
is filtered out for the wave fluctuations because of the Cowling approximation.
We introduce $\boldsymbol{\mathrm{V}}$ the velocity field, $\boldsymbol{\mathrm{B}}$ the magnetic field, $\rho$ the density, $P$ the gaseous pressure, $\Phi$ the gravitational potential, and $\mathcal{G}$ the gravitational constant. We consider waves propagating in the bi-layer set-up presented in \ref{subsec:hyp_mag}. We introduce the background toroidal field $\mathbf{B_{0}^{\boldsymbol{\varphi}}}$ and zonal flow $\rm \mathbf{V}_{0}$. Linear perturbations $(\mathbf{b},\mathbf{u})$ around this large-scale quantities are considered, with:
\begin{equation}
    \mathbf{B}(\mathbf{r}, t) = \mathbf{B}_{0}^{\boldsymbol{\varphi}}(\mathbf{r}, t) + \mathbf{b}(\mathbf{r}, t) \, ,
\end{equation}
\begin{equation}
    \mathbf{V}(\mathbf{r}, t) = \mathbf{V}_{0}(\mathbf{r}, t) + \mathbf{u}(\mathbf{r}, t) \, .
\end{equation}
$t$ being the time and $\boldsymbol{r} = (r,\theta,\varphi)$ the spherical coordinates with the unit vectors \{$\widehat{\boldsymbol{e}}_{r},\widehat{\boldsymbol{e}}_{\theta},\widehat{\boldsymbol{e}}_{\varphi}$\}. \\
\noindent The large scale background quantities are:
\begin{equation}
    \mathbf{B}_{0}^{\boldsymbol{\varphi}} = \sqrt{\mu_{0} \bar{\rho}}r\sin\theta\omega_{\rm A, zone} \widehat{\mathbf{e}}_{\varphi}
\end{equation}
and  
\begin{equation}
    \mathbf{V}_{0} = r\sin\theta\Omega_{\rm zone} \widehat{\boldsymbol{e}}_{\varphi} \, ,
\end{equation}
\noindent with:
\begin{equation}
    \omega_{\rm A, zone} = \begin{cases} \omega_{\rm A,core} & \mbox{if } r < R_{\mathrm{core}} \\ \omega_{\rm A,env} & \mbox{if } r > R_{\mathrm{core}} \end{cases} \, ,
\end{equation}
and
\begin{equation}
\Omega_{\rm zone} = \begin{cases} \Omega_{\mathrm{core}} & \mbox{if } r < R_{\mathrm{core}} \\ \Omega_{\mathrm{env}} & \mbox{if } r > R_{\mathrm{core}} \end{cases} \, .
\end{equation}
We thus consider the rotation frequency and the Alfvén frequency to be discontinuous at the boundary between the core and the envelope as in Fig.~\ref{fig:sketch_mag}. We place ourselves in the interior of each zone, in which we treat both quantities as being uniform. As in \citetalias{Barrault2025ConstrainingDips}, we quantify differential rotation with the parameter $\alpha_{\rm rot}$ such that:
\begin{equation}
    \alpha_{\rm rot} = \Omega_{\rm env}/\Omega_{\rm core}
\end{equation}
\indent The method held to derive the momentum equation is similar to the one provided in \cite{Mathis2011Low-frequencyField} and detailed for consistency in Appendix \ref{App:deriv}. We only highlight the expansion of the relevant quantities here. All scalar fields $X \equiv (\rho, \Phi, P)$ are expressed as a sum of an hydrostatic term $\bar{X}$ and a fluctuation $\tilde{X}$, with:
\begin{equation}
    X(r,\theta,\varphi,t) = \bar{X}(r) + \tilde{X}(r,\theta,\varphi,t) \, .
\end{equation}
The scalar quantities $\tilde{X}$ and the vectorial fields $\boldsymbol{x}$ are expanded as:
\begin{equation}
\tilde{X} = \sum_{\sigma_{\rm in}, m}X'(r,\theta) e^{i(m\varphi + \sigma_{\rm in} t)} \, ,
\end{equation}
\begin{equation}
\boldsymbol{x} = \sum_{\sigma_{\rm in}, m}\boldsymbol{x}'(r,\theta) e^{i(m\varphi + \sigma_{\rm in} t)} \, ,
\end{equation}
with $\sigma_{\rm in}$ the wave angular frequency in an inertial frame. We also define as in \citetalias{Barrault2025ConstrainingDips} the Doppler-shifted local angular frequencies $\sigma_{\rm zone} = \sigma_{\rm in} + m\Omega_{\rm zone}$ and the spin parameters $s_{\rm zone} = 2\Omega_{\rm zone}/\sigma_{\rm zone}$. We adopt the convention of $m<0$ for prograde modes and $m>0$ for retrograde modes.\\
We further define magnetic local wave frequencies $\sigma_{\rm M, zone}$ such that $\sigma_{\rm M, zone}^{2} = \sigma_{\rm zone}^{2}-m^{2}\omega_{\rm A,zone}^{2}$ and magnetic spin parameters $s_{\rm M, zone} = 2\Omega_{\rm zone}/{\sigma_{\rm M,zone}}$, which reduce to the spin parameters in the absence of magnetic fields.\\
We define the magnetic pressure:
\begin{equation}
    P_{\rm M} = \frac{\mathbf{B}^{2}}{\mu_{0}}
\end{equation}
and its linear fluctuation:
\begin{equation}
    \tilde{P}_{\rm M} = \frac{\mathbf{B_{0}^{\varphi}}\cdot \mathbf{b}}{\mu_{0}} \,
\end{equation}
from which the total pressure fluctuation reads
\begin{equation}
\tilde{\Pi} = \tilde{P} +\tilde{P}_{\rm M} \, .
\end{equation}
We define the quantity
\begin{equation}
\tilde{W} = \frac{\tilde{\Pi}}{\bar{\rho}}+ \tilde{\Phi} \, .
\end{equation}
\citet{Mathis2011Low-frequencyField} derived the linearised momentum equation:
\begin{equation}
-\mathcal{A}\boldsymbol{\xi}' + i\mathcal{B}\widehat{\mathbf{e}}_{z} \times \boldsymbol{\xi}' = -\boldsymbol{\nabla} W' + \frac{\rho'}{\bar{\rho}^{2}} \boldsymbol{\nabla} \bar{P} - \Pi'\frac{\boldsymbol{\nabla}\bar{\rho}}{\bar{\rho}^{2}} \, ,
\label{eq:momentum_general}
\end{equation}
where:
\begin{equation}
    \mathcal{A} = \sigma_{\rm M,zone}^{2} = \sigma_{\mathrm{zone}}^{2} - m^{2}\omega_{\rm A,zone}^{2} \, ,
\end{equation}
\begin{equation}
    \mathcal{B}= 2(\Omega_{\mathrm{zone}}\sigma_{\mathrm{zone}} - m\omega_{\rm A,zone}^{2}) \, .
\end{equation}
This equation will be further used for the convective core and the radiative envelope. For a wave propagation in the presence of a magnetic field, we need $\mathcal{A} = \sigma_{\rm M}^{2} > 0$. The magnetic field acts as a filter for low-frequency waves, highlighting our need to work in the regime of an intermediate magnetic field so that $m$-$g$-$i$ modes propagate to the edge of the core. It is worth-noting that this limit is different from the one derived in the case of the \citet{Fuller2015AsteroseismologyStars} type mechanism. We will further discuss this in Section \ref{sec:discuss}.
We define the quantity
\begin{align}
    \nu_{\rm M,zone} = \mathcal{B}\mathcal{A}^{-1} = s_{\mathrm{zone}}\frac{1-2ms_{\mathrm{zone}}\mathrm{Le_{zone}}^{2}}{1-m^{2}\mathrm{Le_{zone}}^{2}{s_{\mathrm{zone}}}^{2}} \, ,
    \label{eq:def_nuzone}
\end{align}

\noindent where we have introduced the Lehnert number of the region:
\begin{equation}
    \mathrm{Le}_{\rm{zone}} = \frac{\omega_{\rm A,zone}}{2\Omega_{\rm zone}} \, ,
\end{equation}
which compares the relative strength of the Lorentz and Coriolis terms \citep{Lehnert1954MagnetohydrodynamicForce.}. We will see that the Lehnert number is the main quantity governing the effect of magnetic fields on the interaction between the convective core $m\text{-}i$ modes and the radiative envelope $m$-$g$-$i$ modes. Even though $\nu_{\rm M}$ was referred to as the magnetic spin parameter in \citet{Dhouib2022DetectingField}, we define it the magnetic structural parameter in this work, to avoid confusion with other quantities. \\
We illustrate in Fig.\ref{fig:field-amp} the background magnetic field profile obtained with different Lehnert numbers in the core or in the envelope for the fast-rotating ZAMS model. With a fixed Lehnert number, due to the density profile of $\gamma$-Dor stars, this field peaks in the radiative envelope, near the convective core. Due to its dependence in $r$, the field decreases approaching the center of the core, being null at the center.

\subsection{Envelope $m$-$g$-$i$ modes}\label{subsec:mdb}
\subsubsection{System of equations under the TARM}
In the framework of a strongly stratified envelope ($N \gg 2\Omega_{\rm env}$), the radial Lagrangian displacement is much smaller than the horizontal one. We can thus neglect the latitudinal component of the rotation vector, hence the radial component of the Coriolis acceleration. This allows us to operate a separation of the radial and the horizontal dynamics. This is known as the Traditional Approximation of Rotation (TAR), which has been extensively used in geophysics and astrophysics for high stratification regime \citep[see e.g.][]{Eckart1960HydrodynamicsAtmospheres,Bildsten1996OceanStars,Lee1997Low-frequencyDependence}. In a magnetic context, one can extend this approximation if $N \gg \omega_{\rm A, env}$, and the latitudinal component of the Lorentz force can also be neglected. \citet{Mathis2011Low-frequencyField} describes this so-called Traditional Approximation of Rotation and Magnetism (TARM).
Under the TARM, the horizontal structure of $m$-$g$-$i$ modes is treated by solving the Laplace tidal equation (LTE):
\begin{equation}
    \mathcal{L}_{\nu_{\rm M,env}}[\Theta_{k}^{m}(\mu;\nu_{\rm M,env})] = -\Lambda_{k}^{m}(\nu_{\rm M,env})\Theta_{k}^{m}(\mu;\nu_{\rm M,env}) \, ,
\end{equation}
with $\mu = \cos \theta$, and the Laplace tidal operator:
\begin{align}
    \mathcal{L}_{\nu_{\rm M,env}}& = \frac{\mathrm{d}}{\mathrm{d}\mu}\left(\frac{1-\mu^{2}}{1-\nu_{\rm M,env}^{2}\mu^{2}}\frac{\mathrm{d}}{\mathrm{d}\mu}\right)\nonumber\\ &- \frac{1}{1-\nu_{\rm M,env}^2\mu^2}\left(\frac{m^{2}}{1-\mu^{2}}+m\nu_{\rm M,env}\frac{1+\nu_{\rm M,env}^2\mu^2}{1-\nu_{\rm M,env}^2\mu^2}\right) \, .
\end{align}
$\Theta_{k}^{m}$ are the renowned Hough functions \citep{Hough1898V.Equations}. They depend on two different integer numbers: $m$ and $k$. They reduce to a Legendre polynomial in the absence of rotation or magnetism, the angular degree $l$ of the polynomial being related to the index $k$ in this case via $k = l-|m|$ \citep{Lee1997Low-frequencyDependence}. We highlight that in this particular magnetic context, the quantity $\nu_{\rm M,env}$ holds the frequency dependence of the angular structure of $m$-$g$-$i$ modes. This is why we called this quantity magnetic structural parameter. We can then expand the perturbations quite similarly to \citetalias{Barrault2025ConstrainingDips} as:
\begin{equation}
X'(r,\theta) = \sum_{k} X'_{k,m}(r)\Theta_{k}^{m}(\mu;\nu_{\rm M,env}) 
\end{equation}

\noindent and

\begin{equation}
\xi'_{r}(r,\theta) = \sum_{k} \xi'_{r;k,m}(r)\Theta_{k}^{m}(\mu;\nu_{\rm M,env}) \, .
\end{equation}

We see the influence of magnetism on the angular structure of modes. Notably, the limit of the equatorial belt in which sub-inertial $g$-$i$ modes are trapped under the TAR is changed in the magnetic case: this co-latitude is $\theta_{c} = \cos^{-1}(1/|\nu_{\rm M, env}|)$. The system of radial ordinary differential equations allowing us to get the radial dependence of $m$-$g$-$i$ modes is derived using the TARM following \citet{Mathis2011Low-frequencyField}:

\begin{align}
    \frac{\mathrm{d}W'_{k,m}}{\mathrm{d}r} = \frac{N^{2}}{\bar{g}}W'_{k,m} + & \frac{1}{r^{2}}(\sigma_{\rm M, env}^{2} - N^{2})(r^{2}\xi'_{r;k,m}) \nonumber \\ & - \frac{1}{\Gamma_{1}}\frac{\mathrm{d}\ln \bar{P}}{\mathrm{d}r}\frac{P'_{\mathrm{M};k,m}}{\bar{\rho}}
    \label{eq:pressure}
\end{align}

\begin{align}
    \frac{\mathrm{d}}{\mathrm{d}r}(r^{2}\xi'_{r;k,m}) = & \left[\frac{\Lambda_{k}^{m}(\nu_{\rm M,env})}{\sigma_{\rm M, env}^{2}} - \frac{\bar{\rho}r^{2}}{\Gamma_{1}\bar{P}}\right]W'_{k,m} \nonumber \\ &- \frac{1}{\Gamma_{1}\bar{P}}\frac{\mathrm{d}\bar{P}}{\mathrm{d}r}(r^{2}\xi'_{r;k,m}) + \frac{r^{2}}{\Gamma_{1}\bar{P}}P'_{\mathrm{M};k,m} \, .
    \label{eq:displacement}
\end{align}
We here described the TARM in a the framework of a uniform Alfvén frequency. The case of a field profile with a radially varying Alfvén frequency has been treated in \citet{Dhouib2022DetectingField}: the Laplace tidal operator would hold a radial dependence, and so would the generalized Hough functions obtained. We point out that the Hough functions derived with this (uniform) Alfvén frequency hypothesized in this work correspond to the ones computed with the Alfvén frequency at the core-to-envelope boundary in the more realistic field profile of \citet{Dhouib2022DetectingField}. In other words, the dip formation mechanism is sensitive to the field intensity at the base of the radiative envelope and does not depend on the Alfvén frequency profile in the rest of the envelope.

\subsubsection{JWKB analysis}

We adopt the anelastic approximation, which acts as a filter on the acoustic waves. We thus hypothesize that the frequency of the sub-inertial waves never equates the Lamb frequency $\tilde{S}$ in the propagation cavity, and both the upper and lower limits $r_{a}$ and $r_{b}$ are defined by the condition $\sigma_{\rm M,env} = N$. We verify this point in Appendix \ref{Appendix:models}. For the Kelvin mode $(m=-1)$, the frequency is so low that we are always in the regime of $\sigma_{\rm M,env} \ll \{N, \tilde{S}\}$ in the cavity. This allows us to neglect in the system given by Eqs.~\eqref{eq:pressure} and \eqref{eq:displacement} the terms scaling as $c_{S}^{2} = \Gamma_{1}\bar{P}/\bar{\rho}$. Under this approximation, we define the two following variables:
\begin{equation}
    v = \bar{\rho}^{1/2} \frac{\sigma_{\rm M,env}}{\sqrt{\Lambda_{k}^{m}(\nu_{\rm M,env})}} r^{2} \xi'_{r;k,m}
    \label{eq:def_v}
\end{equation}
and

\begin{equation}
    w = \left[\frac{\bar{\rho} r^{2}}{N^{2}}\right]^{1/2} W'_{k,m} \, .
    \label{eq:def_w}
\end{equation}

\noindent We can show, in a manner described in e.g. \citet{Press1981RadiativeInteriors} and \citet{Unno1989NonradialStars}, that the two variables follow the approximate equations:
\begin{equation}
    \frac{d^{2}v}{dr^{2}} + k_{r}^{2}v \simeq 0 \, ,
\end{equation}

\begin{equation}
    \frac{d^{2}w}{dr^{2}} + k_{r}^{2}w \simeq 0  \, ,
\end{equation}

\noindent with 
\begin{equation}
    k_{r}^{2} = \left(\frac{N^{2}}{\sigma_{\rm M, env}^{2}}-1\right)\frac{\Lambda_{k}^{m}(\nu_{\rm M,env})}{r^2} \, .
    \label{eq:wavenum}
\end{equation}
This wavenumber is similar to the one derived in the hydrodynamical case \citepalias{Tokuno2022AsteroseismologyOscillations,Barrault2025ConstrainingDips}. It is slightly different due to the anelastic approximation used ab initio in this work, whereas it is applied later on in \citetalias{Tokuno2022AsteroseismologyOscillations}. The imprint of magnetism is seen in the dependence of the LTE eigenvalue $\Lambda_{k}^{m}$ on the magnetic structural parameter $\nu_{\rm M, env}$, reducing to the usual spin parameter in the non-magnetic case, and on the frequency $\sigma_{\rm M, env}$ reducing to $\sigma_{\rm env}$ in the hydrodynamical one. \\
\indent We now follow a JWKB analysis, provided that $N$ and $\tilde{S}$ are much larger than $\sigma_{\rm M, env}$ in the propagation cavity, except near the turning points. Calculations below are pursued in the framework of a continuous $N$ profile at $r_{a}$, while they differ when considering a discontinuous $N$ profile. We will further comment this point in section~\ref{subsec:summary}. We obtain for $r\ll r_b$:
\begin{equation}
    v \simeq \frac{1}{\sqrt{|k_{r}|}}\left(\frac{3}{2}\left|\int_{r_a}^{r}|k_{r}|\mathrm{d}r\right|\right)^{1/6}[a\mathrm{Ai}(\zeta_{1}) + b\mathrm{Bi}(\zeta_{1})]
\end{equation}
and for $r \gg r_{a}$:
\begin{equation}
    w \simeq \frac{1}{\sqrt{|k_{r}|}}\left(\frac{3}{2}\left|\int_{r}^{r_{b}}|k_{r}|\mathrm{d}r\right|\right)^{1/6}[c\mathrm{Ai}(\zeta_{2}) + d\mathrm{Bi}(\zeta_{2})] \, ,
\end{equation}

\noindent $\rm Ai(\zeta)$ and $\rm Bi(\zeta)$ being the Airy functions of the first and second kind, solutions to the differential equation:
\begin{equation}
    \frac{\mathrm{d}^{2}y}{d\zeta^{2}}+\zeta y = 0 \, .
\end{equation}
Within this convention, Ai is exponentially decaying to $-\infty$ while Bi is diverging \citep{Unno1989NonradialStars}. We have introduced:
\begin{equation}
\zeta_{1} = \mathrm{sgn}(k_{r}^{2})\left(\frac{3}{2}\left|\int_{r_a}^{r}|k_{r}|\mathrm{d}r\right|\right)^{2/3}
\end{equation}
and
\begin{equation}
\zeta_{2} = \mathrm{sgn}(k_{r}^{2})\left(\frac{3}{2}\left|\int_{r}^{r_b}|k_{r}|\mathrm{d}r\right|\right)^{2/3} \, .
\end{equation}

\noindent Just as in \citetalias{Tokuno2022AsteroseismologyOscillations}, the Lagrangian pressure perturbation (hence $w$) should decay exponentially for $r\gg r_{b}$, then we must impose $d=0$. Further, using Eq.~\eqref{eq:displacement} in the anelastic limit, and the definitions of $w$ and $v$ given respectively in Eq.~\eqref{eq:def_w}) and Eq.~\eqref{eq:def_v}, we have $|k_{r}w| \simeq |\mathrm{d}v/\mathrm{d}r|$ for $r_{a}\ll r\ll r_{b}$. We thus derive the matching condition:
\begin{equation}
a=-c\sin{B} \, 
\end{equation}
and
\begin{equation}
b=-c\cos{B} \, ,
\end{equation}

\noindent with
\begin{align}
    B = \int_{r_a}^{r_b}k_{r}\mathrm{d}r -\frac{\pi}{2} & \simeq \int_{r_a}^{r_b}\frac{N}{\sigma_{\rm M, env}} \frac{\sqrt{\Lambda_{k}^{m}(\nu_{\rm M,env})}}{r}\mathrm{d}r -\frac{\pi}{2} \nonumber \\ &= \frac{\pi^{2}s_{\rm M, env}}{\Omega_{\rm env} \Pi_{\rm 0,M}} -\frac{\pi}{2} \, ,
\end{align}
where we have adopted the low-frequency limit of the vertical wavenumber.
The asymptotic period spacing modified by magnetism reads:
\begin{equation}
    \Pi_{\rm 0,M} = \frac{2\pi^{2}}{\sqrt{\Lambda_{k}^{m}(\nu_{\rm M, env})}}\left(\int_{r_a}^{r_b}N\frac{\mathrm{d}r}{r}\right)^{-1} \, .
\end{equation}
First, the effect of magnetism arises directly through the dependence of the eigenvalue of the LTE on the magnetic structural parameter $\nu_{\rm M,env}$ and no longer on the spin parameter $s_{\rm env}$ as in the hydrodynamical case. As Kelvin modes have have no latitudinal node ($k=0$,thus $l=|m|$), the azimuthal displacement is dominant over the latitudinal one. One can show that this hierarchy leads to the eigenvalue $\Lambda_{0}^{m}$ reaching asymptotically\footnote{This can be seen using equation (5) of \citet{Townsend2020ImprovedEquations}: if the azimuthal displacement dominates the latitudinal one, then the bracketed term at the RHS side of the equation must be small, leading to $\Lambda_{0}^{m}\approx m^{2}$.} the value $m^{2}$. The effect of the variation of $\Lambda_{0}^{m}$ with magnetism will thus be limited for Kelvin modes of high spin parameter that we are interested in. The effect is however predominant in other types of $m$-$g$-$i$ modes for which the LTE eigenvalue varies appreciably with the spin parameter \citep{Townsend2003AsymptoticStars,Townsend2020ImprovedEquations}. Second, magnetism acts indirectly through the potential variation of the shape of $N$ and the location of the lower and upper turning point. Even though this effect could be sizeable for highly magnetized stars, the regime of intermediate fields considered in this analysis would not appreciably change the structure of the star, since the background gaseous pressure $\bar{P}$ is predominant over the magnetic one $\bar{P}_{\rm M}$, except near the stellar surface \citep{Duez2010EffectSun}. Likewise, a high intensity field could result in an oblateness of the star due to the different magnetic tension at the pole and at the equator, but this effect is underdominant compared to the one of rotation \citep{Fuller2023LinkingStars}, even in the case of stars with a strong magnetic field intensity at the surface, which are not $\gamma$-Dor pulsators.

\subsubsection{Approximated solutions near the core}

We get an expression of $k_{r}^{2}$ near the lower-end of the $g$-mode cavity $r=r_{a}$ (see Fig.~\ref{fig:sketch_mag}):
\begin{equation}
    k_{r}^{2} \simeq \frac{\mathrm{d}k_{r}^{2}}{\mathrm{d}r}\Big|_{r=r_{a}} (r-r_{a}) \simeq \left[\frac{\Lambda_{k}^{m}(\nu_{M,env})}{r^{2} \sigma_{\rm M,env}^{2}}\frac{\mathrm{d}N^{2}}{\mathrm{d}r}\right]_{r=r_{a}}(r-r_{a})
\end{equation}
if we consider that the gradient of $N^{2}$ is predominant over the other terms in the development of $\mathrm{d}k_{r}/\mathrm{d}r$ near $r_{a}$.

We can define the same small parameter $\epsilon$ as in the non-magnetic case:
\begin{equation}
    \epsilon = \left(\frac{r_{\rm a}}{4\Omega_{\rm env}^{2}}\frac{\mathrm{d}N^{2}}{\mathrm{d}r}\Big|_{r=r_{\rm a}}\right)^{-1/3}
\end{equation}

\noindent that leads to:
\begin{equation}
    k_{r}^{2} \simeq \frac{\Lambda_{k}^{m}(\nu_{\rm M,env})s_{\rm M, env}^{2}}{\epsilon^{3} r_{\rm a}^{3}}(r-r_{\rm a}) \, .
\end{equation}

The following calculations are just the same as \citetalias{Tokuno2022AsteroseismologyOscillations} (from their Eqs. 24 to 26). We have to take a closer look at the equivalent of their (27): in the low frequency regime, for $\sigma_{\rm M,env} \ll N$, the magnetic pressure term in \eqref{eq:displacement} is negligible compared to $\Lambda_{k}^{m}(\nu_{\rm M,env})/\sigma_{\rm M,env}^{2}W'_{k,m}$ (see Appendix \ref{App:hierarchy}).\\
Therefore, we obtain:
\begin{equation}
\frac{\mathrm{d}}{\mathrm{d}r}(r^{2}\xi'_{r;k,m}) = \left[\frac{\Lambda_{k}^{m}(\nu_{\rm M,env})}{\sigma_{\rm M,env}^{2}}\right]W'_{k,m} -\frac{1}{\Gamma_{1}}\frac{\mathrm{d} \ln \bar{P}}{\mathrm{d}r}(r^{2}\xi'_{r;k,m}) \, .
\end{equation}
With the definition of $v$, we get:
\begin{equation}
    W'_{k,m} = \left(\frac{\sigma_{\rm M, env}}{\sqrt{\Lambda_{k}^{m}(\nu_{\rm M,env})\bar{\rho}}}\right)\left[\frac{\mathrm{d}v}{\mathrm{d}r}-\left(\frac{1}{2}\frac{\mathrm{d} \ln \bar{\rho}}{\mathrm{d}r} - \frac{1}{\Gamma_{1}}\frac{\mathrm{d} \ln \bar{P}}{\mathrm{d}r}\right)v\right] \, .
\end{equation}
From these expressions, we write:
\begin{equation}
    \frac{\xi'_{r;k,m}}{r}\Bigg|_{r=r_a} \simeq Q \epsilon X_{k}^{m}(s_{\rm M, env}) 
    \label{eq:xi_ra}
\end{equation}
and
\begin{equation}
    W'_{k,m}|_{r=r_a} \simeq Q r_{a}^{2} \sigma_{\rm M, env}^{2} Y_{k}^{m}(s_{\rm M,env}) \, ,
    \label{eq:w_ra}
\end{equation}
where we have introduced very similarly to \citetalias{Tokuno2022AsteroseismologyOscillations} the following functions:
\begin{align}
    X_{k}^{m}(s_{\rm M,env}) = & \Lambda_{k}^{m}(\nu_{\rm M,env})^{1/6}s_{\rm M,env}^{2/3} \nonumber \\ & \times \sin\left(\frac{\pi^{2}s_{\rm M,env}}{\Omega_{\rm env}\Pi_{\rm 0,M}}-\frac{\pi}{6}\right)
    \label{eq:def_X}
\end{align}
and
\begin{align}
    Y_{k}^{m}(s_{\rm M,env}) = \alpha &\Lambda_{k}^{m}(\nu_{\rm M,env})^{-1/2}s_{\rm M,env}^{4/3} \nonumber \\ & \times
    \sin\left(\frac{\pi^{2}s_{\rm M,env}}{\Omega_{\rm env}\Pi_{\rm 0,M}}-\frac{5\pi}{6}\right) \, ,
    \label{eq:def_Y}
\end{align}
with $\alpha \simeq 0.73$, whose exact value is given in Appendix \ref{App:notations}. There is a slight change in $Q$ due to our different definition of $w$ compared to \citetalias{Tokuno2022AsteroseismologyOscillations} that does not matter in our calculations since $Q$ is a common linear multiplicative term.\\
\indent The expressions for the linear perturbation to the Lagrangian displacement and the total dynamical pressure are now determined through respectively Eq.~\eqref{eq:xi_ra} and Eq.~\eqref{eq:w_ra} at the lower boundary of the $m$-$g$-$i$ modes cavity. Yet, the coupling of the envelope $m$-$g$-$i$ modes and core $m\text{-}i$ modes is happening at $R_{\rm core}$. However, due to the high value of the $N$ gradient from $R_{\rm core}$ to $r_{a}$, the core-to-envelope boundary is close to the lower-turning point of $m$-$g$-$i$ modes: $R_{\rm core} \approx r_{a}$. In this case, the mode structure at $r_{a}$ is close to the one at $R_{\rm core}$. We however need to check this affirmation.
We thus adapt the estimation of the ratios (34) and (35) of \citetalias{Tokuno2022AsteroseismologyOscillations} to the magnetized case: here $N^{2}|_{r=R_{\rm core}}=0$ and $N^{2}|_{r=r_a}=\sigma_{\rm M,env}^{2} $. Then:
\begin{equation}
    \frac{R_{\rm core}}{r_a} \sim 1-\frac{\epsilon^3}{s_{\rm M,env}^{2}}
\end{equation}
and
\begin{equation}
    \frac{v|_{r=R_{\rm core}}}{v|_{r=r_a}} \sim 1+O\left(\frac{\epsilon^2}{s_{\rm M,env}^{4/3}}\right)
\end{equation}
This expansion can be readily seen as follows: if the $N$ gradient at $r_{a}$ is small, hence $\epsilon$ is high, then at fixed $s_{\rm M,env}$, $r_{a}$ moves away from $R_{\rm core}$. Conversely, at fixed $\epsilon$, hence $N$ gradient, the lower the frequency of the mode in the frame co-rotating with the envelope is (thus higher $s_{\rm M,env}$), the lower is the difference between $r_{a}$ and $R_{\rm core}$. Summing up, using the expansion of $v$ and $w$ at $r_{a}$ for the coupling at $R_{\rm core}$ is only valid for low $\epsilon$ and high $s_{\rm M, env}$. We will stay in this regime for the rest of this work.

\subsection{Core $m\text{-}i$ modes: derivation of Bryan solutions}\label{subsec:core_modes}

We now consider convective core magneto-inertial ($m\text{-}i$) modes. To get an analytical modeling of core $m\text{-}i$ modes similar to the Bryan solutions obtained in the hydrodynamical case \citep{Bryan1889TheEllipticity} and stay in an analytical framework for core $m\text{-}i$ modes, we assume there a uniform core density. We point out the importance of this approximation, the core density stratification being the main source of the variation of the spin parameter at which the interaction occurs in the solid-body, non-magnetic study of \citet{Ouazzani2020FirstRevealed}. Indeed, as computed in \citet{Wu2005OriginModes}, the pure inertial mode spin parameter decreases with an increasing steepness of the core density profile. \citet{Ouazzani2020FirstRevealed} noticed this effect, highly correlated to the age of the star on the MS: the more evolved the star is, the steeper the gradient of core density, decreasing the spin parameter of the pure inertial mode, hence the location of the dip in period. However, we decided not to take into account this effect in our study, focusing first on the effect of magnetism only on the location and morphology of the dip. \\
\indent We revert back to the momentum equation \eqref{eq:momentum_general} used before further development using the TARM. In a non-stratified core, the TARM drops and one has to consider the set of equations with a full treatment of both the Coriolis acceleration and the Lorentz force. We base our study on the one given in \citet{Malkus1967HydromagneticWaves}, adapting Bryan solutions to the case of our magnetic set-up, with a uniform Alfvén frequency in the core. Literature results such as the establishment of the linear system of equations and the main properties of core modes are only recalled in the main text below, while being developed in detail in Appendix \ref{App:deriv} for readability.\\
We consider the momentum equation assuming a uniform density:
\begin{equation}
-\mathcal{A}\boldsymbol{\xi}' + i\mathcal{B}\widehat{\mathbf{e}}_{z} \times \boldsymbol{\xi}'  +\boldsymbol{\nabla} W' = 0 \label{eq:momentum_inertial}
\end{equation}
Along with this hypothesis of mean density in the core, the anelastic approximation which is assumed for both the core and the envelope reverts back to an hypothesis of incompressibility in the convective core. In an incompressible framework, we retrieve the Bryan solutions \citep{Bryan1889TheEllipticity}, which consists in solutions separated in ellipsoidal coordinates \citep{Wu2005OriginModes}.
The corresponding expressions at the boundary of the core are:
\begin{equation}
    \frac{\xi'_{r;l,m}(r)}{r}\Bigg|_{r=R_{\rm core}} \propto C_{l}^{m}(1/\nu_{\rm M, core})P_{l}^{m}(\mu)
    \label{eq:xi_mi}
\end{equation}
and 
\begin{equation}
    W'_{l,m}(r)|_{r=R_{\rm core}} \propto R_{\rm core}^{2}\sigma_{\rm M,core}^{2}P_{l}^{m}(1/\nu_{\rm M,core})P_{l}^{m}(\mu) \, .
    \label{eq:w_mi}
\end{equation}

\noindent We defined $C_{l}^{m}$ as in \citet{Ouazzani2020FirstRevealed} and \citetalias{Barrault2025ConstrainingDips}:
\begin{equation}
 C_{l}^{m}(x) = x \left(\frac{dP_{l}^{m}(x)}{dx} - \frac{m}{1-x^{2}}P_{l}^{m}(x)\right) \, .
 \label{eq:condition}
\end{equation}

\noindent We consider isolated $m\text{-}i$ modes, that we will further make interact with $m$-$g$-$i$ modes. For $m\text{-}i$ modes, the condition of $\xi'_{r} = 0$ at the core boundary, translates into $C_{l}^{m}(1/\nu_{\rm M,core})=0$. This is an eigenvalue problem that sets the values of $\nu_{\rm M,core}$ for each isolated $(l,m)$ $m\text{-}i$ mode, $\nu_{\rm M, core}^*$.\\
\indent We see that with the considered magnetic configuration, the MHD problem is very similar to the hydrodynamical one: the spin parameter controlling the frequency dependence of the variables of interest is replaced by the magnetic structural parameter. The critical latitudes at $\mu = 1/s_{\rm core}$ in the hydrodynamical case, typical of inertial modes, are shifted by the additional influence of the Lorentz force in such configuration, appearing now at latitudes $\mu = 1/\nu_{\rm M, core}$. We see that as in the case of the TARM in the radiative envelope, the parameter $\nu_{\rm M,core}$ specific to the core controls the angular structure of the $m\text{-}i$ modes, hence advocating for our denomination as the magnetic structural parameter.

\noindent Then the whole solution of the variables of interest near the core is the following sum:
\begin{equation}
\frac{\xi'_{r}}{r}\Big|_{r=R_{\rm core}} = \sum_{l}b_{l}C_{l}^{m}(1/\nu_{\rm M, core})\tilde{P}_{l}^{m}(\mu)
\end{equation}

\noindent and:
\begin{equation}
W'|_{r=R_{\rm core}} = \sigma_{\rm M,core}^{2} R_{\rm core}^{2}\sum_{l}b_{l}P_{l}^{m}(1/\nu_{\rm M, core})\tilde{P}_{l}^{m}(\mu)
\end{equation}
with $\tilde{P}_{l}^{m}$ the normalized Legendre polynomial:
\begin{equation}
    \tilde{P}_{l}^{m} \equiv \sqrt{\frac{(2l+1)(l-m)!}{2(l+m)!}}P_{l}^{m}(x)\, ,
\end{equation}
and $b_{l}$ constant factors. \\

\subsection{Impact of the Alfvén frequency profile}
The field configuration taken throughout this work of a toroidal field with bi-layer Alfvén frequency is a strongly simplified modelling of the reality. MHD simulations of convective cores have shown mixed toroidal-poloidal character for the dynamo-generated magnetic fields, on an extended range of spatial scales \citep{Brun2005SimulationsAction,Featherstone2009EffectsStars,Augustson2016TheStars}. We chose to consider a simpler configuration to remain analytical, keeping in mind that the seismic probe would be sensitive to the largest spatial scales of the magnetic fields due to the low degree of the inertial mode.
Considering a core magnetic field of varying Alfvén frequency would break the analytical character of the obtained solutions. Terms composed of the gradient of the Alfvén frequency appear at the RHS of Eq.~\eqref{eq:momentum_inertial}. The solutions become non-separable in ellipsoidal coordinates. No analytical solution can be derived and the equation must be solved numerically. The condition of null radial Lagrangian displacement at the boundary evolves from the one given by Eq.~\eqref{eq:condition}, resulting in a different eigenvalue problem and a $\nu_{\rm M,core}^*$ shifted from the one obtained with a constant Alfvén frequency in the core.\\
Numerical work is then needed to solve the eigenvalue problem in the presence of a radially varying Alfvén frequency, and derive the angular structure of the isolated $m\text{-}i$ mode in a more realistic magnetic field configuration. This can be done by using a spectral code, for instance DEDALUS \citep{Burns2020Dedalus:Methods}, using its eigenvalue solver. When both $\nu_{\rm M,core}^*$ and the angular structure of the mode are obtained, we can rewrite the quantities of interest on the basis of Legendre polynomials, as suggested by \citetalias{Tokuno2022AsteroseismologyOscillations} to treat a non-uniform density in the hydrodynamical case. Eqs.~\eqref{eq:xi_mi} and \eqref{eq:w_mi} are then respectively replaced by:
\begin{equation}
\frac{\xi'_{r}}{r}\Big|_{r=R_{\rm core}} \propto \sum_{l}g_{l}(\nu_{\rm M, core})\tilde{P}_{l}^{m}(\mu)
\end{equation}

\noindent and:
\begin{equation}
W'|_{r=R_{\rm core}} \propto \sigma_{\rm M,core}^{2} R_{\rm core}^{2}\sum_{l}h_{l}(\nu_{\rm M, core})\tilde{P}_{l}^{m}(\mu)
\end{equation}
The reasoning could then be pursued with this numerical solutions. However, we choose to stay in the framework of Bryan solutions for this work as it is our present aim to set up an analytical laboratory for the comprehension of the effect of magnetism on the dip profile.

\section{Coupling equation and approximate analytical profiles for the dip structure}\label{der:cont_brunt}
Having now derived the core and envelope oscillation modes structure, we consider in this section their coupling through the core-to-envelope boundary. We here recall that all relevant parameters and notations are regrouped and explained in Appendix \ref{App:notations}.
\subsection{Matching quantities in the magnetic set-up}

Compared to the hydrodynamical, solid-body rotating case, the matching of the quantities at the interface leading to mode coupling would rigorously comprise an influence of the background Lorentz force. This point is discussed in Appendix \ref{App:hierarchy}. Since we neglected the non-sphericity of the hydrostatic background as in \citet{Mathis2011Low-frequencyField}, for consistency we consider $\mathrm{d} \bar{P}_{\rm tot}/\mathrm{d}r \simeq -\bar{\rho}\bar{g}$, $-\bar{g}\widehat{\boldsymbol{e}}_{r}$ being the local self-gravity acceleration.\\
The Lagrangian perturbation of the total pressure is written:
\begin{align}
    \delta P_{\rm tot} = & P_{\rm tot}' + \frac{\mathrm{d} \bar{P}_{\rm tot}}{\mathrm{d}  r} \xi'_{r} \\
     = & \bar{\rho}W'-r\bar{\rho}{\bar{g}}\left(\frac{\xi'_{r}}{r}\right) \, .
\end{align}
We ensure the continuity of $\xi_{r}'$ and $\delta P_{\rm tot}$.
If the background density $\bar{\rho}$ is continuous at the boundary, the continuity of $\delta P_{\rm tot}$ is equivalent to the one of $W'$, the Eulerian dynamical pressure perturbation.

The matching equations at $R_{\rm core}$ equivalent to those of \citetalias{Barrault2025ConstrainingDips} are the following:
\begin{align}
    \sum_{k}a_{k}\epsilon &X_{k}^{m}(s_{\rm M, env})\Theta_{k}^{m}(\mu;\nu_{\rm M, env}) \nonumber \\ &= \sum_{l}b_{l}C_{l}^{m}(1/\nu_{\rm M, core})\tilde{P}_{l}^{m}(\mu)
    \label{eq:match_displacement}
\end{align}
and:
\begin{align}
    \sigma_{\rm M, env}^{2}\sum_{k}&a_{k}Y_{k}^{m}(s_{M, \rm env})\Theta_{k}^{m}(\mu;\nu_{\rm M, env}) \nonumber \\ &= \sigma_{\rm M, core}^{2}\sum_{l}b_{l}P_{l}^{m}(1/\nu_{\rm M, core})\tilde{P}_{l}^{m}(\mu) \, .
    \label{eq:match_pressure}
\end{align}
We project Eq.~\eqref{eq:match_displacement} on the Hough functions basis. We get: 
\begin{equation}
    a_{k}\epsilon X_{k}^{m}(s_{\rm  M, env}) = \sum_{l}b_{l}C_{l}^{m}(1/\nu_{\rm  M, core})c_{k,l} \, ,
\end{equation}
where we have defined:
\begin{equation}
    c_{k,l} = \int_{-1}^{1}\Theta_{k}^{m}(\mu;\nu_{\rm M, env })\tilde{P}_{l}^{m}(\mu)\mathrm{d}\mu \, .
    \label{def:ckl}
\end{equation}
After projecting Eq.~\eqref{eq:match_pressure} on Hough functions, we obtain the matrix equation:
\begin{equation}
[\mathcal{M}(s_{\rm M,env},\nu_{\rm M, core}) - \epsilon \mathcal{N}(s_{\rm M, env},\nu_{\rm M, core})]\vec{b} = \vec{0} \, ,
\label{eq:matrix_cont}
\end{equation}
$\vec{b}$ being the column vector of the terms $b_{l}$. The matrices are defined as:
\begin{equation}
    [\mathcal{M}]_{k,l} = c_{k,l}\sigma_{\rm M, env}^{2}Y_{k}^{m}(s_{\rm M, env})C_{l}^{m}(1/\nu_{\rm M, core})
\end{equation}
and
\begin{equation}
    [\mathcal{N}]_{k,l} = c_{k,l}\sigma_{\rm M, core}^{2}X_{k}^{m}(s_{\rm M,  env})P_{l}^{m}(1/\nu_{\rm M, core}) \, .
\end{equation}
For $\vec{b}$ not to be trivial, we have the following condition:
\begin{equation}
    \text{det}[\mathcal{M}(s_{\rm M, env},\nu_{\rm M,core})-\epsilon \mathcal{N}(s_{\rm M, env},\nu_{\rm M, core})] = 0 \, .
\end{equation}
In a similar manner to the one adopted in \citetalias{Tokuno2022AsteroseismologyOscillations} and \citetalias{Barrault2025ConstrainingDips}, we consider only the under-dominant term of order $O(\epsilon)$ in this determinant, corresponding to a core $m\text{-}i$ mode of index ($l,m$) interacting with a series of $m$-$g$-$i$ modes of index ($k,m$). We refer to \citetalias{Barrault2025ConstrainingDips} for the extended reasoning behind this approximation. Appendix D of \citet{Galoy2024PropertiesStars} tackles a more rigorous multi-mode interaction extended to the non-negligible geometrical factors, in the case of a non-magnetic uniformly rotating star. We let the application of this method to the magnetic case for future studies, while verifying a posteriori in subsection~\ref{subsec:degeneracy} that the geometrical factor for the dominant mode interaction considered here does not vary appreciably in the range of Lehnert number contrasts explored in this work.
Isolating the interacting under-dominant term leads to the following simplified condition:
\begin{equation}
    \frac{\sigma_{\rm M, env}^{2}Y_{k}^{m}(s_{\rm M,env})C_{l}^{m}(1/\nu_{\rm M, core})}{\sigma_{\rm M, core}^{2}X_{k}^{m}(s_{\rm  M, env})P_{l}^{m}(1/\nu_{\rm M, core})} \simeq \epsilon \, .
    \label{eq:gen_coupl}
\end{equation}
We identify the similarity in the structure of this coupling equation to Eq.(28) in \citetalias{Barrault2025ConstrainingDips}. This is not a coincidence, as the choice of a toroidal field in our set up makes the Lorentz force having a similar mathematical form as the Coriolis acceleration.
In the framework of a solid-body rotating model with a continuous Alfvén frequency, the coupling equation simplifies to:
\begin{equation}
    \frac{Y_{k}^{m}(s_{\rm M})C_{l}^{m}(1/\nu_{\rm M})}{X_{k}^{m}(s_{\rm M})P_{l}^{m}(1/\nu_{\rm M})} \simeq \epsilon \, .
\end{equation}
Even though this equation holds similarities with the hydrodynamical counterpart (Eq 56 in \citetalias{Tokuno2022AsteroseismologyOscillations}), we emphasize here the difference between the magnetic structural parameter $\nu_{\rm M}$, controlling the frequency dependency of the Hough functions and the eigenvalue of the LTE $\Lambda_{k}^{m}$ in the TARM as well as the Bryan solutions, and the magnetic spin parameter $s_{\rm M}$. This is a key distinction with the work of \citetalias{Tokuno2022AsteroseismologyOscillations}, even in the case of uniformly rotating stars with continuous Alfvén frequency. Indeed, the non-magnetic equivalent of both the magnetic structural and spin parameter is the hydrodynamic spin parameter. This is where the formalism of \citetalias{Barrault2025ConstrainingDips} is mandatory to use, as even in this simplified magnetic case the quantities to match in the coupling equation do not depend on the same frequency variable. \\
\indent Reverting back to the general case, we define 
\begin{equation}
    F_{l}^{m}(\nu_{\rm M, core}) \equiv -\frac{C_{l}^{m}(1/\nu_{\rm M, core})}{P_{l}^{m}(1/\nu_{\rm M, core})} \, .
\end{equation}
Using this definition and those of $X_{k}^{m}$ and $Y_{k}^{m}$ given by Eqs \eqref{eq:def_X} and \eqref{eq:def_Y} in the coupling equation Eq~\eqref{eq:gen_coupl}, we obtain:
\begin{align}
    F_{l}^{m}(\nu_{\rm M, core})\frac{\sqrt{3}}{2}&\frac{\alpha}{\Lambda_{k}^{m}(\nu_{\rm M, \rm env})^{2/3}}s_{\rm M,  env}^{2/3}\frac{\sigma_{\rm M, env}^2}{\sigma_{\rm M, core}^{2}} \nonumber \\ & \left[\cot\Big{(}\frac{\pi^2s_{\rm M, env}}{\Omega_{\rm env}\Pi_{\rm 0,M}}-\frac{\pi}{6}\Big{)}+\frac{1}{\sqrt{3}}\right] \simeq \epsilon \, .
    \label{eq:synth_coupling}
\end{align}
\indent We further seek to retrieve the inertial dip profile in the magnetic case assuming approximations that are close to the ones made in \citetalias{Barrault2025ConstrainingDips}, to get a better understanding of the effects of magnetism. The coupling equation \eqref{eq:synth_coupling} is completed by relations linking the magnetic spin and structural parameters $s_{\rm M, zone}$ and $\nu_{\rm M,zone}$ based on the unicity of the frequencies in the inertial frame from both sides of the boundary and the knowledge of the rotation rates $\Omega_{\rm zone}$ and the Alfvén frequencies $\omega_{\rm A,zone}$. We define for this purpose the following functions:

\begin{equation}
    u_{\rm core} \colon s_{\rm M, core} \mapsto \nu_{\rm M, core} \, ,
    \label{fund:numc_smc}
\end{equation}
\begin{equation}
    u_{\rm env} \colon s_{\rm M, env} \mapsto \nu_{\rm M, env} \, ,
    \label{fund:nume_sme}
\end{equation}
and
\begin{equation}
    G_{\rm M} \colon s_{\rm M, core} \mapsto s_{\rm M, env} \, .
    \label{fund:sme_smc}
\end{equation}
These functions are computed by ensuring the constant value of the frequency in the inertial frame $\sigma_{\rm in}$. In the general magnetic, differentially rotating case, the equations linking the magnetic structural parameters and spin parameters are quadratic, whereas in the non-magnetic case the equation $s_{\rm env} = G(s_{\rm core})$ was linear, and an analytical expression for $G$ could be derived. Practically, we compute these functions and their gradients numerically.
In the limiting case of differential rotation without magnetism, $G_{\rm M}$ simplifies to the function $G$ used in \citetalias{Barrault2025ConstrainingDips}.

\subsection{Derivation of analytical profiles for the dip structure for the uniform and bi-layer cases}

We further consider in this section derivations of the dip profile in a uniformly rotating star with a uniform Alfvén frequency, a case that can be considered as the simplest extension of \citetalias{Tokuno2022AsteroseismologyOscillations}'s model. The magnetic field would be strong enough to completely suppress differential rotation in the convective core \citep{Brun2005SimulationsAction} and in the radiative zone \citep{Ferraro1937TheField,Gaurat2015EvolutionZone,Moyano2023AngularPulsators}. It would also perfectly connect the convective core and the radiative envelope: \citet{Augustson2016TheStars} (see their Fig.~6), have clearly established that compared to a hydrodynamical case, taking into account a magnetic field would result in a nearly solid-body rotating star. \\
We consider as well the more general differentially-rotating, bi-layer Alfvén frequency model, which can be seen as a generalisation of \citetalias{Barrault2025ConstrainingDips}'s model. The field would be strong enough to inhibit differential rotation in the two zones, but would not perfectly connect the core and the envelope, leading to a discontinuity in the rotation rate and the Alfvén frequency.

\subsubsection{Case of solid-body rotation and uniform Alfvén frequency}

We treat first the case of a solid-body rotating star with a uniform Alfvén frequency, highlighting the impact of the most simple magnetic configuration when compared to the non-magnetic case treated by \citetalias{Tokuno2022AsteroseismologyOscillations}. We will refer in that case to the function $u$ without subscript, linking $\nu_{\rm M}$ to $s_{\rm M}$. In this case, the coupling equation reads:
\begin{align}
    F_{l}^{m}(\nu_{\rm M})\frac{\sqrt{3}}{2} \frac{\alpha}{\Lambda_{k}^{m}(\nu_{\rm M})^{2/3}}s_{\rm M}^{2/3} \nonumber \left[\cot\Big{(}\frac{\pi^2s_{\rm M}}{\Omega\Pi_{0,\rm M}}-\frac{\pi}{6}\Big{)}+\frac{1}{\sqrt{3}}\right] \simeq \epsilon \, .
    \label{eq:coupling_cont}
\end{align}
Let us expand the prefactor around the zero of $F_{l}^{m}$, the magnetic structural parameter of the $m\text{-}i$ mode $\nu_{\rm M}^{*}$:
\begin{align}
    & F_{l}^{m}(\nu_{\rm M})\frac{\sqrt{3}}{2} \frac{\alpha}{\Lambda_{k}^{m}(\nu_{\rm M})^{2/3}}u^{-1}(\nu_{\rm M})^{2/3} \simeq \nonumber \\ &
    \left[\frac{\mathrm{d}F_{l}^{m}}{\mathrm{d}\nu_{\rm M}}\frac{\sqrt{3}}{2} \frac{\alpha}{(\Lambda_{k}^{m}(\nu_{\rm M}))^{2/3}}(u^{-1}(\nu_{\rm M}))^{2/3}\right]_{\nu_{\rm M}^{*}}(\nu_{\rm M}-\nu_{\rm M}^{*}) \, .
\end{align}

\noindent Defining the structure factor with solid-body rotation and uniform Alfvén frequency $V_{\rm M}$ as :
\begin{equation}
    V_{\rm M} = -\left[\frac{\mathrm{d}F_{l}^{m}}{\mathrm{d}\nu_{\rm M}}\frac{\sqrt{3}}{2} \frac{\alpha}{(\Lambda_{k}^{m}(\nu_{\rm M}))^{2/3}}(u^{-1}(\nu_{\rm M}))^{2/3}\right]_{\nu_{\rm M}^{*}} \, ,
\end{equation}
we can make the comparison with the non-magnetic, solid-body rotating structure factor $V$ of \citetalias{Tokuno2022AsteroseismologyOscillations}:
\begin{equation}
    V_{\rm M}/V = \left[\left(\frac{u^{-1}(\nu_{\rm M})}{\nu_{\rm M}}\right)^{2/3}\right]_{\nu_{\rm M}^{*}} \, .
\end{equation}
We define the magnetic control parameter $\Gamma_{\rm M} = \dfrac{3\pi\epsilon}{4\Omega V_{\rm M}}$.
The coupling equation with $V_{\rm M}$ reads:
\begin{equation}
    -V_{\rm M}(\nu_{\rm M}-\nu_{\rm M}^{*}) \nonumber \left[\cot\left(\frac{\pi^2s_{\rm M}}{\Omega\Pi_{\rm 0,M}}-\frac{\pi}{6}\right)+\frac{1}{\sqrt{3}}\right] \simeq \epsilon \, .
\end{equation}
The same reasoning that in \citetalias{Barrault2025ConstrainingDips} or \citetalias{Tokuno2022AsteroseismologyOscillations}, detailed in Appendix \ref{App:Mod_lorentz}, leads to:
\begin{equation}
    \frac{1}{\Delta \mathrm{P_{M}}} - \frac{1}{\Pi_{0,\rm M}} \simeq \frac{\dfrac{\Gamma_{\rm M}}{\pi}\dfrac{\mathrm{d}u}{\mathrm{d}s_{\rm M}}\bigg|_{\bar{s}_{\rm M}}}{\left(\left(\mathrm{P_{M}}-\mathrm{P}_{\rm M}^{*}\right)\dfrac{\mathrm{d}u}{\mathrm{d}s_{\rm M}}\bigg|_{\frac{\bar{s}_{\rm M}+s^{*}_{\rm M}}{2}} + \dfrac{\Gamma_{\rm M}}{\sqrt{3}}\right)^{2}+\Gamma_{\rm M}^{2}} \, ,
    \label{eq:final_dip_cont}
\end{equation}
with: $\rm P_{\rm M} = \pi s_{\rm M}/\Omega$, $\rm P_{\rm M}^{*} = \pi s_{\rm M}^{*}/\Omega = \pi u^{-1}(\nu_{\rm M}^{*})/\Omega$.

\subsubsection{General solution: bi-layer rotation rate and Alfvén frequency}\label{dip_cont}

We move now to a case in which the magnetic fields are of significant strength from both sides of the boundary to ensure a bi-layer rotation, but are discontinuous at $R_{\rm core}$. This configuration is a first analytical model allowing for different magnetic field strengths in both zones, as it would be the case if different generation and sustaining mechanisms are at play in the convective core and the radiative envelope, respectively a convective dynamo-generation in the core and a Tayler-Spruit mechanism or a fossil field in the envelope. This case of a discontinuous Alfvén frequency would be favoured if a strong core dynamo field would reduce the amount of core-to-envelope boundary mixing: the fields would be connecting on a lengthscale lower than the local wavelength of the mode, which would effectively probe a discontinuous drop of the Alfvén frequency.\\
\indent In this case, the coupling equation reads:
\begin{align}
    F_{l}^{m}(\nu_{\rm M, core})& \frac{\sqrt{3}}{2} \frac{\alpha}{\Lambda_{k}^{m}(\nu_{\rm  M, env})^{2/3}}s_{\rm M,env}^{2/3}\frac{s_{\rm M, core}^{2}}{s_{\rm M, env}^{2}}\alpha_{\rm rot}^{2} \nonumber \\ &\times \left[\cot\left(\frac{\pi^2s_{\rm M, env}}{\Omega\Pi_{\rm 0,M}}-\frac{\pi}{6}\right)+\frac{1}{\sqrt{3}}\right] \simeq \epsilon \, .
    \label{eq:coupl_disc}
\end{align}
We recall $\alpha_{\rm rot} = \Omega_{\rm env}/\Omega_{\rm core}$. Expanding $F_{l}^{m}$ around its zero in $\nu_{\rm M, core}$:
\begin{align}
     & F_{l}^{m}(\nu_{\rm M, core}) \frac{\sqrt{3}}{2} \frac{\alpha(G_{\rm M}\circ u_{\rm core}^{-1}(\nu_{\rm M, core}))^{2/3}}{\Lambda_{k}^{m}((u_{\rm env}\circ G_{\rm M} \circ u^{-1}_{\rm core})(\nu_{\rm  M, core}))^{2/3}} \nonumber \\ &\times \frac{u_{\rm core}^{-1}(\nu_{\rm M, core})^{2}\alpha_{\rm rot}^{2}}{(G_{\rm M}\circ u_{\rm core}^{-1}(\nu_{\rm M, core}))^{2}} \nonumber \\
     &\simeq \left[\frac{\mathrm{d}F_{l}^{m}}{\mathrm{d}\nu_{\rm M, core}} \frac{\sqrt{3}}{2} \frac{\alpha(G_{\rm M}\circ u_{\rm core}^{-1}(\nu_{\rm M, core}))^{2/3}}{\Lambda_{k}^{m}(u_{\rm env}\circ G_{\rm M} \circ u^{-1}_{\rm core}(\nu_{\rm M, core}))^{2/3}}\right. \nonumber \\
     & \left.\times\frac{(u_{\rm core}^{-1}(\nu_{\rm M, core}))^{2}\alpha_{\rm rot}^{2}}{(G_{\rm M}\circ u_{\rm core}^{-1}(\nu_{\rm M, core}))^{2}} \right]_{\nu_{\rm M,core}^{*}} (\nu_{\rm M, core}-\nu_{\rm M, core}^{*}) \, ,
\end{align}
$\circ$ being the composition operator. We define in this case the structure factor $V_{\rm M,\neq}$ as :
\begin{align}
    V_{\rm M, \neq}  = & -\left[\frac{\mathrm{d}F_{l}^{m}}{\mathrm{d}\nu_{\rm M,core}} \frac{\sqrt{3}}{2} \frac{\alpha(G_{\rm M}\circ u_{\rm core}^{-1}(\nu_{\rm M,core}))^{2/3}}{\Lambda_{k}^{m}(u_{\rm env}\circ G_{\rm M} \circ u^{-1}_{\rm core}(\nu_{\rm M,core}))^{2/3}}\right. \nonumber \\
    & \left. \times \frac{(u_{\rm core}^{-1}(\nu_{\rm M,core}))^{2}}{(G_{\rm M}\circ u_{\rm core}^{-1}(\nu_{\rm M,core}))^{2}}\alpha_{\rm rot}^{2}\right]_{\nu_{\rm M, core}^{*}} \, .
\end{align}
The correction to the solid-body rotating, non-magnetic case is:
\begin{align}
    V_{\rm M,\neq}/V = & \left[\left(\frac{\Lambda_{k}^{m}(\nu_{\rm  M,core})}{\Lambda_{k}^{m}(u_{\rm env}\circ G_{\rm M} \circ u^{-1}_{\rm core}(\nu_{\rm M, core}))}\right)^{2/3} \right. \nonumber \\ 
    & \left. \times \left(\frac{G_{\rm M}\circ u_{\rm core}^{-1}(\nu_{\rm M,core})}{\nu_{\rm M,core}}\right)^{2/3} \right. \nonumber \\
    & \left. \times \left(\frac{u_{\rm core}^{-1}(\nu_{\rm M, core})}{{G_{\rm M}\circ u_{\rm core}^{-1}(\nu_{\rm M, core})}}\right)^{2} \alpha_{\rm rot}^{2}\right]_{\nu^{*}_{\rm M,core}} \, .
\end{align}
The coupling equation is:
\begin{equation}
    -V_{\rm M,\neq}(\nu_{\rm M,core}-\nu_{\rm  M, core}^{*}) \left[\cot\Big{(}\frac{\pi^2s_{\rm M,env}}{\Omega_{\rm env}\Pi_{0,\rm M}}-\frac{\pi}{6}\Big{)}+\frac{1}{\sqrt{3}}\right] \simeq \epsilon
    \label{eq:bef_lor} \, .
\end{equation}
Defining in that case the magnetic control parameter $\Gamma_{\rm M, \neq}= \dfrac{3\pi\epsilon}{4\Omega_{\rm env} V_{\rm M,\neq}}$, we get after some calculations detailed in Appendix \ref{App:Mod_lorentz} the following expression:
\begin{align}
    & \frac{1}{\Delta \mathrm{P_{M}}} - \frac{1}{\Pi_{0,\rm M}} \simeq \nonumber \\
    & \frac{\dfrac{\Gamma_{\rm M,\neq}}{\pi}\dfrac{\mathrm{d}u_{\rm core}\circ G_{\rm M}^{-1}}{\mathrm{d}s_{\rm M, env}}\bigg|_{\bar{s}_{\rm M, env}}}{\left((\mathrm{P_{M}}-\mathrm{P}_{\rm M}^{*})\dfrac{\mathrm{d}u_{\rm core}\circ G_{\rm M}^{-1}}{\mathrm{d}s_{\rm M,env}}\bigg|_{\frac{\bar{s}_{\rm M, env}+s^{*}_{\rm M, env}}{2}} + \dfrac{\Gamma_{\rm M,\neq}}{\sqrt{3}}\right)^{2}+\Gamma_{\rm M,\neq}^{2}} \, ,
    \label{eq:final_profile}
\end{align}
where $\rm P_{M} = \pi s_{\rm M,env}/\Omega_{\rm env}$, $\rm P_{M}^{*} = \pi s_{\rm M, env}^{*}/\Omega_{\rm env} = \pi (G_{\rm M} \circ u_{\rm core}^{-1})(\nu_{\rm M, core}^{*})/\Omega_{\rm env}$.

\subsection{From magnetic variables to quantities in the co-rotating frame}
\label{subsec:mag_to_corot_var}
Expressions Eqs.~\eqref{eq:final_dip_cont} and \eqref{eq:final_profile} are quite comparable to respectively (65) of \citetalias{Tokuno2022AsteroseismologyOscillations} and (37) of \citetalias{Barrault2025ConstrainingDips}. Yet, these expressions are written using the magnetic variables $\rm P_{M}$ and $\rm \Delta P_{M}$, and need further manipulation to account for envelope magnetism and revert back to $\rm P$ and $\Delta \rm P$, respectively the period and the period-spacing of the modes in the frame co-rotating with the envelope. For a mode of radial order $n$ and of magnetic period $\mathrm{P}_{\mathrm{M},n}$, the period in the co-rotating frame is:
\begin{equation}
    \mathrm{P}_{n}= \frac{\mathrm{P}_{\mathrm{M},n}}{\sqrt{1+\left(\dfrac{m\omega_{\rm A,env}\mathrm{P}_{\mathrm{M},n}}{2\pi}\right)^{2}}} \, .
\end{equation}
Differentiating this relation, to retrieve a modified dip profile, this holds approximately for the period-spacing:
\begin{equation}
    \Delta \mathrm{P}_{\mathrm{co}}= \frac{\Delta \mathrm{P}_{\mathrm{M}}}{\left(1+\left(\dfrac{m\omega_{\rm A,env}\mathrm{P}_{\mathrm{M}}}{2\pi}\right)^{2}\right)^{3/2}} \, .
\end{equation}
Envelope magnetism, in this particular configuration, adds a curvature to the period-spacing pattern. Particularly, the hydrodynamic spin parameter $s^{*}_{\rm env}$ of the pure inertial mode in the frame co-rotating with the envelope is further shifted by the envelope magnetism towards:
\begin{equation}
    s^{*}_{\rm env} = \frac{s_{\rm M,env}^{*}}{\sqrt{1 + m^{2}\mathrm{Le_{\rm env}^{2}}s_{\rm M, env}^{*2}}} \, .
\end{equation}

\subsection{Summary of the expressions}\label{subsec:summary}

Our calculations were pursued in the framework of a continuous $N$ at the core boundary. If the wavelength of the $m\text{-}g\text{-}i$ mode is superior to the lengthscale at which $N$ varies at the boundary, a formalism treating $N$ as being discontinuous is more relevant, as discussed in both \citetalias{Tokuno2022AsteroseismologyOscillations} and \citetalias{Barrault2025ConstrainingDips}. We did not treat this case in the core of the text in this work for conciseness. Calculations are similar to the former case, except for some specificities detailed in Appendix~\ref{App:dip_disc}. \\
\indent We provide for readability in Appendix~\ref{App:summary} the table~\ref{tab:gen_expression}, which gathers the expressions for the coupling equation to solve numerically as well as expressions for the approximate Lorentzian profile derived in the cases of (1) solid-body rotation and uniform Alfvén frequency and (2) solid-body rotation and bi-layer Alfvén frequency, in the framework of a continuous $N$ in the region $[R_{\rm core},r_{a}]$. The same is also given in Table~\ref{tab:gen_expression} in the framework of a discontinuous $N$ at $R_{\rm core}$.

\section{Results and Discussion}\label{sec:discuss}
We consider for the following discussion and examples the interaction between $(k=0,m=-1)$ envelope $m\text{-}g\text{-}i$ modes and the $(l=3,m=-1)$ m-i modes, except in section~\ref{subsec:lift:deg}. This envelope Kelvin mode series is the most frequent in the spectrum of $\gamma$-Dor analyzed by \citet{Li2020Gravity-modeKepler}, and this particular interaction creates dips in a range of spin parameters observable with the 4-years long baseline of the \textsl{Kepler} mission. Dips observed by \citet{Saio2021RotationModes} resulted from this interaction. For this particular case, $F_{l}^{m}$ reads:
\begin{equation}
    F_{3}^{-1}(\nu) = -\frac{\nu^{2}-10\nu-15}{(\nu+1)(\nu^{2}-5)} \, ,
\end{equation}
and the magnetic structural parameter of the core mode is $\nu_{\rm M,core}^{*} \simeq 11.3245$.
\subsection{Effect of magnetism on the morphology of the dip}
\subsubsection{Case 1: model with uniform rotation rate and Alfvén frequency}
\begin{figure}
    \centering
    \includegraphics[width=\linewidth]{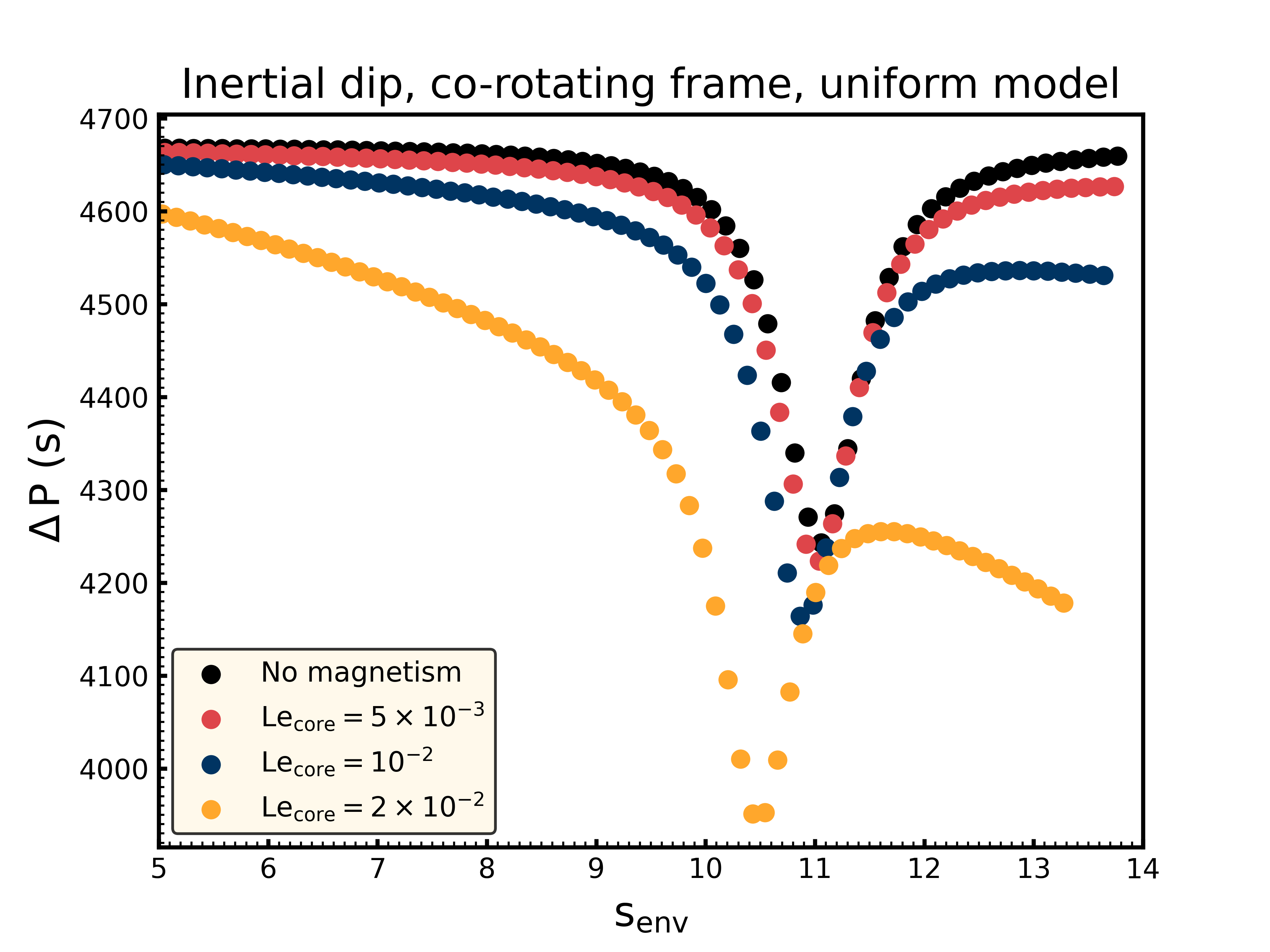}
    \caption{PSPs containing inertial dips obtained by solving Eq.~\eqref{eq:coupling_cont}, from a situation with no magnetic fields (black) to a $\rm Le=2\times 10^{-2}$. The coupling parameter is fixed at $\epsilon =1.5\times 10^{-2}$, the buoyancy travel time $\Pi_{0,\rm M} = 4670\rm s$, the rotation rate $\Omega=2\pi\times 1.14 \, \rm c.d^{-1}$, parameters adapted to the $im$ model.}
    \label{fig:cont_dip}
\end{figure}
First, we study the case 1 where the star is in solid-body rotation with uniform Alfvén frequency. We take a closer look at the coupling equation Eq.~\eqref{eq:coupling_cont}, an we compare it to Eq.(57) of \citetalias{Tokuno2022AsteroseismologyOscillations}. With the magnetic variables $s_{\rm M}$ and $\nu_{\rm M}$, the two equations display the same structure. The difference lies in the dependence of the angular structure of the modes (Hough functions for envelope $m$-$g$-$i$ modes, Legendre polynomial for the Bryan solution of core $m\text{-}i$ modes) on the magnetic structural parameter $\nu_{\rm M}$, and not on the magnetic spin parameter $s_{\rm M}$, while without magnetism, the two quantities both revert back to the hydrodynamic spin parameter $s$. \\
\indent For the considered mode interaction, the equations determining the spin parameter of the mode in the core evolved from the hydrodynamical case to the magnetic one: the condition $\nu^{*} = 11.3245$ leads to a reduced hydrodynamical envelope spin parameter $s^{*}$ of the pure inertial mode. We plot in Fig.~\ref{fig:cont_dip} inertial dips in the frame co-rotating with the envelope obtained by solving numerically the coupling equation Eq.~\eqref{eq:coupling_cont}, for a fixed value of the coupling parameter $\epsilon$ and different Lehnert numbers. The dip is effectively brought to lower periods due to magnetism. In this figure, the additional curvature brought by envelope magnetism is retrieved, as explained in subsection \ref{subsec:mag_to_corot_var}, and modelised in more complex magnetic topologies \citep{Dhouib2022DetectingField,Lignieres2024PerturbativeModes,Rui2024AsteroseismicStars}. \\
\indent This additional curvature of the period-spacing further brings the dip to lower periods in the co-rotating frame, or spin parameters. The problem is thus to disentangle the core magnetic field contribution in this dip shift. 

\subsubsection{Model with uniform rotation rate and bi-layer Alfvén frequency}
\label{subsec:results_bilayer}

\begin{figure}
    \centering
    \includegraphics[width=1.05\linewidth]{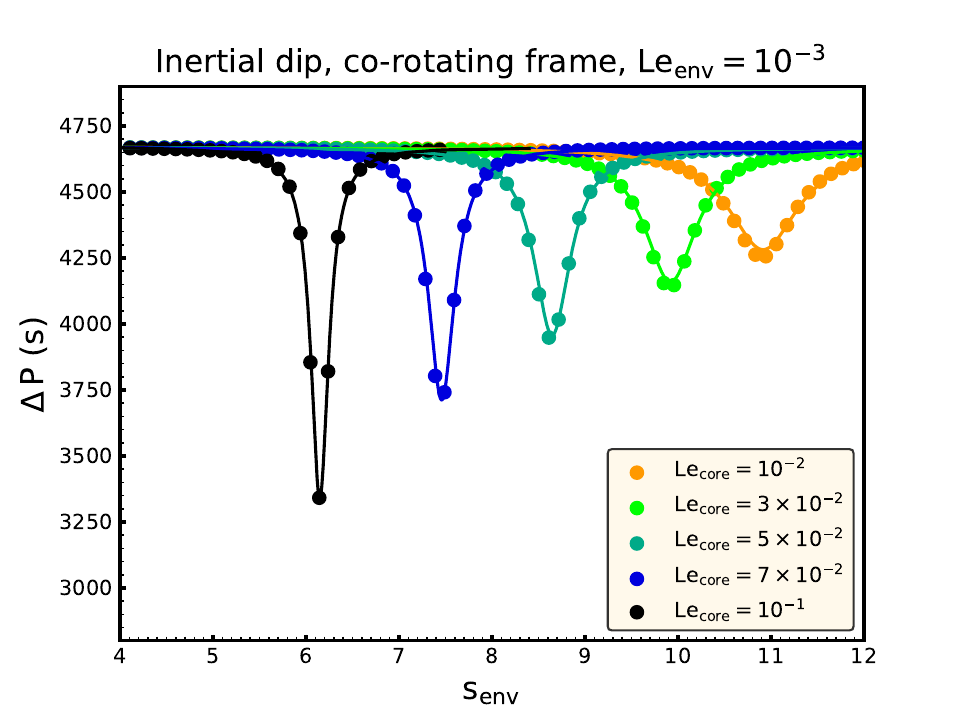}
    \caption{Inertial dips overplotted for different core Lehnert number for a fixed envelope Lehnert number $\rm Le_{env} = 10^{-3}$ and coupling parameter $\epsilon = 1.5 \times 10^{-2}$, in a uniformly rotating star, in the framework of a continuous $N$ at $r_{a}$. Dots are obtained by solving Eq.~\eqref{eq:coupling_cont}, and continuous line by applying the dip profile given by Eq.~\eqref{eq:coupl_disc}.}
    \label{fig:disc_Alfv}
\end{figure}

We move to the case of a bi-layer Alfvén frequency, to finely exhibit the contribution of core and envelope magnetism, respectively. We stay in the framework of a solid-body rotating star, to isolate the magnetic effects from the one of the Doppler shift brought by differential rotation, highlighted in \citetalias{Barrault2025ConstrainingDips}.\\
We show the evolution of the dip shape and location for a fixed envelope Lehnert number $\rm Le_{env}=10^{-3}$ with values of $\epsilon$, $\Omega_{\rm env}$ and $\Pi_{0,\rm M}$ adapted to the $im$ model, and an evolving $\rm Le_{core}\in \{1\times 10^{-2},3\times 10^{-2},5\times 10^{-2},7\times 10^{-2},1\times 10^{-1} \}$ in Fig.~\ref{fig:disc_Alfv}. These Lehnert numbers corresponding to magnetic field mid-core values of $\{0.29,0.86,1.4,20,29\}\rm MG$ at the equator. We plot the numerical solutions of the coupling equation \eqref{eq:coupl_disc} as dots, and superimpose Lorentzian profiles described by Eq.~\eqref{eq:final_profile}. We first verify the compliance of the Lorentzian profile with the numerical computation for a small $\epsilon$. If $\epsilon$ becomes non negligible, terms of order $\epsilon^{2}$ that are neglected in our derivation of the dip profile play a role and make the dip profile shallower than the numerically-computed dip, as already seen in \citetalias{Tokuno2022AsteroseismologyOscillations}. For this envelope Lehnert number, chosen to avoid the regime where the envelope $m$-$g$-$i$ modes are suppressed, the magnetic curvature seen in Fig.~\ref{fig:cont_dip} is indistinguishable from a flat baseline. \\
\indent The dip is shifted to lower envelope spin parameters, which is due to the already mentioned fact that $s_{\rm M,env}^*$ decreases from a fixed value $\nu_{\rm M,core}^*$ with core magnetism. The dip is getting thinner as the core magnetism increases. This is comparable to the differentially-rotating, hydrodynamic case analyzed in \citetalias{Barrault2025ConstrainingDips}. This can be seen physically if we analyze the variation of the function $G_{\rm M}^{-1}$, the contrast between $s_{\rm M,core}$ and $s_{\rm M,env}$ as a function of the Lehnert numbers of both the core and the envelope. In the case of a uniformly rotating star, this reads:
\begin{equation}
    s_{\rm M, core} = G_{\rm M}^{-1}(s_{\rm M,env}) = \frac{s_{\rm M, env}}{\sqrt{1-m^2(\rm Le_{core}^2 - Le_{env}^2)s_{\rm M,env}^{2}}} \, .
    \label{eq:smcore}
\end{equation}
Given that the envelope modes are equally spaced in magnetic spin parameter $s_{\rm M,env}$, we can see that the action of a magnetic field stronger in the core than in the envelope ($\rm Le_{core}>Le_{env}$) is to increase $s_{\rm M,core}$. From the core magnetic frame where the magnetic frequency variables are used, the spacing between the $m$-$g$-$i$ modes is increased with enhanced magnetic contrast. This is absolutely relatable to the differentially rotating, hydrodynamical situation \citepalias{Barrault2025ConstrainingDips} in which the Doppler shift decreases the density of $g$-$i$ modes seen from the frame co-rotating with the core.\\
Differentiating Eq.~\eqref{eq:smcore}, we get the following expression for the density of $m$-$g$-$i$ modes seen in the core magnetic frame:
\begin{equation}
    \frac{\mathrm{d}s_{\rm M,core}}{s_{\rm M,core}} = \frac{\mathrm{d}s_{\rm M,env}}{s_{\rm M,env}} \times \frac{1}{1-m^2(\rm Le_{core}^2 - Le_{env}^2)s_{M,env}^2} \, .
    \label{eq:dsmcore}
\end{equation}
We see that this magnetic Doppler shift is thus decreasing the density of the envelope modes seen from the core magnetic frame.
The core $m\text{-}i$ mode can couple significantly with less modes in its vicinity. The dip with increasing $\rm Le_{core}$ for a fixed $\rm Le_{env}$ contains thus less modes, and is deepened, with only a few modes influenced by the core $m\text{-}i$ mode. \\
One can see considering Eqs.~\ref{eq:smcore} and \ref{eq:dsmcore} that the magnetic effects are only significant if the product $\rm Le_{\rm zone}^2s_{\rm M,zone}^2$ is inferior to 1, but not negligibly small. This translates to $\omega_{\rm A,zone}^2$ being smaller, but non negligible compared to  $\sigma_{\rm zone}^2$: the lower the mode frequency, the stronger the magnetic effect. Since the $m\text{-}i$ mode has a fixed value of $\nu_{\rm M,core}^*$, its frequency is increasing for faster rotating stars. This means that the magnetic effects will be sensible on the interaction for a lower magnetic field amplitude for slower rotators. We further discuss this point in sub-section \ref{subsec:degeneracy}.\\

\subsection{Core magnetic field measurability from the study of the dip}
\label{subsec:detect:mag}

\begin{figure*}
    \centering
    \includegraphics[trim = 1cm 5cm 0 5cm,clip,width = 1.1\linewidth]{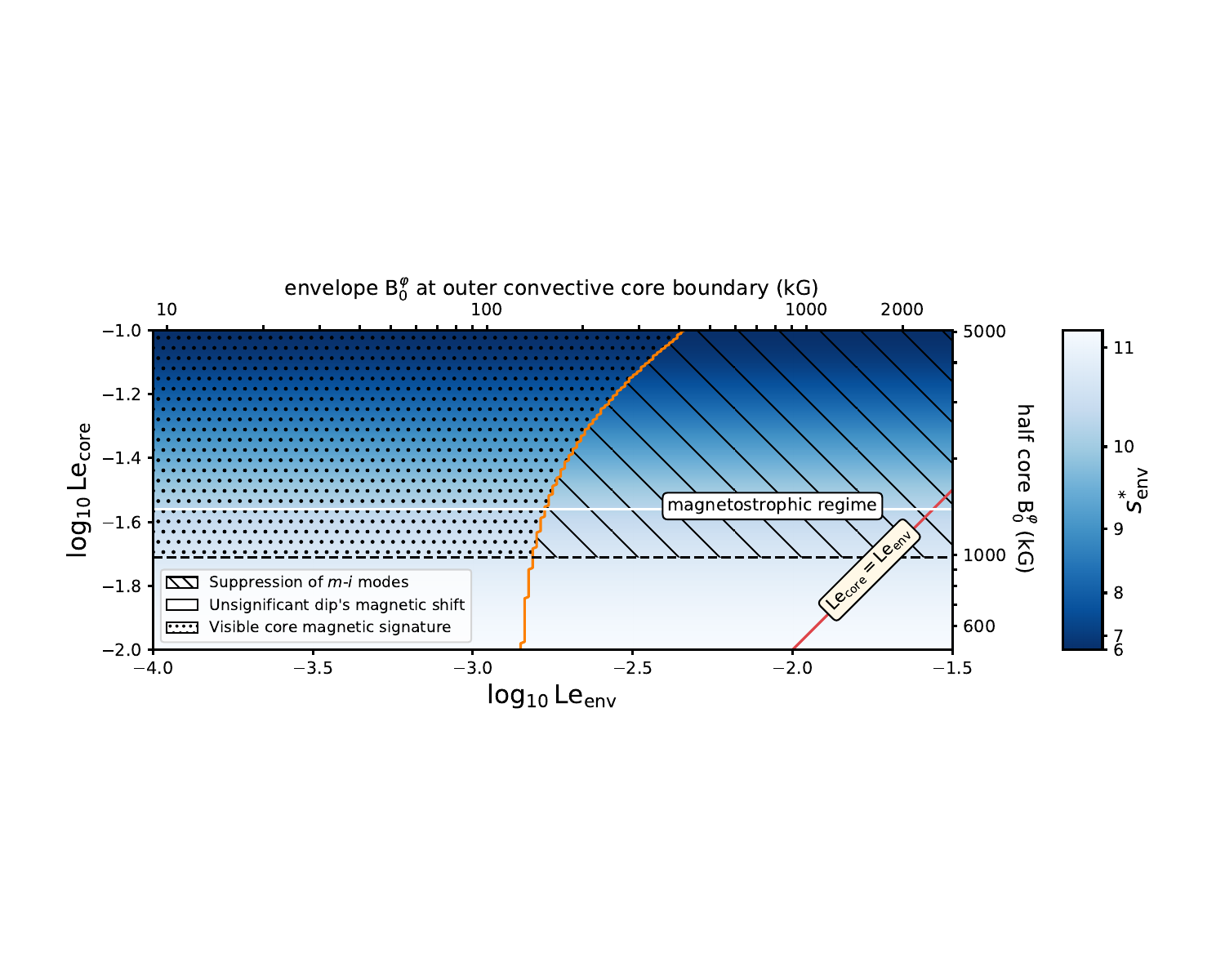}
    \caption{Ranges of core and envelope magnetism for which a measurement of core magnetic field is considered accessible from the dip study, for model $fz$, represented as a dotted region in the $\{\rm Le_{core}, Le_{env}\}$ parameter space. The figure is coloured by the value of the envelope hydrodynamic spin parameter corresponding to the pure inertial mode, $s_{\rm env}^{*}$. Core and envelope Lehnert numbers are related to respectively the half-core magnetic field, and the magnetic field at the outer core-to-envelope boundary. Above the hatched line, the variation of the $s_{\rm env}^{*}$ is considered significantly deviating from its value in the solid-body, hydrodynamic case. On the right of the orange line, the hatched region shows the parameter space for which modes at $s_{\rm env}^*$ could be suppressed. The red continuous line shows the locus of equal Lehnert numbers in both zones. The white horizontal line shows an estimation of the magnetic field in the magnetostrophic regime obtained from the MESA model. The super-equipartition regime does not appear for the range of considered $\rm Le_{core}$.}
    \label{fig:detect_zone_ZAMS}
\end{figure*}
\begin{figure*}
    \centering
    \includegraphics[trim = 1cm 5cm 0 5cm,clip,width = 1.1\linewidth]{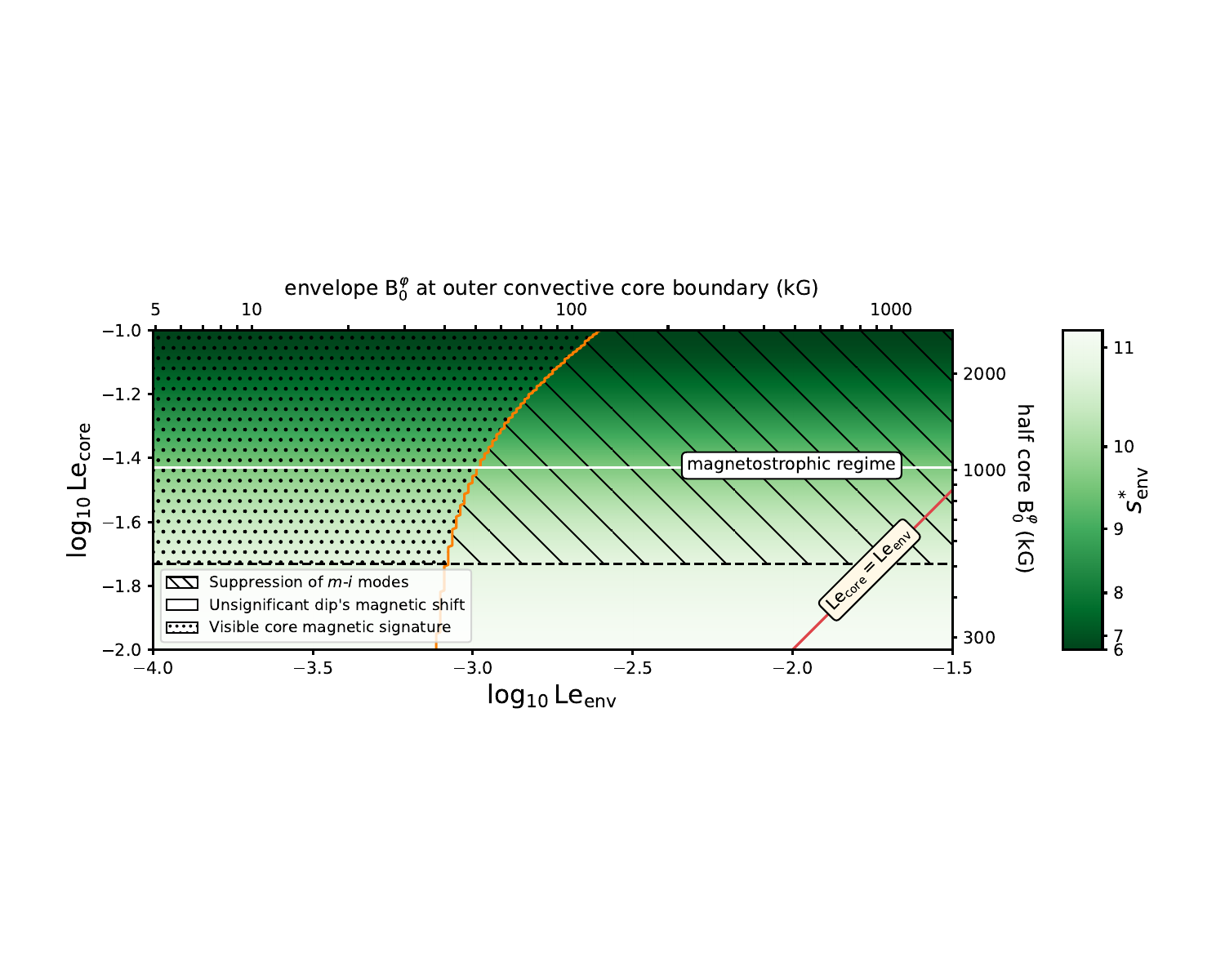}
    \caption{Same as Fig.\ref{fig:detect_zone_ZAMS}, for model $iz$.}
    \label{fig:detect_zone_ZAMS_slow}
\end{figure*}
\begin{figure*}
    \centering
    \includegraphics[trim = 1cm 5cm 0 5cm,clip,width = 1.1\linewidth]{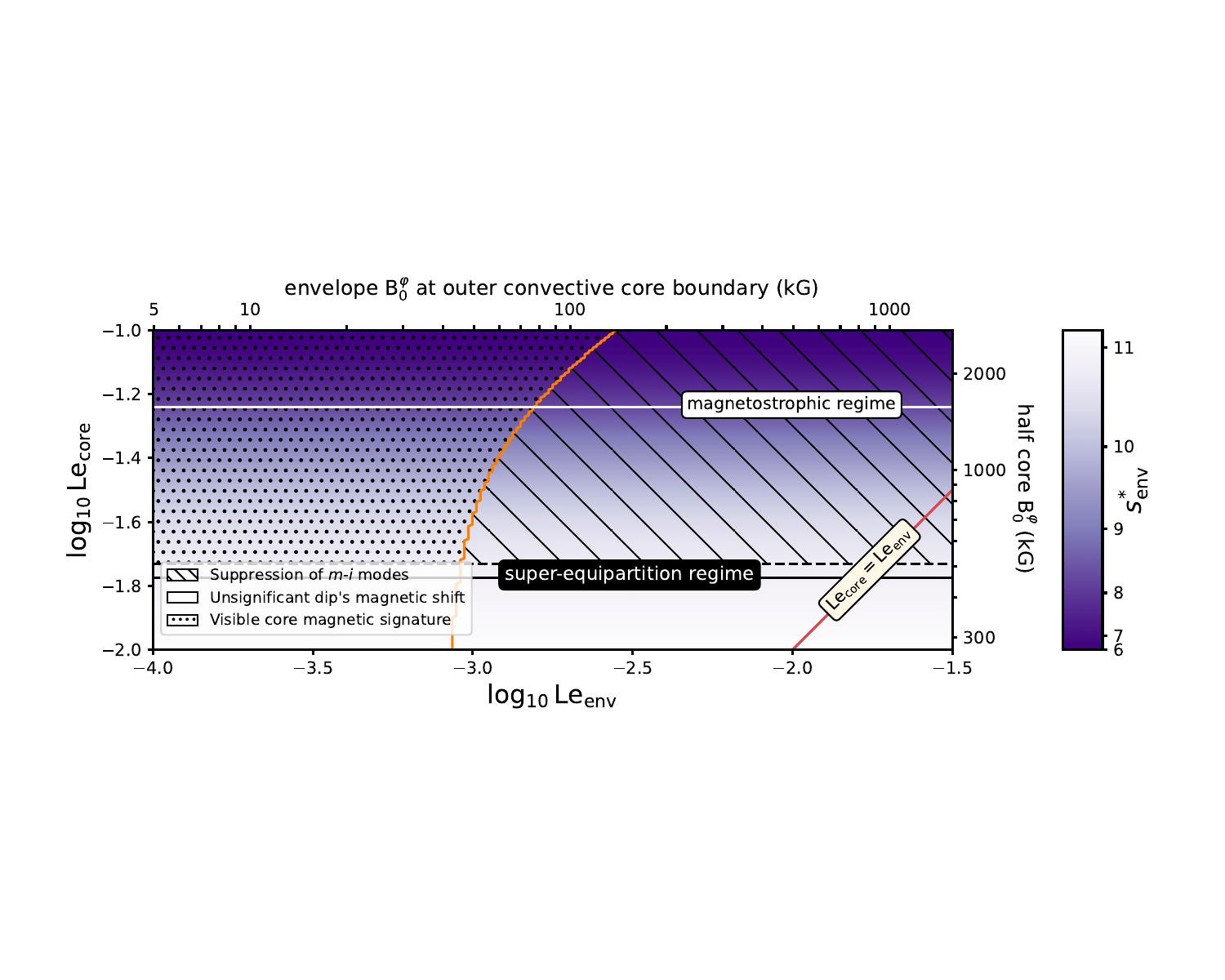}
    \caption{Same as Fig.\ref{fig:detect_zone_ZAMS}, for model $im$. The black continuous horizontal line shows an estimation of the magnetic field in the super-equipartition regime obtained from the MESA model.}
    \label{fig:detect_zone_TAMS_slow}
\end{figure*}
We present the evolution of the envelope (hydrodynamical) spin parameter at which the core $m\text{-}i$ mode appears, $s_{\rm env}^*$, as a function of $\rm Le_{core}$ in Figs.~\ref{fig:detect_zone_ZAMS}, \ref{fig:detect_zone_ZAMS_slow} and \ref{fig:detect_zone_TAMS_slow}, corresponding to respectively the $fz$, $iz$ and $im$ models. While the morphology of the dip is modified by an envelope magnetism, as it makes the density of $m$-$g$-$i$ modes vary around the period at which the $m\text{-}i$ mode appears, $s_{\rm env}^{*}$ is unsensitive to envelope magnetism and provides a window on the core only: regardless of envelope magnetism, if $m$-$g$-$i$ modes of spin parameters around $s_{\rm env}^{*}$ are present in the PSP, then they can interact with the magneto inertial mode. In the model considering a continuous $N$ at the core-to-envelope boundary, we have $s_{\rm env}^* \neq s_{\rm env, min}$, the spin parameter at which the dip reaches its minimum, the latter depending on envelope magnetism and stratification. \\
\indent Degeneracy with core-to-envelope differential rotation put aside (later discussed in subsection \ref{subsec:degeneracy}), the inference of a convective core magnetic field from the study of the inertial dip can be hindered by two processes: 1) a non-significant shift of the dip structure due to magnetism compared to a solid-body rotating, hydrodynamical situation, and 2) a suppression of $m$-$g$-$i$ modes close to the core due to a strong magnetism. These two processes limit the range of core and envelope magnetism potentially inferred from the dip study to a regime in which core magnetic fields are sufficiently strong to have a significant impact on the envelope spin parameter $s_{\rm env}^{*}$ at which the core $m\text{-}i$ mode would appear, and envelope magnetic fields lower than the limit at which they would suppress $m$-$g$-$i$ modes. We show in Figs.~\ref{fig:detect_zone_ZAMS}, \ref{fig:detect_zone_ZAMS_slow} and \ref{fig:detect_zone_TAMS_slow} an estimate of the extent of these ranges for the three considered models as a dotted region. In each figure, the considered ranges of core and envelope Lehnert numbers are translated to respectively the mid-core value and the near-core equatorial values of the magnetic field.

\subsubsection{Non-significant shift of the dip due to weak core magnetic field}
For each figure, we estimate a lower limit on the difference between $s_{\rm env}^{*}$ and the solid-body, non-magnetic value of 11.3245 that could be considered significant enough in data. This corresponds to the horizontal dashed lines in the figures. Above, the difference is considered high enough for a significant deviation from the solid-body, non magnetic model. The main uncertainty hindering this inference is the one on the near-core rotation rate, estimated by fitting the PSP in the inertial frame with the TAR. An estimation of this uncertainty is made using an average of the uncertainties derived by \citet{Li2020Gravity-modeKepler} in a range of near-core rotation rates $\Omega_{\rm nc}$ surrounding the one of the models, by $\pm 0.1\Omega_{\rm nc}$. \\

\subsubsection{Suppression of $m\text{-}g\text{-}i$ modes by strong envelope magnetic field}
We then aim to estimate a limit in the envelope magnetic field above which the vertical propagation would be inhibited. In this situation, the $m$-$g$-$i$ waves would be suppressed before reaching the core and would not form modes seen in the PSP. In our model with a pure toroidal field, we saw that the $m$-$g$-$i$ mode become evanescent is $\sigma_{\rm env}<\omega_{\rm A,env}$. The envelope Alfv\'en frequency would thus be the relevant cut-off frequency in our framework. However, we seek to give more accurate limits on a potential cut-off frequency, by considering a suppression mechanism by the radial component of the field the modes are most sensitive too, explored by \citet{Rui2023GravityFields}.\\
\indent In both the field generation mechanisms that we described, i.e. the fossil field scenario and the Tayler-Spruit mechanism, a radial field should be present alongside a toroidal field, even though the toroidal component can dominate. \citet{Braithwaite2009AxisymmetricComponents} stated that the ratio between the energy in the poloidal component over the total energy has to be superior than $10^{5}$ for a fossil field to be stable, and \citet{Petitdemange2024TaylerSpruitLayers} exhibit a ratio of $10^{-2}$ to $10^{-1}$ of the radial component of the field compared to the toroidal one. Yet, we saw in Fig.\eqref{fig:prop-diag} that in $\gamma$-Dor interiors $N/\omega_{\rm A, env}>10^{3}$. The negligibility of $\omega_{\rm A}$ computed with the toroidal field in our framework means that the frequency at which the modes are suppressed by a \citet{Rui2023GravityFields}'s like mechanism, $\omega_{\rm B} = \sqrt{N\omega_{\mathrm{A},r}}$ (with $\omega_{\mathrm{A},r}$ the Alfv\'en frequency computed with the radial component of the field), is still superior to $\omega_{\rm A}$ given the toroidal-to-radial component ratios considered in both mechanisms. \\
\indent To give a conservative estimate of the limit envelope magnetic field that would lead to $m$-$g$-$i$ mode suppression, we thus choose to consider the frequency $\omega_{\rm B} = \sqrt{\bar{N}\omega_{\rm A,env}}$, with $\bar{N}$ a mean value of $N$ probed by the $m$-$g$-$i$ modes in the JWKB limit given in Appendix~\ref{Appendix:models}. If $s_{\rm env}^{*}$ exceeds $2\Omega_{\rm env}/\omega_{B}$, it means that the mode would be suppressed in a situation where the radial component of the magnetic field has the same amplitude as the toroidal one. This limit is marked by an orange line on the three figures. One could also consider the maximum value of $N$, $N_{\rm max}$ reached close to the core-to-envelope boundary. This value deviates more from the averaged value as the star ages, because of the receeding convective core. It is unclear if the $m$-$g$-$i$ mode energy would actually be completely tamed by the local $N$ spike near the core, or if it would result in only a partial suppression, as seen observationally in the case of RGB stars \citep[e.g][]{Garcia2014StudyModes}. For these reasons, as well as the previously explained uncertainty on the suppression mechanism when a dominantly toroidal magnetic field is considered, this line should be considered with care, and can only be regarded as a rough estimate of this limit field.\\
\indent We see that in none of the models, the line of $\rm Le_{core} = Le_{env}$ lies in the dotted region: a magnetic field with uniform $\omega_{\rm A}$ which would have a detectable shift of $s^{*}_{\rm env}$ would imply a suppression of $m$-$g$-$i$ modes by the envelope magnetic field.

\subsubsection{Comparison to expectations for core and envelope magnetic fields}
In Figs.~\ref{fig:detect_zone_ZAMS}, ~\ref{fig:detect_zone_ZAMS_slow} and ~\ref{fig:detect_zone_TAMS_slow}, we plot in horizontal lines an estimate of the core field strength obtained in a magnetostrophic regime, where the Lorentz force balances the Coriolis acceleration. Given the fast rotation of the $\gamma$-Dors we are interested in, the Rossby numbers of the convective flow are very low, of the order of $10^{-4}$ to $10^{-3}$ in all the considered models. This advocates for a regime of super-equipartition, with the ratio of magnetic to kinetic energy densities scaling as $a+b\rm Ro^{-1}$. This relation is considered in several simulations listed in \citet{Augustson2019APenetration}. However, we note that their minimal Rossby number is superior to the Rossby numbers estimated in our case, using the Mixing Length Theory from our MESA models. \citet{Featherstone2009EffectsStars} found a magnetic energy density of 10 times the kinetic energy density in the case of a $2.0 \, \rm M_{\odot}$ star rotating 10 times slower than the slowest of our cases. We thus retained this estimate for a lower limit on the core dynamo-generated magnetic field to be expected. This limit is too low to appear on Figs.~\ref{fig:detect_zone_ZAMS} and ~\ref{fig:detect_zone_ZAMS_slow} but plotted as a black vertical line in Fig.~\ref{fig:detect_zone_TAMS_slow}. The mid-core magnetic field amplitude in the regime of equipartition $\rm B_{equi}$ is given in Table~\ref{tab:models}. With the strong rotation rate at play in the considerd $\gamma$-Dors, core dynamo should reach the regime of magnetostrophy, with flows heavily structured by rotation and magnetism. We retain this field $\rm B_{ms}$ as an order-of-magnitude estimate of the dynamo-generated magnetic field in our models.\\
\indent A magnetostrophic regime would correspond to a $s_{\rm env}^*$ of 10.26 in the $fz$ model, of 9.63 in the $iz$ model, of 8.41 in the $im$ model. We see that even though the magnetostrophic field is higher for the fast rotating model, it would be harder to probe than in the intermediate-rotating model, as we showed that the magnetic effect is depending on the Lehnert number: a fast rotation requires a strong magnetic field for its effects to be noticeable. At comparable rotation rates, the $im$ model shows a more noticeable effect of magnetism on $s_{\rm env}^*$ due to the increased convective velocities computed in the MLT for its core.\\
\indent As for the envelope, an estimation of Tayler-Spruit generated toroidal magnetic fields is uneasy to get from first principles, as it is a highly non-linear process. In MHD simulations, \citet{Petitdemange2023Spin-downLayers} obtained a toroidal magnetic field of the order of $40 \, \rm kG$ for a set of parameters in a weakly stratified layer $N/\Omega\approx 1.24$, centered at the middle of their modelled radiative zone. In a strongly stratified regime, with $N/\Omega\approx 50$, the magnetic field could reach $2\, \rm MG$ close at the core-to-envelope boundary, but for a high Rossby number inapplicable to our situation. \citet{Petitdemange2024TaylerSpruitLayers} showed that a number of simulations showed a magnetostrophic regime of the dynamo, i.e. the ratio of the magnetic energy on the kinetic energy scales as the inverse of the Rossby number. Given the rotation rates of the three models, we hypothesize that the Tayler-Spruit generated magnetic field will be stronger for the $fz$ model compared to the $iz$ and the $im$. From the point of view of the mode suppression, stars with intermediate rotation stand as the best target for the inertial dip study. We however prefer not to state on the establishment of this dynamo in our modelled stars and on the field amplitude expected in their radiative zone. \\

\subsubsection{Impact of uncertainties on envelope stratification}

It remains to discuss if one can actually measure $s_{\rm env}^{*}$, due to uncertainties on the near-core stratification. First, as shown in Table.~\ref{tab:gen_expression} and already discussed in the hydrodynamical case by \citetalias{Barrault2025ConstrainingDips}, a continuous $N$ profile at the boundary induces a phase shift of the $m$-$g$-$i$ mode compared to a discontinuous $N$, which makes the location of the minimum of the dip shifted from this value $s_{\rm env}^{*}$. Indeed in the case of a discontinuous $N$, we have $r_{a}=R_{\rm core}$ and the minimum of the dip is at $s_{\rm env}^{*}$. If one aims to measure $s_{\rm env}^{*}$, there is thus an uncertainty coming from an indetermination of the near-core profile of $N$. Second, chemical modulations present in the PSP will have an impact on the dip fitting procedure in a way described in Appendix G of \citetalias{Barrault2025ConstrainingDips}.
A precise determination of $s_{\rm env}^{*}$ would then benefit from a precise modelling of radiative zone's features and properties.

\subsection{Degeneracy between differential rotation and magnetism}
\subsubsection{Effect on the spin parameter of the dip}
\label{subsec:degeneracy}
\begin{figure}
    \centering
    \includegraphics[width=1.05\linewidth]{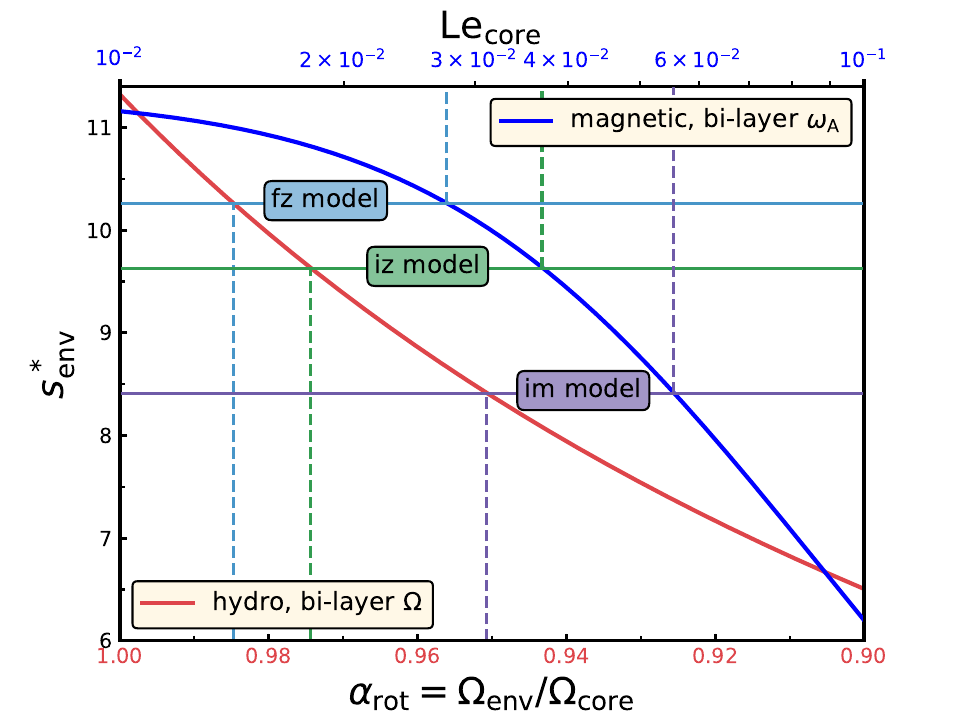}
    \caption{Red: Variation of $s^{*}_{\rm env}$ in the hydrodynamic, differentially-rotating case with respect to the amount of differential rotation. Blue: the MHD, bi-layer Alfvén frequency case with respect to the core Lehnert number. Considered core Lehnert numbers translate to mid-core magnetic fields range from 0.5 to 5 $\rm MG$ in the $fz$ model, and 0.25 to 2.5 $\rm MG$ in the $im$ model. Horizontal thin lines represent the value of $s_{\rm env}^*$ expected for each of the models in the magnetostrophic regime.
    In the $im$ model, a mid-core equatorial magnetic field in the magnetostrophic regime of $1.63 \, \rm MG$, corresponding to $\rm Le_{\rm core} = 5.6\times 10^{-2}$, would shift the dip to $s_{\rm env}^* = 8.41$ in a solid-body rotating, magnetic model. This is the same effect as a differential rotation of $\alpha_{\rm rot} = 0.951$ in the differentially rotating, hydrodynamic model.}
    \label{fig:degen}
\end{figure}

We saw in subsection \ref{subsec:results_bilayer} that the effect of a bi-layer Alfvén frequency is quite comparable to the signature of differential rotation. The local magnetic frequency of the $m\text{-}i$ mode is increased in the envelope compared to the core if $\rm Le_{core}>Le_{\rm env}$, which is comparable to a Doppler shift brought by increased rotation in the core compared to the envelope. To give further perspective, we compare the expression of $G_{\rm M}$ to its hydrodynamical equivalent $G$ used in \citetalias{Barrault2025ConstrainingDips}:
\begin{equation}
    s_{\rm M, env} = G_{\rm M}(s_{\rm Mcore}) = \frac{s_{\rm M,core}}{\sqrt{1+m^{2}(\mathrm{Le_{core}^{2}} - \mathrm{Le_{env}^{2}}) s_{\rm M, core}^{2}}}
\end{equation}
and
\begin{equation}
    s_{\rm env} = G(s_{\rm core}) = \frac{\alpha_{\rm rot}s_{\rm core}}{1+\frac{m}{2}(\alpha_{\rm rot} - 1) s_{\rm core}} \, ,
\end{equation}
with $\alpha_{\rm rot} = \Omega_{\rm env}/\Omega_{\rm core}$. The key difference is that at the zeroth order in respectively $(\mathrm{Le_{core}^{2}} - \mathrm{Le_{env}^{2}})s_{\rm M,core}^2$ and $(\alpha_{\rm rot} - 1) s_{\rm core}$, $s_{\rm M,env}\propto s_{\rm M,core}$, while $s_{\rm env}\propto \alpha_{\rm rot}s_{\rm core}$. This implies that shifts due to a limited rotation contrast will only be mimicked by a strong magnetic contrast, if the magnetic field is high enough to ensure that $\rm Le_{core}^2s_{\rm M,core}^2$ is non negligibly small compared to unity. \\
\indent We investigate the degeneracy of the signature of differential rotation and magnetism, beginning first with the sole effect on $s_{\rm env}^*$, the hydrodynamical envelope spin parameter at which the $m\text{-}i$ mode (or pure inertial mode in the hydrodynamical case) appears. In Fig.~\ref{fig:degen}, we plot this quantity, invariant with $\rm Le_{env}$, as a function of the amount of differential rotation in the hydrodynamical model, and of the core Lehnert number. As expected with the qualitative reasoning previously held, the signature of a core magnetic field would be only significant for relatively high amplitudes: 1 \% of core-to-envelope differential rotation would have the same effect as a Lehnert number of $2.2\times 10^{-2
}$, which corresponds to an equatorial mid-core magnetic field of $1.1 \, \rm MG$ in the $fz$ model, $600 \, \rm k G$ in the $iz$ model, and $620 \, \rm k G$ in the $im$ model. \\
Due to the competition between magnetism and the establishment of differential rotation previously explained, it is unlikely that a differential rotation of about $10 \%$ between the core and the envelope will be present at the same time as a core magnetic field at a $\rm MG$ scale. In MHD simulations of a 2.0 $\rm{M_{\odot}}$ A-type star, \citet{Featherstone2009EffectsStars} found a core-to-envelope differential rotation of less than 1 \%. Considering the expected shift in $s_{\rm M,env}^{*}$ given for the three models, this amount of differential rotation would make the detection of magnetic fields uncertain in the case of the $fz$ model: the magnetic effect would be of the same order of magnitude as the differential rotation effect for the estimated core magnetic field in the magnetostrophic regime, and underdominant for a super-equipartion regime. In the $im$ model, the magnetic effect would be on the contrary predominant over the differential rotation effect in a magnetostrophic regime.\\
\indent We conclude by verifying one of the hypotheses inherent to our model: the non-negligibility of the geometrical factor $c_{k,l}$ defined in Eq.~\eqref{def:ckl}. For the range of $\rm Le_{core}$ considered and a low envelope Lehnert number $\rm Le_{env} \approx 10^{-3}$ required for the modes not to be suppressed by the radial component of the magnetic field, the envelope magnetic structural parameter controlling the angular part of Bryan solutions for $m$-$g$-$i$ modes varies in the interval $\nu_{\rm M,env}^* \in [6.2,11.2]$ for $\rm Le_{core} \in [10^{-2},10^{-3}]$. These values are included in the interval of $s_{\rm env}^*$ controlling the angular part of Bryan solutions in the hydrodynamical case analyzed in \citetalias{Barrault2025ConstrainingDips}. The reasoning held in their Appendix F thus holds for the present study and $c_{k,l}$ in our present situation is not negligibly small compared to unity either. \\

\subsubsection{Does the dip shape give additional contraints?}
\begin{figure}
    \centering
    \hspace*{-0.7cm}
    \includegraphics[width=1.2\linewidth]{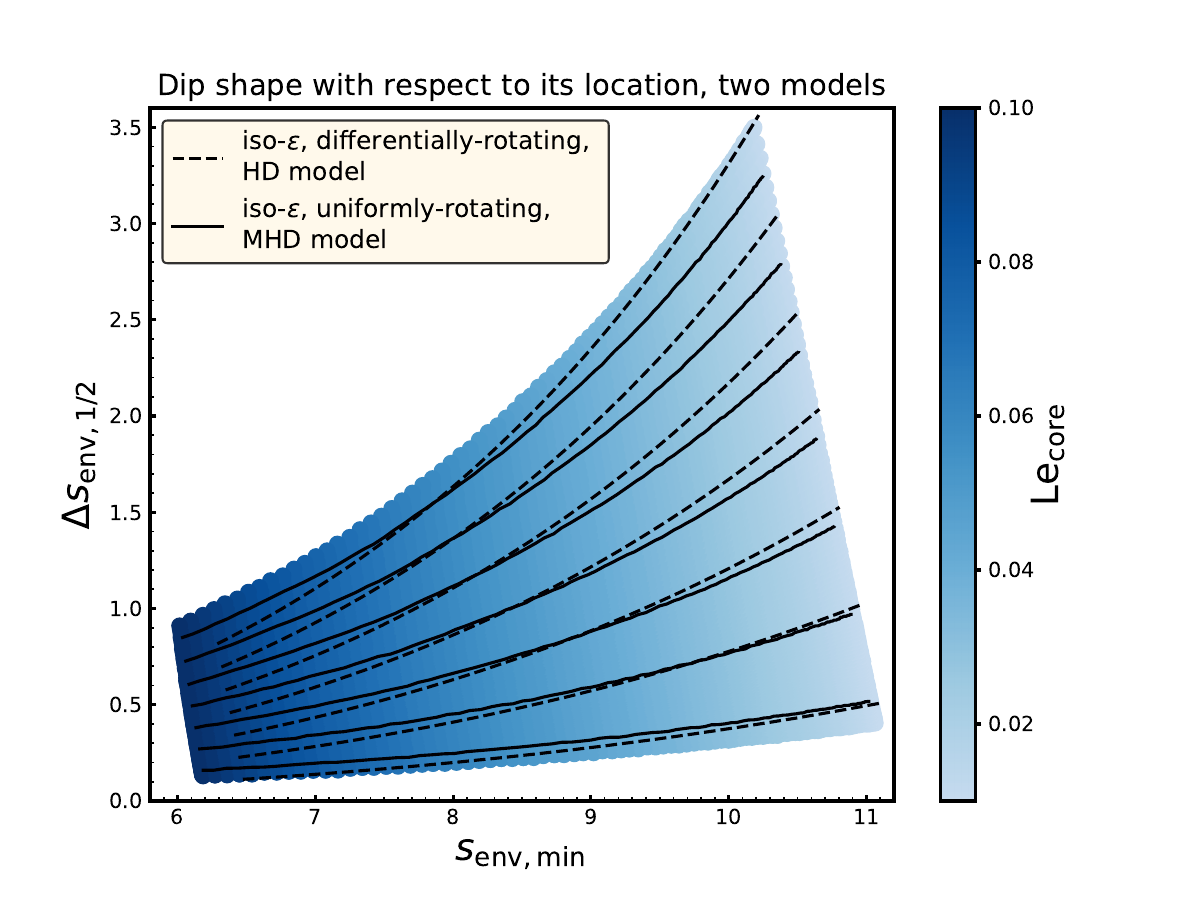}
    \caption{Half-width of the dip $\Delta s_{\rm env,1/2}$ and its minimum spin parameter $s_{\rm env,min}$ for $\rm Le_{core}\in [10^{-2};10^{-1}]$ and a fixed $\rm Le_{\rm env} = 10^{-3}$. $\epsilon$ varies between $6\times 10^{-3}$ and $6\times 10^{-2}$, small enough values allowing us to use the Lorentzian profile derived for this analysis. Thick lines are iso-$\epsilon$ in the uniformly rotating, magnetic case. Dashed line are the same obtained from the hydrodynamic, differentiallty-rotating model.}
    \label{fig:diff_rot_mag}
\end{figure}
So far, we have only considered the variation of $s_{\rm env}^*$, the envelope spin parameter corresponding to the core mode, in the uniformly rotating, bi-layer Alfv\'en frequency framework, and in the bi-layer rotating, hydrodynamical one. We saw that an uncertainty on the near-core stratification translates into an uncertainty on the measurement of $s_{\rm env}^*$, and that there is a strong degeneracy between the hydrodynamical and the MHD cases, that we can try to alleviate considering the antagonist effect of strong magnetic fields and differential rotation.\\
\indent The position of the minimum of the dip $s_{\rm env,min}$, offset from $s_{\rm env}^*$ in the continuous $N$ framework, and equal to it in the discontinuous framework, is not the only observable related to the dip structure. As we saw in section~\ref{subsec:results_bilayer}, the dip gets thinner with increasing $\rm Le_{core}$, so a decrease of the half-width of the dip, $\Delta s_{\rm env,1/2}$ can be a hint for an increase of $\rm Le_{core}$. However, both an increase of core magnetism and core rotation compared to the envelope, have a similar effect on this half-width, as it is the case on $s_{\rm env}^*$. One has also to consider the effect of the coupling parameter $\epsilon$, an increase of which would widen the dip and displace it towards low spin parameters. This latter effect is nevertheless underdominant compared to differential rotation, as argued in \citetalias{Barrault2025ConstrainingDips}. The latter work studied the potential measurement of both the differential rotation $\alpha_{\rm rot}$ and the coupling parameter $\epsilon$ from the dip study. We see that magnetism here complicates the study and adds an extra-layer of degeneracy. 
A way to partially lift it is to provide constraints on this parameter $\epsilon$ by stellar modelling, as done by \citet{Aerts2023ModeStars} for a sample of $\gamma$-Dor and SPB stars. However, $\epsilon$ measured this way suffers from the propagation of uncertainties on stellar modelling such as the core-to-envelope boundary mixing, which would change the shape of $N$ near-core, key parameter for the mode interaction throughout the boundary. Uncertainty also comes from the use of either a continuous or a discontinuous $N$ at the boundary in our models \citepalias{Barrault2025ConstrainingDips}. Due to the receding convective core during the MS evolution leaving a strong gradient of near-core molecular weight and hence a sharp increase of $N$ near-core, the latter model would be more applicable for stars approaching the TAMS, so both models would not be equally favoured during the evolution, but an uncertainty would still remain. \\
\indent We plot in Fig.~\ref{fig:diff_rot_mag} a map for which the dip's minimum $s_{\rm env, min}$ and its half-width $\delta s_{\rm env, 1/2}$ are plotted and coloured by their core Lehnert number in the interval $\rm Le_{core} \in [10^{-2};10^{-1}]$, for a fixed value $\rm Le_{env} = 10^{-3}$. In this regime, the envelope magnetic field would not likely cause $m$-$g$-$i$ mode suppression. Lines of iso-$\epsilon$ are overplotted as thick black lines on this figure. We retrieve the fact that at equal $\epsilon$, both the half-width and the minimum of the dip decrease with increasing core magnetism. This behaviour is the same as for differential rotation \citepalias[see Fig.4 of][]{Barrault2025ConstrainingDips}. We overplot on this figure the same lines of iso-$\epsilon$ obtained in the differentially rotating hydrodynamical framework from \citetalias{Barrault2025ConstrainingDips} as dashed lines, with $\alpha_{\rm rot}\in [0.90;1]$ and the same range of $\epsilon$ as for the magnetic case. We see that even though these lines are not overlying, which is expected due to the different expressions of $s_{\rm env,min}$ and $\Delta s_{\rm env, 1/2}$ in the MHD or the hydrodynamical framework, the whole region of $(s_{\rm env,min},\Delta s_{\rm env, 1/2})$ obtained with the magnetic model is spanned by the hydrodynamical model. If no constraint is given on $\epsilon$, a complete degeneracy is observed: the $(s_{\rm env,min},\Delta s_{\rm env, 1/2})$ given by some combination $(\alpha_{\rm rot}, \epsilon_{\rm rot})$ in the hydrodynamical model can be obtained by a combination $(\mathrm{Le_{core}}, \epsilon_{\rm M})$ in the MHD model. Even if a constraint is given on $\epsilon$, the uncertainty on its determination would be too high to safely exclude one of the models. As there would be an additional uncertainty due to the core density stratification not considered in the present model, and an uncertainty on the continuity of $N$ at the boundary, we can state that differential rotation and core magnetism in the considered toroidal configuration are inseparable from a study of the inertial dip between $(l=3,m=-1)$ core $m\text{-}i$ mode and the $(k=0,m=-1)$ envelope Kelvin mode series. \\

\subsection{Lifting degeneracies from the study of additional $g$-$i$ mode series}\label{subsec:lift:deg}
We foresee a promising lead in using inertial dips appearing for different envelope mode series than the $(k=0,m=-1)$ one extensively analyzed in the present work. \citet{Galoy2024PropertiesStars} analyzed two other types of inertial dips for $(k=0,m=-2)$ and $(k=1,m=0)$ series appearing at spin parameters potentially observable in $\gamma$-Dor stars: around $s_{\rm env, min} = 8.5$ for the former and $s_{\rm env, min} = 2.2$ for the latter one. Especially, a dip in the $(k=0,m=-2)$ mode series would be a prime target for our purposes: we saw that the effect of magnetism scales as $m^{2}$, whereas the effect of differential rotation scales as $m$ only. This property has already been used in the context of RGB stars \citep{Loi2020EffectStars,Bugnet2021MagneticFrequencies,Mathis2021ProbingMagneto-asteroseismology,Li2022MagneticStars,Mathis2023AsymmetriesFields,Das2024UnveilingStars}: magnetism provokes asymmetry in the rotational triplets of mixed modes, an effect that can only be attributed to the magnetic field if one considers the first-order effect of rotation. In a similar manner, integrating two dips in the study would contribute to a lift of the degeneracy between differential rotation and core magnetism.
From the study of one inertial dip in the $(k=0,m=-1)$ mode series, we saw that there is degeneracy with a value of $\alpha_{\rm rot}$ in the hydrodynamical model, and a value of $\rm Le_{\rm core}$ in the MHD one. From the study of the dip in $(k=0, m=-2)$ mode series, there would be an other degeneracy between an $\alpha_{\rm rot}$ and a $\rm Le_{\rm core}$. Due to the difference between the differential rotation effect, doubled from the $m=-1$ to the $m=-2$ mode series, and the magnetic effect quadrupled between the same series, only one model will be consistent with both observations.\\
\indent Yet, no detection of dips in the $(k=0,m=-2)$ mode series has ever been made. 
The number of detection of dips in $(k=0,m=-1)$ mode series, the most frequent PSP pattern found in $\gamma$-Dor stars \citep{Li2020Gravity-modeKepler}, is expected to grow. The sample consists as for now of 16 stars visually selected by \citet{Saio2021RotationModes}, without systematic study of the most extended up-to-date \citet{Li2020Gravity-modeKepler}'s sample. We hereby point the interest in analyzing both mode series when available: 42 \% of the stars analyzed by \citet{Li2020Gravity-modeKepler} containing a $(k=0,m=-1)$ PSP contain as well a $(k=0,m=-2)$ one (see their Fig.7).

\subsection{Limitation of the scope}\label{subsec:limit}
\subsubsection{Limitations shared with the hydrodynamic case}
This model shares the limitation exposed in subsection 4.7 of \citet{Barrault2025ConstrainingDips}. We neglected in both works the variation of the buoyancy travel time $\Pi_{\rm 0,M}$, we considered only the dominant mode interaction and thus stayed in the framework of low $\epsilon$ values and disregarded the variation of the geometrical parameter $c_{k,l}$. We hypothesized a sphericity of the core and of the entire star. These hypotheses taken in the hydrodynamical model are consistent with the presence of magnetism, as the magnetic field strengths taken in this work give rise to Lorentz forces only weakly influencing the equilibrium structure, especially in the core and the inner radiative zone (see Appendix~\ref{App:hierarchy}). This is less true for outer regions, for which a magnetic field could displace the upper turning point of the $m$-$g$-$i$ modes $r_{b}$ from their non-magnetic equivalents. The overall sphericity of the outer regions of the star could be attained, leaving an imprint on the angular structure and spectrum of $m$-$g$-$i$ modes \citep{Dhouib2021ThePlanets}. This effect would however be negligible compared to the flattening due to the centrifugal acceleration \citep{Duez2010EffectSun}, which itself has no impact on the coupling problem due to the concentration of the $g\text{-}i$ mode energy at the base of the radiative envelope, but would induce a small shift in the $g\text{-}i$ mode frequencies \citep{Dhouib2021ThePlanets}.Additionally, contrary to their chemically peculiar equivalent Ap stars, $\gamma$-Doradus likely possess a weak surface field, still undetected up-to-date \citep{Henriksen2023RotationalRotation}, which would not change the geometry of the outer layers of the star.\\
\subsubsection{Limitations due to the considered field}
\indent The profile and geometry of the field taken stand out as an obvious limitation of our model. The toroidal field with bi-layer Alfv\'en frequency is taken in this work as a laboratory towards the comprehension of the magnetic effects and their resemblance, or dissemblance with differential rotation. The toroidal hypothesis is in line with a domination of the toroidal component in the dynamo-generated core magnetic field due to fast rotation, as well as for the mainly toroidal field hypothesized in the Tayler-Spruit mechanism. Yet, a radial component of the magnetic field will supposedly have a stronger effect than the toroidal component on the envelope $m$-$g$-$i$ mode, since their displacement is mainly transverse. Second, the geometry of the dynamo-generated core magnetic field would be far more complex, and the influence of a poloidal component would have to be studied. Our study would still be useful in the framework of a realistic field by providing an estimate of the impact of its toroidal component. We expect the impact of a complex field on the dip structure to be qualitatively in line with our findings based on this simple field profile, as what has been shown for the magnetic shift of $g\text{-}i$ modes with more complex geometries \citep{Rui2024AsteroseismicStars,Lignieres2024PerturbativeModes} than the purely toroidal one \citep{Dhouib2022DetectingField}. \\
We nevertheless adopted this strongly simplified radial profile especially to maintain the analytical character of the core $m\text{-}i$ modes in the form of Bryan solutions. As argued in subsection~\ref{subsec:core_modes}, computing the eigenfrequencies and the structure of core $m\text{-}i$ modes in a more general magnetic configuration and core density gradient, with a (supposedly columnar) differential rotation would require the use of a spectral code. It would allow for a fine comprehension of the contributions of each effects along the aging of the star, which builds up an increasing core density gradient. The subsequent integration in the model developed in this work would be the scope of an entire separate semi-analytical work, while our present study aims at providing an analytical comprehension of the dip formation.\\
As for the hypothesis of constant Alfvén frequency in the envelope, this could be lifted in future works by considering $m$-$g$-$i$ modes in the more general framework of \citet{Dhouib2022DetectingField} applying a TARM with a radially varying profile of the Alfv\'en frequency, or model $m$-$g$-$i$ modes with a different magnetic field, taking the recent works of \citet{Rui2024AsteroseismicStars} and \citet{Lignieres2024PerturbativeModes} as a start.\\

\subsubsection{Potential of direct numerical calculations}
The efforts to better understand the inertial dip and its intricate dependencies will also come the refinements of stellar oscillation code, as done by \citet{Ouazzani2020FirstRevealed}, \citet{Saio2021RotationModes} and \citet{Galoy2024PropertiesStars} in a hydrodynamic framework. The extension of the codes used in these works to a MHD framework shows a great potential that will allow to probe more realistic field geometries and topologies. \\
Improvement is also expected in the treatment of the transition region from the core to the envelope. As \citetalias{Tokuno2022AsteroseismologyOscillations} argues based on \citet{Saio2021RotationModes}, overshooting with a radiative gradient in the transition zone would result in a multiplicity of dips, whereas convective penetration with an adiabatic gradient would not. In a magnetic context, this would be the precise zone where field lines from the core and the envelope might connect, and the dip could show probing power of this poorly constrained phenomenon, as well as some hint on the type of transport at play in this intermediate zone.

\subsection{Best targets for measurements of core or envelope magnetism}
From our work, the measurement of core magnetism by the dip study and envelope magnetism by the curvature of the PSP and the suppression of modes appear difficult to perform at the same time. Indeed, the inertial dip in the $(k=0,m=-1)$ Kelvin mode series appears at relatively high spin parameters ($s_{\rm env}^* > 8$ for a realistic magnetic field in the $im$ model, see Fig.~\ref{fig:degen}). The PSP thus needs not to be cut at high mode periods by the action of an envelope magnetic field on $g$-$i$ modes. This results in an undistinguishable signature of the envelope magnetism on the PSP curvature in Fig.~\ref{fig:disc_Alfv}. On the other hand, envelope magnetism will be most detected if the PSP is curved and cut at an unusually low spin parameter in the PSP, in this case the dip would not appear in the PSP. Additionally, we saw that realistic core magnetic fields will most likely leave an imprint on the dip in the $im$ model, aged and with intermediate rotation. The more the star ages on the MS, the more modulated the PSP is, due to the building of chemical stratification. The measurement of an additional curvature due to envelope magnetism will be made uneasy in a PSP with modulations. Moreover, an aged star will present a higher near-core $N$ due to a strong molecular weight gradient left by the receding convective core. The coupling parameter $\epsilon$ would be lower for such stars compared to ZAMS ones, thus the dip would be deeper and more distinguishable.\\
\indent We observed that in our magnetic set-up, a core magnetism would be degenerate with a core-to-envelope differential rotation. A curvature in the PSP would not help to distinguish magnetic fields from differential rotation either, as it would not have a sizeable imprint on the PSP if the inertial dip is observed. This means that the results on the measurement of differential rotation by \citet{Saio2021RotationModes} could be reinterpreted in light of these elements. This work derived a differential rotation by fitting a numerically-computed PSP with a dip profile to 16 stars observed by \citet{Li2020Gravity-modeKepler}, with different hypotheses on the core-to-envelope overshooting. From our results, the differential rotation rates observed could be upper limits, as core magnetism would further shift the dip towards low periods. \\
We hypothesize that core magnetism could be an explanation for stars having unusually high core-to-envelope differential rotation detected \citep[see Fig. 14 of][]{Saio2021RotationModes}. This is the case of KIC05985441 with a detected $\alpha_{\rm rot} = \Omega_{\rm env}/\Omega_{\rm core} = 0.85$ in the model without overshooting and an inertial dip at $s_{\rm env, min}\approx 5.6$ and KIC04390625 with $\alpha_{\rm rot} = 0.92$ in the model without overshooting and an inertial dip at $s_{\rm env, min}\approx 6.6$. We can evaluate an upper limit on the magnetic field that would result in such shifts with our solid-body rotating, magnetic model. For KIC05985441, the dip was expected with solid-body rotation from \citet{Saio2021RotationModes}'s calculations at $s_{\rm env,min}\approx 9.9$. To reach the observed $s_{\rm env, min}$, our magnetic model with uniform rotation requires a mid-core equatorial magnetic field of $B_{0,\rm max}^{\varphi}|_{R_{\rm core}/2}\approx 5.2 \, \rm MG$. This can be considered as an upper limit, since differential rotation was not considered. If now we hypothesize a differential rotation of $\alpha_{\rm rot} \approx 0.93$, which would make KIC05985441 join the bulk of stars of this evolutionary stage analyzed by \citet{Saio2021RotationModes}, the magnetic field required to shift the dip to the observed value with the combined effect of differential rotation would be $B_{0}^{\varphi}|_{R_{\rm core}/2}\approx 2.8 \, \rm MG$. As for KIC04390625, the observed spin parameter of the dip would be reached from the solid-body estimation of $s_{\rm env,min}\approx 8.6$ with a magnetic field of $B_{0,\rm max}^{\varphi}|_{R_{\rm core}/2}\approx 3.3 \, \rm MG$. \\
\indent Of course, these estimations cannot be taken as measurements and are only presented as examples for the application of the model developed in the present work. Pursuing a data-based study demands further developments integrating with an ab initio approach all the effects responsible for a shift of the dip structure with a more realistic magnetic topology, whereas our computations only considered the non-magnetic effects through the calculations of \citet{Saio2021RotationModes} with their specific grid of models.

\section{Conclusion}
In this study, we explore the coupling between envelope and core modes in intermediate to fast rotating $\gamma$-Doradus stars in the case where magnetism is taken into account, for the first time. In this framework, sub-inertial envelope magneto-gravito-inertial modes can propagate to the convective core, with a magneto-inertial character. The interaction with core modes leaves an imprint on the period-spacing pattern of envelope modes, creating a so-called inertial dip with a Lorentzian shape. We build on the previous analytical works of \citet{Tokuno2022AsteroseismologyOscillations} and \citet{Barrault2025ConstrainingDips} to exhibit the dependencies of the dip location and shape on both core and envelope magnetism. Having a window on core magnetism with the inertial dip study would give an unprecedented view on the innermost magnetic activity of intermediate-mass stars, which is key to better understand angular momentum transport throughout the evolution of the star, chemical mixing, compact objects formation, among other uncertainties in modern stellar physics. \\
\indent In this first work on the inertial dip formation in a magnetic framework, we aim to stay in an analytical framework to provide a fine understanding of the magnetic effects, comparing them to core-to-envelope differential rotation previously explored with the same model. To achieve this, several hypotheses need to be applied that make the model depart from the most realistic cases. We thus highlight that this first study can be understood as a laboratory towards the fine comprehension of the influence of magnetism on the inertial dip formation. Especially, we consider a toroidal field with a bi-layer Alfvén frequency in this work, allowing a contrast of magnetic field strengths from both sides of the convective core - radiative envelope boundary, and to build on previous analytical works of \citet{Malkus1967HydromagneticWaves} for the core and \citet{Mathis2011Low-frequencyField} for the envelope. \\
\indent We show that while envelope magnetism adds an additional curvature to the Period-Spacing Pattern, as already investigated in more realistic configurations by \citet{Dhouib2022DetectingField}, \citet{Lignieres2024PerturbativeModes} and \citet{Rui2024AsteroseismicStars}, core magnetism tends to lower the spin parameter at which the inertial dip appears and to make it deeper and thinner with increasing strength. We draw a parallel to the effect of differential rotation, as the magnetic field induces a frequency shift of modes from both sides of the boundary comparable to a Doppler effect. However, the effect of differential rotation is noticeable for a few percents of core-to-envelope rotation contrast, while the magnetic effect only arises if the Lorentz force is non negligible compared to the Coriolis acceleration. Thus, at fixed magnetic field strength, the effect of magnetism will be more important on the inertial dip if the rotation rate is lower. \\
\indent We estimate the shift of the spin parameter of the inertial dip for three different models belonging to the $\gamma$-Doradus instability strip, varying the rotation rate and the mass, hence evolutionary stage since the low-mass end of the instability strip gathers young stars and the high mass end evolved stars on the Main Sequence \citep{Bouabid2013EffectsStars}. We estimate that core fields in a typical magnetostrophic regime expected could be probed by the inertial dip for the intermediate rotators, while the effect for the fast rotator would be hard to disentangle from a small amount of differential rotation still to be expected from realistic MHD simulations \citep{Featherstone2009EffectsStars}. We constrain a range of envelope fields for which this interaction could occur by estimating a limit from which the envelope field would suppress gravito-inertial modes before they reach the core: about $50 \, \rm kG$ for intermediate rotators and $110 \, \rm kG$ for the fast rotator. \\
\indent This range of allowed envelope fields would leave a small imprint on the curvature of the period-spacing pattern of gravito-inertial modes. The effect of the core magnetic field is found to be degenerate with the effect of differential rotation from the study of the inertial dip in the $(k=0,m=-1)$ gravito-inertial mode series, which is not surprising due to the similar mathematical nature of the Lorentz force and the Coriolis acceleration in such framework. We foresee that the study of inertial dips in the $(k=0,m=-2)$ series would help to lift the degeneracy, due to the quadratic dependence of the magnetic effect on the azimuthal number $m$, to compare with the linear effect of differential rotation. As in the case of mixed pressure-gravity modes in red giant branch stars \citep{Das2024UnveilingStars}, the dip study would benefit from a multi-mode analysis, which would be an even more difficult task than in the latter case due to the scarcity of dips detections in the $(k=0,m=-1)$ mode series, let alone the non detection of them in the $(k=0,m=-2)$ one. \\
\indent This study, together with \citet{Barrault2025ConstrainingDips} focusing on differential rotation, can be understood as an exploration of different magnetic regimes. \citet{Barrault2025ConstrainingDips} explored the regime of magnetic field low enough to allow a significant core-to-envelope differential rotation to occur. The rotation effects would then be prominent on the inertial dip. The present works investigates the regime of intermediate field, strong enough to inhibit largely differential rotation. The magnetic effect can then be prominent over the differential rotation effect. However, this study does not tackle the regime of magnetic field strong enough to suppress modes from both sides of the boundary. While this mechanism in the envelope has been theoretically investigated \citep{Rui2023GravityFields} and successfully applied to infer a lower limit on the radial field strength in the SPB star HD 43317 \citep{Lecoanet2022AsteroseismicHD43317}, it would benefit from the study of predominantly toroidal configurations expected in the context of a generation by a Tayler-Spruit mechanism. As for the convective core, no study has been made to our knowledge on the potential conversion of inertial waves to slow Alfvénic waves in a strong field regime. Based on the field strength required for such a phenomenon, the dip region in which gravito-inertial modes would penetrate the core would appear as a gap in the period-spacing pattern. We let this promising study for future works, allowing us to complete our exploration of the field regimes expected in the interior of $\gamma$-Doradus stars. \\
\indent This work marks only the beginning of the exploration of the magnetic effects on inertial dips formation. From a theoretical point of view, we need to integrate the influence of core density stratification contributing in the shift of the dip's spin parameter to low periods \citep{Ouazzani2020FirstRevealed}, and a multi-mode interaction developed in Appendix D of \citet{Galoy2024PropertiesStars}. Significant developments will also arise from the use of numerical calculations as explored in \citet{Ouazzani2020FirstRevealed}, \citet{Saio2021RotationModes} and \citet{Galoy2024PropertiesStars}, with an extension to more realistic magnetic configurations. From a data-oriented point of view, a systematic search of inertial dips in gravito-inertial mode series of intermediate to fast $\gamma$-Doradus stars is now desirable, to build up on the first sample of \citet{Saio2021RotationModes}. All in all, the study of the inertial dip is still at its infancy, and holds great promise for future inference of convective core properties, in a manner quite comparable to the discovery of mixed pressure-gravity modes in red giants, which has allowed our understanding of their innermost layers. In the long term, detections of internal magnetic fields on the main sequence from the dip study will be integrated with measurement in their evolved counterparts to form a coherent picture for the evolution of internal magnetism along the ageing of intermediate-mass stars. \\

\begin{acknowledgements}
We thank the referee for their comments and suggestions which allowed us to improve the quality of this manuscript.
L. Barrault and L. Bugnet gratefully acknowledge support from the European Research Council (ERC) under the Horizon Europe programme (Calcifer; Starting Grant agreement N$^\circ$101165631). S. Mathis acknowledges support from the PLATO CNES grant at CEA/DAp. S. Mathis and J.S.G. Mombarg acknowledge support from the European Research Council through HORIZON ERC SyG Grant 4D-STAR 101071505. While partially funded by the European Union, views and opinions expressed
are however those of the authors only and do not necessarily reflect those of the European Union or the European Research Council. Neither the European Union nor the granting authority can be held responsible for them. L. Barrault thanks T. Van Reeth and C. Aerts for their invaluable teachings. The authors thank also the members of the Asteroseismology and Stellar Dynamics group of the Institute of Science and Technology Austria (ISTA) for very useful discussion: A. Cristea, L. Einramhof, K. M. Smith, S. Torres.
\end{acknowledgements}

\bibliographystyle{aa} 
\bibliography{references_mag}
\begin{appendix} 
\longtab[1]{
\begin{landscape}
\begin{longtable}{lrcrrrrrrrrl}
...
\end{longtable}
\end{landscape}
}

\section{Relevant physical quantities of the considered models}\label{Appendix:models}
Table \ref{tab:models} summarizes the physical parameters relevant to the study for the three considered models. We here give the precise definition of all the listed quantities. We introduce $N_{\rm max}$ the value of the near-core Brunt-Väisälä near-core. It depends on the prescription of core-to-envelope boundary mixing. We define $\bar{N}$ an average value of the Brunt-Väisälä angular frequency experienced by $g$-$i$ modes such that:
\begin{equation}
    \bar{N} = \frac{\displaystyle\int_{r_a}^{r_{b}}\frac{N}{r}\mathrm{d}r}{\displaystyle \int_{r_a}^{r_{b}}\frac{\mathrm{d}r}{r}} \, .
\end{equation}
In the same manner, we define $\Omega_{\rm nc}$ the rotation rate probed by $g$-$i$ modes in the JWKB limit:
\begin{equation}
    \Omega_{\rm nc} = \frac{\displaystyle\int_{r_a}^{r_{b}}\Omega\frac{N}{r}\mathrm{d}r}{\displaystyle \int_{r_a}^{r_{b}}\frac{N}{r}\mathrm{d}r}
\end{equation}
This value is heavily weighted by the near-core rotation rate due to the near-core spike in $N$ and the denominator in $r$. This corresponds to $\Omega_{\rm env}$ in our bi-layer model. \\
We estimate the core magnetic field using different hypotheses on the force balance at play in stellar interiors. First, if the Lorentz force balances the Coriolis acceleration, the magnetostrophic regime is reached. An order of magnitude reasoning requires to estimate $l_{\rm B}$, a typical distance of variation of the magnetic field. If the dynamo-generated magnetic field shows small structures determined by convection, then this distance $l_{\rm B}$ should equate the typical variation lengthscale of the velocity field \citet{Augustson2019APenetration}. However, we hypothesized in our framework a large-scale background magnetic field in the core. A precise computation of the Lorentz force (see Appendix~\ref{App:hierarchy}) leads for this type of field at the equator to $l_{\rm B} = r/2$, r being the distance from the center of the star. The magnetic field in the magnetostrophic regime thus reads: \\
\begin{equation}
    B_{\rm ms} \simeq \sqrt{\mu_{0}\bar{\rho}r\Omega_{\rm core}v_{\rm conv}} \, .
\end{equation}
Second, in an equipartition regime in which the magnetic energy density equates the kinetic energy density:
\begin{equation}
    B_{\rm equi} \simeq \sqrt{\mu_{0}\bar{\rho}}v_{\rm conv} \, .
\end{equation}
We provide an estimate of those fields, along with a computation of the Rossby number $\mathrm{Ro} = \dfrac{v_{\rm conv}}{2\Omega_{\rm core}l_{\rm conv}}$ at mid convective core in Table \ref{tab:models}.

\section{Notations and conventions used in this work}
\label{App:notations}
We provide for readability in Table~\ref{tab:parameter} a list of all of the symbols used throughout this manuscript. In panel 2 of this table, the subscript 'zone' is either 'core' or 'env' depending on the considered zone, respectively the convective core and the radiative envelope. All variables in the panel "Parameters of the interaction" have their tilde counterpart when considering a discontinuous $N$ at the boundary.

\begin{table*}
\begin{center}
\begin{tabular}{|c c|} 
 \hline
 Symbol & Signification  \\ [0.5ex] 
 \hline\hline
 \multicolumn{2}{|c|}{\textbf{Physical background quantities}}  \\ 
 \hline
 $\mathrm{\boldsymbol{B}}_{0}^{\varphi}$ & Background toroidal magnetic field   \\

 $\mathrm{\boldsymbol{V}}_{0}$ & Background velocity field   \\

 $\bar{\rho}$ & Background density  \\
  $\bar{P}$ & Background gaseous pressure   \\

 $\bar{P}_{\rm M}$ & Background magnetic pressure  \\ 
     $N$ & Brunt-Väisälä angular frequency \\
     $\tilde{S}$ & Lamb angular frequency modified by rotation \\  
     $c_{S}$ & Sound speed \\
     $\bar{g}$ & Local gravity acceleration \\
 \hline
 \multicolumn{2}{|c|}{\textbf{Constants}} \\
 \hline
 $\mathcal{G}$ & Gravitational constant \\
 $\mu_{0}$ & Vacuum permeability \\
 $\Gamma_{1}$ & First adiabatic index \\
 $\alpha$ & $3^{1/3} \Gamma(2/3)/\Gamma(1/3)\approx 0.73$ \\
 \hline
 \multicolumn{2}{|c|}{\textbf{Angular frequencies and related quantities}}  \\
 \hline
 $\Omega_{\rm zone}$ & Rotation rate of the considered zone \\
 $\omega_{\rm A, zone}$ & Alfvén angular frequency in the zone \\
    $\sigma_{\rm in}$ & Wave angular frequency in an inertial frame \\
    $\sigma_{\rm zone}$ & Wave angular frequency in an frame co-rotating with the zone \\    
    $\sigma_{\rm M,zone}$ & Wave magnetic angular frequency in a frame co-rotating with the zone \\
    $s_{\rm zone}$ & Hydrodynamic spin parameter of the zone\\
    $s_{\rm M,zone}$ & Magnetic spin parameter of the zone \\
    $\nu_{\rm M, zone}$ & Magnetic structural parameter of the zone \\ 
    $\rm P_{M}$ & Modified magnetic period \\
    $\rm P$ & Period in the frame co-rotating with the envelope \\
    $\rm P_{in}$ & Period in an inertial frame \\
    $\Delta\rm P_{M}$ & Modified magnetic period-spacing \\
    $\Delta\rm P$ & Period-spacing in the frame co-rotating with the envelope \\
    $\Delta\rm P_{in}$ & Period-spacing in an inertial frame \\
    $\alpha_{\rm rot}$ & Differential rotation, $\Omega_{\rm env}/\Omega_{\rm core}$ \\
    $\rm Le_{zone}$ & Lehnert number of the considered zone, $\omega_{\rm A,zone} / 2\Omega_{\rm zone}$ \\
    $G_{\rm M}$ & $s_{\rm M,env} = G_{\rm M}(s_{\rm M, core})$ \\
    $u_{\rm zone}$ & $\nu_{\rm M,zone} = u_{\rm zone}(s_{\rm M, zone})$ \\    
\hline
 \multicolumn{2}{|c|}{\textbf{Angular structure quantities}}  \\
 \hline
 $l$ & Angular degree of the mode \\
 $m$ & Azimuthal order \\
 $k$ & $l-|m|$ in a non-rotating star\\
$\mu$ & $\cos \theta$ \\
$\Theta_{k}^{m}$ & Hough function \\
$\Lambda_{k}^{m}$ & Eigenvalue of the Laplace Tidal Equation \\
$P_{l}^{m}$ & Legendre polynomial \\
$\tilde{P}_{l}^{m}$ & Normalized Legendre polynomial \\
$C_{l}^{m}$ & $x\left(\dfrac{\rm d}{\mathrm{d}x}-\dfrac{m}{1-x^{2}}\right)P_{l}^{m}(x)$ \\
$c_{k,l}$ & Geometrical factor \\
$F_{l}^{m}$ & Phase function of pure inertial modes \\
\hline
\multicolumn{2}{|c|}{\textbf{Parameters of the interaction (continuous $N$)}}  \\
\hline
$\epsilon$ & Coupling parameter\\
$V$ & Structure factor with solid-body rotation and no magnetism\\
$V_{\rm M}$ & Structure factor with solid-body rotation and uniform Alfvén frequency\\
$V_{\rm M, \neq}$ & Structure factor with bi-layer rotation and Alfvén frequency \\
$\Gamma_{\rm M}$ & Magnetic control parameter with uniform Lehnert number\\
$\Gamma_{\rm M, \neq}$ & Magnetic control parameter with bi-layer Lehnert number\\
\hline
\multicolumn{2}{|c|}{\textbf{Magnetic fields estimate}}  \\
\hline
$\rm B_{ms}$ & Estimate of the core magnetic field in a magnetostrophic regime \\
$\rm B_{equi}$ & Estimate of the core magnetic field in a equipartition regime \\
$\rm Ro$ & Rossby number \\
$v_{\rm conv}$ & Typical convective velocity given by the MLT \\
$l_{\rm conv}$ & Typical convective length scale given by the MLT \\
$l_{\rm B}$ & Typical length scale of magnetic field variation \\
\hline
\end{tabular}
\end{center}
\label{tab:parameter}
\end{table*}

\section{Envelope and core modes derivation}\label{App:deriv}
\subsection{Linearisation of the equations and general considered system}
We assume for the rest of this derivation that the background magnetic tension and pressure is small compared to the background gaseous pressure \citep{Duez2010EffectSun}. Additionally, the effect of the centrifugal acceleration is low in the internal layers we are interested in \citep{Ballot2010GravityStars}. Under these conditions, the deformed magneto-hydrostatic equilibrium arising from the action of both the Lorentz force and the centrifugal acceleration reverts to the radially symmetric hydrostatic equilibrium.
All the quantities X are then expressed as a sum of an hydrostatic term $\bar{X}$ and a fluctuation $\tilde{X}$:
\begin{equation}
    X(r,\theta,\varphi,t) = \bar{X}(r) + \tilde{X}(r,\theta,\varphi,t) \, .
\end{equation}

\noindent As for the velocity, its fluctuation reads \citep{Unno1989NonradialStars}:
\begin{equation}
    \mathbf{u}= \partial_{t}\boldsymbol{\xi} + (\mathbf{V}_{0}\cdot \boldsymbol{\nabla}) \boldsymbol{\xi} - (\boldsymbol{\xi} \cdot \boldsymbol{\nabla}) \mathbf{V}_{0} =  (\partial_{t} + \Omega\partial_{\varphi})\boldsymbol{\xi} \, .
\end{equation}

\noindent The linearized induction equation is written:
\begin{equation}
    \mathbf{b} = \nabla \times \left(\boldsymbol{\xi} \times \boldsymbol{B}_{0}^{\varphi}\right)
\end{equation}

\noindent Due to the nearly incompressible character of low-frequency anelastic modes, this holds:
\begin{equation}
    \mathbf{b} = \sqrt{\mu_{0} \bar{\rho}} \omega_{A} \partial_{\varphi} \boldsymbol{\xi} \, .
    \label{eq:ind_lin}
\end{equation}
The linearized momentum equation is:
\begin{align}
    & (\partial_{t} + \Omega \partial_{\varphi})\left[(\partial_{t} + \Omega \partial_{\varphi})\boldsymbol{\xi} + 2\Omega\widehat{\boldsymbol{e}}_{z} \times \boldsymbol{\xi} \right] = \\
    & -\frac{1}{\bar{\rho}}\boldsymbol{\nabla}{\tilde{\Pi}(\boldsymbol{r},t)}-\boldsymbol{\nabla}\tilde{\Phi} + \frac{\tilde{\rho}}{\bar{\rho}^{2}}\boldsymbol{\nabla}{\bar{P}} + \frac{\boldsymbol{F}_{\mathcal{L}}^{\mathrm{Te}}(\boldsymbol{\xi})}{\bar{\rho}} \, ,
\end{align}
with $\Omega\widehat{\boldsymbol{e}}_{z}$ the rotation vector where $\boldsymbol{\hat{e}_{z}} = \cos{\theta}\boldsymbol{\hat{e}_{r}} - \sin{\theta}\boldsymbol{\hat{e}_{\theta}}$.

\noindent We define the total pressure fluctuation, composed of the magnetic and the gaseous terms:
\begin{equation}
\tilde{\Pi} = \tilde{P} +\frac{\mathbf{B}_{0}^{\varphi} \cdot \mathbf{b}}{\mu_{0}} \, .
\end{equation} 
Using Eq.~\eqref{eq:ind_lin}, the wave magnetic tension force is:
\begin{align}
\boldsymbol{F}_{\mathcal{L}}^{\mathrm{Te}}(\boldsymbol{\xi}) & = \frac{1}{\mu_{0}}\left[(\mathbf{B}_{0}^{\varphi}\cdot \nabla)\mathbf{b} + (\mathbf{b}\cdot \nabla)\mathbf{B}_{0}^{\varphi} \right] \\ 
& = \bar{\rho}\omega_{A}^{2}\left[\partial_{\varphi^{2}} \boldsymbol{\xi} + 2\boldsymbol{\hat{e}_{z}} \times \partial_{\varphi}\boldsymbol{\xi}\right] \, .
\end{align}
In this framework, the continuity, and Poisson's equations read:
\begin{align}
    & \tilde{\rho} + \boldsymbol{\nabla} \cdot (\bar{\rho} \boldsymbol{\xi}) = 0 \, , \\
    & \nabla^{2}\tilde{\Phi} = 4\pi G \tilde{\rho} \, .
\end{align}
Finally, the energy equation is written in the adiabatic limit:
\begin{equation}
    \left(\frac{\tilde{P}}{\Gamma_{1}\bar{P}} + \frac{\tilde{\rho}}{\bar{\rho}} \right) + \boldsymbol{\xi}\cdot\left(\frac{1}{\Gamma_{1}}\boldsymbol{\nabla}\ln\bar{P} - \boldsymbol{\nabla}\ln\bar{\rho}\right) = 0 \, .
\end{equation}
\noindent We adopt the same expansions and conventions as in \citet{Mathis2011Low-frequencyField}:
\begin{equation}
\tilde{X} = \sum_{\sigma_{\rm in}, m}X'(r,\theta) e^{i(m\varphi + \sigma_{\rm in} t)}
\end{equation}
\begin{equation}
\boldsymbol{x} = \sum_{\sigma_{\rm in}, m}\boldsymbol{x}'(r,\theta) e^{i(m\varphi + \sigma_{in} t)} \, .
\end{equation}

We introduce the local wave pulsations, related to the pulsations in the inertial frame by: 
\begin{equation}
    \sigma_{\rm zone} = \sigma_{\rm in} + m\Omega_{\rm zone} \, .
\end{equation}

Under this convention, we have for the velocity field:
\begin{equation}
    \mathbf{u}' = i\sigma_{\rm zone} \boldsymbol{\xi}'
\end{equation}
and for the magnetic perturbation:
\begin{equation}
    \mathbf{b}' = im\sqrt{\mu_{0} \bar{\rho}} \omega_{\rm A, zone} \boldsymbol{\xi}' \, .
\end{equation}

\noindent The momentum equation is then \eqref{eq:momentum_general}, which we recall here:
\begin{equation}
-\mathcal{A}\boldsymbol{\xi}' + i\mathcal{B}\widehat{\mathbf{e}}_{z} \times \boldsymbol{\xi}' = -\boldsymbol{\nabla} W' + \frac{\rho'}{\bar{\rho}^{2}} \boldsymbol{\nabla} \bar{P} - \Pi'\frac{\boldsymbol{\nabla}\bar{\rho}}{\bar{\rho}^{2}} \, ,
\label{eq:app_mom_lin}
\end{equation}

\noindent where $\mathcal{A} = \sigma_{M,\mathrm{zone}}^{2} = \sigma_{\mathrm{zone}}^{2} - m^{2}\omega_{A,\mathrm{zone}}^{2}$ and  
$\mathcal{B}= 2(\Omega_{\mathrm{zone}}\sigma_{\mathrm{zone}} - m\omega_{A,\mathrm{zone}}^{2})$.

\subsection{Core modes}\label{App_malk}
We hereby recall \citet{Malkus1967HydromagneticWaves}'s calculations starting from \eqref{eq:app_mom_lin}, adapted to the core's properties. With uniform density and assuming isentropy, it reads:
\begin{equation}
-\mathcal{A}\boldsymbol{\xi}' + i\mathcal{B}\widehat{\mathbf{e}}_{z} \times \boldsymbol{\xi}'  +\boldsymbol{\nabla} W' = 0 \label{eq:momentum_inertial}
\end{equation}
The anelastic approximation we use for both zones reduces to a Boussinesq one for magneto-inertial modes in the case of constant core density considered. In the continuity equation, this is equivalent to incompressibility:
\begin{equation}
\boldsymbol{\nabla} \cdot \boldsymbol{\xi}' = 0
\end{equation}

\noindent We take the divergence of (\ref{eq:momentum_inertial}):
\begin{equation}
  \nabla^{2} W' - i\mathcal{B}\widehat{\mathbf{e}}_{z}\cdot(\boldsymbol{\nabla} \times \boldsymbol{\xi}') = 0  \, ,
  \label{eq:div}
\end{equation}

\noindent and the curl:
\begin{equation}
\mathcal{A}(\boldsymbol{\nabla} \times \boldsymbol{\xi}') + i\mathcal{B}(\widehat{\mathbf{e}}_{z} \cdot \boldsymbol{\nabla})\boldsymbol{\xi}'= 0 \label{eq:curl} \, .
\end{equation}

\noindent We then project \eqref{eq:curl} along $\widehat{\mathbf{e}}_{z}$ and multiply it by $\mathcal{B}$; we obtain:
\begin{equation}
\mathcal{A}\mathcal{B}\widehat{\mathbf{e}}_{z}\cdot(\boldsymbol{\nabla} \times \boldsymbol{\xi}') + i\mathcal{B}^{2}\widehat{\mathbf{e}}_{z} \cdot \frac{\partial{\boldsymbol{\xi}'}}{\partial{z}} = 0 \, .
\label{eq:projection}
\end{equation}
The projection along $\widehat{\mathbf{e}}_{z}$ of \eqref{eq:momentum_inertial} and the derivation with respect to z reads:

\begin{equation}
-\mathcal{A}\widehat{\mathbf{e}}_{z}\cdot\frac{\partial \boldsymbol{\xi}}{\partial z}'+\frac{\partial^{2} W'}{\partial z^{2}} = 0 \, .
\label{eq:div_proj}
\end{equation}
Using \eqref{eq:div}, \eqref{eq:projection} and  \eqref{eq:div_proj} we finally get:

\begin{equation}
      \nabla^{2} W' - \frac{\mathcal{B}^{2}}{\mathcal{A}^{2}}\frac{\partial^{2}W'}{\partial z^{2}} = 0 \, .
      \label{eq:Poincaré}
\end{equation}

As in the case of $m$-$g$-$i$ modes, we retrieve the magnetic structural parameter: 
\begin{equation}
    \nu_{\rm M,core} = \frac{\mathcal{B}}{\mathcal{A}} = \frac{2(\Omega_{\rm core}\sigma_{\rm core} - m\omega_{\rm A, core}^{2})}{\sigma_{\rm core}^{2} - m^{2}\omega_{\rm A,  core}^{2}}
\end{equation}

\noindent and we rewrite it for consistency:
\begin{equation}
    \nu_{\rm M,core} = s_{\mathrm{core}}\frac{1-2ms_{\mathrm{core}}\mathrm{Le_{core}}^{2}}{1-m^{2}\mathrm{Le_{core}^{2}}{s_{\mathrm{core}}}^{2}}
\end{equation}
with the constant Lehnert number $\mathrm{Le_{\rm core}}$ related to the core. 
\noindent In this formalism, we thus retrieve the canonical Poincaré equation for magneto-inertial waves, with the adapted parameter $\nu_{\rm  M,core}$. \\
We define:
\begin{equation}
    \Psi = \frac{1}{\sigma_{\rm M, core}^{2}}W' \, ,
\end{equation}

\noindent Eq.~\eqref{eq:Poincaré} becomes:
\begin{equation}
    \nabla^{2}\Psi - \nu_{\rm M, core}^{2}\frac{\partial^{2}\Psi}{\partial z^{2}} = 0 \, .
\end{equation}

\noindent From the definition of $\Psi$, we get:
\begin{equation}
    \Pi' = \bar{\rho}\sigma_{\rm M, core}^{2}\Psi
\end{equation}

As for the Lagrangian displacement $\boldsymbol{\xi'}$, we start from the momentum equation, written as:
\begin{equation}
    -\boldsymbol{\xi'} + i\frac{\mathcal{B}}{\mathcal{A}}\widehat{\mathbf{e}}_{z}\times \boldsymbol{\xi'} = - \frac{1}{\mathcal{A}}\boldsymbol{\nabla} W' \, .
\end{equation}
From the definitions of $\mathcal{A}$ and $\nu_{M, \mathrm{core}}$ we have:
\begin{equation}
\boldsymbol{\xi'} - i\nu_{\rm M,core}\widehat{\mathbf{e}}_{z}\times \boldsymbol{\xi'} = \boldsymbol{\nabla} \Psi \, .
\label{eq:for_xi}
\end{equation}

\noindent Taking the dot and vectorial product by $\widehat{\mathbf{e}}_{z}$, we obtain:
\begin{equation}
    \widehat{\mathbf{e}}_{z} \cdot \boldsymbol{\xi'} = \frac{\partial \Psi}{\partial{z}}
\end{equation}
and
\begin{equation}
    \widehat{\mathbf{e}}_{z} \times \boldsymbol{\xi'} - i\nu_{\rm M,core}(\widehat{\mathbf{e}}_{z} \cdot \boldsymbol{\xi'})\widehat{\mathbf{e}}_{z} + i\nu_{\rm M,core} \boldsymbol{\xi'}  = \widehat{\mathbf{e}}_{z}\times\boldsymbol{\nabla} \Psi 
\end{equation}

\noindent or

\begin{equation}
    i\nu_{\rm M,core} \boldsymbol{\xi'} = -\widehat{\mathbf{e}}_{z} \times \boldsymbol{\xi'} + i\nu_{\rm M,core}\frac{\partial\Psi}{\partial z}\widehat{\mathbf{e}}_{z} + \widehat{\mathbf{e}}_{z}\times\boldsymbol{\nabla} \Psi \, ,
\end{equation}
respectively.
\noindent Using \eqref{eq:for_xi}, we get:
\begin{align}
    i\nu_{\rm M,core} \boldsymbol{\xi'} = & \frac{1}{i\nu_{\rm M,core}}\boldsymbol{\nabla} \Psi - \frac{1}{i\nu_{\rm M,core}}\boldsymbol{\xi'} \nonumber \\
    & + i\nu_{\rm M,core}\frac{\partial\Psi}{\partial z}\widehat{\mathbf{e}}_{z} + \widehat{\mathbf{e}}_{z}\times\boldsymbol{\nabla} \Psi \, ,
\end{align}

\noindent that we reorganize:
\begin{equation}
    \boldsymbol{\xi'} = \frac{1}{1-\nu_{\rm M,core}^{2}}\left[1 + i\nu_{\rm M,core}(\widehat{\mathbf{e}}_{z}\times) - \nu_{\rm M,core}^{2}\widehat{\mathbf{e}}_{z}\widehat{\mathbf{e}}_{z}\cdot \right]\boldsymbol{\nabla} \Psi \, ,
\end{equation}

\noindent Projecting this operator on $\widehat{\mathbf{e}}_{r}$, we get:
\begin{equation}
    \xi'_{r} = \frac{1}{1-\nu_{\rm M,core}^{2}}\left[\frac{\partial}{\partial r} + \frac{m \nu_{\rm M,core}}{r} - \mu \nu_{\rm M,core}^{2}\frac{\partial}{\partial z}\right]\Psi \, .
\end{equation}

This expression is equivalent to the one studied by \citet{Wu2005OriginModes}, with the magnetic structural parameter $\nu_{\rm M,core}$ replacing the hydrodynamical spin parameter $s_{\rm core}$. We then retrieve Eqs.~\eqref{eq:xi_mi} and \eqref{eq:w_mi}, the expressions of respectively $\boldsymbol{\xi'}$ and $W'$ at the core boundary.\\

\section{Hierarchy of the frequencies and negligibility of the magnetic terms}\label{App:hierarchy}
\FloatBarrier
\begin{figure}
    \centering
    \includegraphics[width=\linewidth]{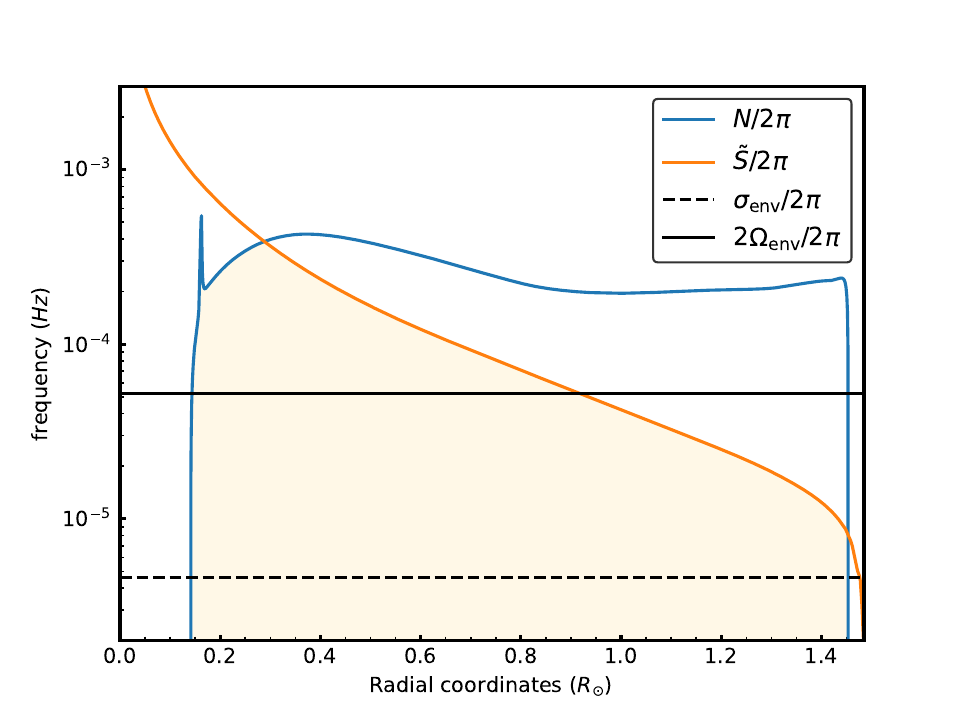}
    \caption{Propagation diagram for the $fz$ model. The $g$-$i$ mode cavity is coloured in beige. The dashed horizontal line is the local angular frequency of a typical mode with $s=10$.}
    \label{fig:prop-diag}
\end{figure}
\subsection{Background rotational and magnetic deformation}
We here verify the accuracy of our hypothesis of low contribution of both the centrifugal acceleration and the Lorentz force compared to the local self-gravity at the bottom of the radiative envelope. First, at the equator, close to the core, the centrifugal acceleration reads $\vec{a}_{\rm centr} = R_{\rm core}\Omega_{\rm env}^2\widehat{\mathbf{e}}_{r}$, which needs to be compared to $\bar{\boldsymbol{g}} = -GM_{\rm core}/R_{\rm core}^{2}\widehat{\mathbf{e}}_{r} = -R_{\rm core}\Omega_{\rm crit}^{2}\widehat{\mathbf{e}}_{r}$, the Keplerian critical rotation rate at the edge of the convective core. As seen in Appendix \ref{Appendix:models}, the ratio $\Omega_{\rm env}/\Omega_{\rm crit}$ is at most $3.6 \%$ for the fastest rotating model. This ensures the negligibility of the centrifugal acceleration in the background hydrostatic equilibrium. \\
As for the background Lorentz force, it reads:
\begin{align}
    \mathbf{f_{lor}}& =  \dfrac{1}{\bar{\rho}\mu_{0}}((\boldsymbol{\nabla}\times \mathbf{B}_{0}^{\varphi}) \times \mathbf{B}_{0}^{\varphi})  \nonumber \\ 
    &= -\left(r\sin \theta \omega_{A}\right)^{2}\left(\left(\frac{2}{r}+\frac{1}{2}\frac{\mathrm{d}\ln \bar{\rho}}{\mathrm{d}r}\right)\widehat{\mathbf{e}}_{r} + \left(\frac{2\cot \theta}{r}\right)\widehat{\mathbf{e}}_{\theta} \right) 
\end{align}
The term $\frac{2}{r}$ dominates over the background density variation $\frac{1}{2}\frac{\mathrm{d}\ln\bar{\rho}}{\mathrm{d}r}$
close to the core, even in the $N$ spike region in our models. An estimate of the background Lorentz force at the equator close to the core is then $\mathbf{f}_{\rm lor}\approx -2 R_{\rm core}\omega_{\rm A,env}^{2}\widehat{\mathbf{e}}_{r}$. We thus have $\left|\frac{\mathbf{f}_{\rm lor}}{\vec{a}_{\rm centr}}\right|\approx 8 \mathrm{Le_{\rm env}^{2}}$. Given the Lehnert numbers considered in this work, the effect of the Lorentz force on the background equilibrium can be as well neglected.

\subsection{Magnetic pressure fluctuation}
In the system of \citet{Mathis2011Low-frequencyField}, some terms have to be carefully taken care of if one has to work outside of the anelastic approximation. The magnetic pressure perturbation can be rewritten as:
\begin{equation}
    P'_{M}=\frac{\mathbf{B}_0^{\varphi}.\mathbf{b}}{\mu_0} = im\bar{\rho}r\sin\theta\omega_{\rm A, env}^{2}\xi'_\varphi \, .
\end{equation}
If we write the magnetic pressure perturbation as function of the total dynamical pressure perturbation:
\begin{equation}
    P'_{\mathrm{M};k,m} = -m\bar{\rho}\left(\frac{\omega_{\rm A, env}}{\sigma_{\rm M, env}}\right)^{2}\sin\theta \, \mathcal{H}_{k,m}^{\varphi}W'_{k,m} \, .
\end{equation}

\noindent We see that in the frequency regime that we are interested in ($\omega_{\rm \rm A, env} \ll \sigma_{\rm M, env}$), the magnetic pressure perturbation is negligible compared to the gaseous one: $P'_{\mathrm{M};k,m}/\bar{\rho} \ll W'_{k,m}$ hence $P'_{\mathrm{M};k,m} \ll P'_{k,m}$. The whole system can be thus rewritten in this limit:

\begin{align}
    \frac{\mathrm{d}W'_{k,m}}{\mathrm{d}r} = &\left(\frac{N^{2}}{\bar{g}}\right)W'_{k,m} 
    +\frac{1}{r^{2}}(\sigma_{\rm M, env}^{2} - N^{2})(r^{2}\xi'_{r;k,m})
\end{align}

\begin{align}
    \frac{\mathrm{d}}{\mathrm{d}r}(r^{2}\xi'_{r;k,m}) = \left[\frac{\Lambda_{k}^{m}(\nu_{\rm M,env})}{\sigma_{\rm M, env}^{2}} - \frac{r^{2}}{c_{S}^{2}}\right] W'_{k,m} \nonumber \\ 
     -\frac{1}{\Gamma_{1}\bar{P}}\frac{\mathrm{d}\bar{P}}{\mathrm{d}r}(r^{2}\xi'_{r;k,m}) \, .
\end{align}

This allows us to derive a wavenumber using the same method as in \citet{Press1981RadiativeInteriors} and \citet{Unno1989NonradialStars}: 
\begin{equation}
    k_r^{2} = \dfrac{(\sigma_{\rm M, env}^{2}-N^{2})\left(\sigma_{\rm M, env}^{2}-\tilde{S}^{2}\right)}{c_{S}^{2}\sigma_{\rm M, env}^{2}} \, ,
\end{equation}
which simpifies to \eqref{eq:wavenum} when adopting the anelastic approximation. We check the hierarchy of frequencies in a typical model shown in Fig.~\ref{fig:prop-diag}.

\section{Derivation of the modified Lorentzian profile}\label{App:Mod_lorentz}
We start with the coupling equation Eq.~\eqref{eq:bef_lor}:
\begin{equation}
    \left[\cot\left(\frac{\pi^2s_{\rm M,env}}{\Omega_{\rm env}\Pi_{0,\rm M}}-\frac{\pi}{6}\right)+\frac{1}{\sqrt{3}}\right] \simeq -\frac{\epsilon/V_{\rm M,\neq}}{\nu_{\rm M, core}-\nu_{\rm M, core}^{*}} \, .
\end{equation}
For simplicity of notations, we drop the subscript env in the frequency-dependent variables: $s_{\rm M,env} \rightarrow s_{\rm M}$ and $\nu_{\rm M,env} \rightarrow \nu_{\rm M}$.
We write Eq.~\eqref{eq:bef_lor} for two neighbouring solutions $\nu_{\rm M, core,1}$ and $\nu_{\rm M,core,2}$ (related respectively to the envelope quantities $\nu_{\rm M,1}$ and $s_{1}$, and $\nu_{\rm M,2}$ and $s_{\rm M,2}$ via the relations described in Eqs. \eqref{fund:numc_smc}, \eqref{fund:nume_sme}, and \eqref{fund:sme_smc} with $\nu_{\rm M,core,2} > \nu_{\rm M, core, 1})$. We assume that the envelope $m$-$g$-$i$ modes are closely spaced, so that only one $m\text{-}i$ mode is non-negligibly matching envelope $m\text{-}g\text{-}i$ modes.
For this reason, the two solutions $s_{\rm M,1}$ and $s_{\rm M,2}$ are on two different branches of the cotangent, and the function $1/(\nu_{\rm M, core} - \nu_{\rm M,core}^*)$ does not vary appreciably between those two solutions. Therefore, we can consider that $\frac{\pi^2(s_{\rm M,2}-s_{\rm M,1})}{\Omega_{\text{env}}\Pi_{\rm 0,M}}-\pi = 2x \ll 1$.
In the case of $\nu_{\rm M, core,1}<\nu_{\rm M,\text{core}}^{*}<\nu_{\rm M, core,2}$ the solutions belong to the same branch of the cotangent, as pointed out by \citetalias{Tokuno2022AsteroseismologyOscillations}. We thus rewrite Eq.~\eqref{eq:bef_lor} as:
\begin{equation}
    \tan\left(\frac{\pi^{2}s_{\rm M}}{\Omega_{\text{env}}\Pi_{\rm 0,M}}\right) \simeq -\frac{1}{\frac{1}{\sqrt{3}} + \frac{\epsilon/V_{\rm M,\neq}}{\nu_{\rm M,core} - \nu_{M,core}^*}}
    \label{eq:tangent}
\end{equation}
and the solutions are on two different branches of the tangent.
Thus, in both cases:
\begin{equation}
    \frac{\pi^2 s_{\rm M,1,2}}{\Omega_{\text{env}}\Pi_{\rm 0,M}} - \frac{\pi}{6} = \frac{\pi^2 \bar{s}_{\rm M}}{\Omega_{\text{env}}\Pi_{\rm 0,M}} - \frac{\pi}{6} \pm \frac{\pi}{2} \mp x
\end{equation}
with $\bar{s}_{\rm M} = \frac{s_{\rm M,1}+s_{\rm M,2}}{2}$.
The resulting system is thus, noting $\bar{S} = \frac{\pi^2 \bar{s}_{\rm M}}{\Omega_{\text{env}}\Pi_{\rm 0,M}} - \frac{\pi}{6}$ and using a first order Taylor expansion:
\begin{align}
    \begin{cases}
        \cot(\bar{S}) - x(1+\cot^{2}(\bar{S})) +1/\sqrt{3}&\simeq -\frac{\epsilon/V_{\rm M,\neq}}{\nu_{\rm M, core,1}-\nu_{\rm M,core}^{*}}\\
        \cot(\bar{S}) + x(1+\cot^{2}(\bar{S})) +1/\sqrt{3}&\simeq -\frac{\epsilon/V_{\rm M,\neq}}{\nu_{\rm M, core,2}-\nu_{\rm M,core}^{*}}
    \end{cases}
    \, .
\end{align}
Further manipulations lead to:
\begin{align}
    &\left[1+\Big(\frac{\epsilon /V_{\text{diff}}}{\overline{\nu_{\rm M, core}}-\nu_{\rm M,core}^{*}}+\frac{1}{\sqrt{3}}\Big)^{2}\right] \nonumber \\
    & \times \left[\frac{\pi^{2}(s_{\rm M,1}-s_{\rm M,2})}{\Omega_{\text{env}} \Pi_{\rm 0, M}}-\pi\right]\simeq -\frac{\epsilon}{V_{\rm M,\neq}}\frac{\nu_{\rm M, core, 1}-\nu_{\rm M,core, 2}}{(\overline{\nu_{\rm M, core}}-\nu_{\rm M,core}^{*})^{2}} \, .
    \label{eq:inter}
\end{align}
We hereby aim to complement Appendix C of \citetalias{Barrault2025ConstrainingDips} stating that we do not require $\overline{\nu_{\rm M,core}}-\nu_{\rm M,core}^* \gg \delta\nu_{\rm M,core}$ in the derivation of this equation, with $\overline{\nu_{\rm M,core}} = (\nu_{\rm M,core,1} + \nu_{\rm M,core,2})/2$ and $\delta\nu_{\rm M,core} = (\nu_{\rm M,core,1} - \nu_{\rm M,core,2})/2$ \footnote{Indeed, performing the Taylor expansion of Eq.~\eqref{eq:tangent} from $\nu_{\rm M,core}^*$ holds the same result for solutions such that $\nu_{\rm M,core,2}<\nu_{\rm M, core}^{*}<\nu_{\rm M,core,1}$.}.
We then rewrite Eq.~\eqref{eq:inter} exhibiting only dependencies on the envelope magnetic spin parameter, expanding the function $u_{\rm core} \circ G_{\rm M}^{-1}$ at the middle points $\overline{s}_{\rm M}$ and $(\bar{s}_{\rm M}+s^{*}_{\rm M})/2$, with $s_{\rm M}^{*} = (G_{\rm M} \circ u_{\rm core}^{-1})(\nu_{\rm M,core}^{*})$, for the numerator and the denominator of the RHS of Eq.\eqref{eq:inter}, respectively. This reads:
\begin{align}
    &\left[1+\left(\frac{\epsilon /V_{\rm M,\neq}}{\bar{s}_{\rm M}-s^{*}_{\rm M}}\left(\frac{\mathrm{d}u_{\rm core}\circ G_{\rm M}^{-1}}{\mathrm{d}s}\Big|_{\frac{\bar{s}_{\rm M}+s^{*}_{M}}{2}}\right)^{-1}+\frac{1}{\sqrt{3}}\right)^{2}\right] \nonumber \\
    & \times\left[\frac{\pi^{2}(s_{\rm M,1}-s_{\rm M,2})}{\Omega_{\text{env}} \Pi_{0,\rm M}}-\pi\right] \nonumber \\
    & \simeq -\frac{\epsilon}{V_{\rm M,\neq}}\frac{s_{\rm M,1}-s_{\rm M,2}}{(\bar{s}_{\rm M}-s^{*}_{\rm M})^{2}}\left(\frac{\mathrm{d}u_{\rm core}\circ G_{\rm M}^{-1}}{\mathrm{d}s_{\rm M}}\Big|_{\frac{\bar{s}_{\rm M}+s^{*}_{\rm M}}{2}}\right)^{-1} \\
    & \times\left(\frac{\mathrm{d}u_{\rm core}\circ G_{\rm M}^{-1}}{\mathrm{d}s_{\rm M}}\Big|_{\bar{s}_{\rm M}}\right) \, .
\end{align}
This equation is finally rearranged to give Eq.~\eqref{eq:final_profile}.

\section{Derivation of the coupling equation and the dip shape for discontinuous magnetic fields, rotation rates and densities at the core-to-envelope boundary}
\label{App:dip_disc}

As argued in \citetalias{Tokuno2022AsteroseismologyOscillations} and \citetalias{Barrault2025ConstrainingDips}, if the radial wavelength of the $m$-$g$-$i$ mode is higher than the typical lengthscale at which $N$ varies, a model treating a discontinuous jump of $N$ is better suited than the one considering a continuous $N$ considered in Section~\ref{der:cont_brunt}. This model would be favoured as the star ages and its convective core recedes, leaving a strong near-core chemical stratification. We parametrize the background density and the $N$ jump as such:
\begin{equation}
    \begin{cases}
        \displaystyle \lim_{r \to R_{\rm core}^-}  \bar{\rho} = \bar{\rho}_{\rm core} \\
        \displaystyle \lim_{r \to R_{\rm core}^+} \bar{\rho} = \bar{\rho}_{\rm core} + \Delta\bar\rho = \bar{\rho}_{\rm env}
    \end{cases}
\end{equation}
and
\begin{equation}
    \begin{cases}
        \displaystyle \lim_{r \to R_{\rm core}^-}  N = 0 \\
        N|_{R_{\rm core}} = +\infty\\
        \displaystyle  \lim_{r \to R_{\rm core}^+}  N = N_{0}
    \end{cases}
    \, ,
\end{equation}
where $R_{\rm core}^-$ and $R_{\rm core}^+$ are respectively the lower and upper limits of the core-to-envelope boundary. In this set-up, we can consider either a first-order discontinuity of the background density, i.e. $\Delta\bar\rho\neq0$, or a second-order, i.e. $\Delta\bar\rho= 0$ but $N$ is discontinuous.\\
The vertical wavenumber at the upper edge of the boundary is in the anelastic approximation:
\begin{equation}
    k_{r}^{2} = \dfrac{\Lambda_{k}^{m}(\nu_{\rm M,env})s_{\rm M,env}^{2}}{4\tilde{\epsilon}^{2}R_{\rm core}^{2}} \, ,
\end{equation}
where $\tilde{\epsilon}=\dfrac{\Omega_{\rm env}}{N_{0}}$. \\
A JWKB analysis leads in the same way as computed by \citetalias{Tokuno2022AsteroseismologyOscillations} to the following expressions for the terms in the expansion of the Lagrangian radial displacement perturbation and the total pressure perturbation:
\begin{equation}
    \dfrac{\xi_{r;k,m}^{'}}{r}\Bigg|_{R_{\rm core}^{+}}\simeq Q\tilde{\epsilon}\tilde{X}_{k}^{m}(s_{\rm M,env})
\end{equation}
and 
\begin{equation}
    W'_{k,m}\big|_{R_{\rm core}^{+}}\simeq QR_{\rm core}^{2}\sigma_{\rm M,env}^{2}\tilde{Y}_{k}^{m}(s_{\rm M,env}) \, ,
\end{equation}
where $Q$ is a common linear multiplicating term and
\begin{align}
     \tilde{X}_{k}^{m}& (s_{\rm M,env}) =2\Lambda_{k}^{m}(\nu_{\rm M,env})^{1/4} \nonumber \\
    & \times s_{\rm M,env}^{1/2}\sin\left(\frac{\pi^{2}s_{\rm M,env}}{\Omega_{\rm env}\Pi_{0,\rm M}}-\frac{\pi}{4}\right) \, ,
\end{align}
\begin{align}
    \tilde{Y}_{k}^{m}&(s_{\rm M,env}) = -\Lambda_{k}^{m}(\nu_{\rm M,env})^{-1/4} \nonumber \\
    & \times s_{\rm M,env}^{3/2}\cos\left(\frac{\pi^{2}s_{\rm M,env}}{\Omega_{\rm env}\Pi_{0,\rm M}}-\frac{\pi}{4}\right) \, .
\end{align}
In this framework, ensuring the continuity of both the Lagrangian perturbation of the total pressure and the Lagrangian displacement is not equivalent to ensuring the continuity of the Eulerian total pressure perturbation and the Lagrangian displacement.\\
\indent Neglecting again the centrifugal acceleration and the background Lorentz force, that would alter the sphericity of the region, we use the expression of the local background self-gravity $\bar{g} = \mathcal{G}M_{\rm core}/R_{\rm core}^{2}$. The expressions of the Lagrangian total pressure from both sides of the boundary are:
\begin{align}
     \bar{\rho}_{\rm env}&\sum_{k}a_{k}\left[\sigma_{\rm M,env}^{2}R_{\rm core}^{2}\tilde{Y}_{k}^{m}(s_{\rm M,env})\right. \nonumber \\
    & \left. -\left(\frac{\mathcal{G}M_{\rm core}}{R_{\rm core}^{2}} \right)\tilde{\epsilon}\tilde{X}_{k}^{m}(s_{\rm M, env})\right] \times \Theta_{k}^{m}(\mu;\nu_{\rm M,env})
\end{align}
from the envelope, and
\begin{align}
     \bar{\rho}_{\rm core}&\sum_{k}b_{l}\left[\sigma_{\rm M,core}^{2}R_{\rm core}^{2}P_{l}^{m}(1/\nu_{\rm M,core})\right. \nonumber \\
    & \left. -\left(\frac{\mathcal{G}M_{\rm core}}{R_{\rm core}^{2}} \right)C_{l}^{m}(1/\nu_{\rm M, core})\right] \times \tilde{P}_{l}^{m}(\mu)
\end{align}
from the core.\\
Those two quantities are made equal to ensure the continuity of the total pressure perturbation. As for the Lagrangian displacement, the continuity reads:
\begin{align}
    \sum_{k} & a_{k}\epsilon \tilde{X}_{k}^{m}(s_{\rm M,env})\Theta_{k}^{m}(\mu;\nu_{\rm M,env}) \nonumber \\
    & = \sum_{l}b_{l}C_{l}^{m}(1/\nu_{\rm M,core})\tilde{P}_{l}^{m}(\mu)
\end{align}
We obtain a similar matrix equation as Eq.~\eqref{eq:matrix_cont}:
\begin{equation}
    [\mathcal{\tilde{M}}(s_{\rm M, env},\nu_{\rm M,core})-\tilde{\epsilon}\tilde{\mathcal{N}}(s_{\rm M,env},\nu_{\rm M,core})]\vec{b} = \vec{0}
\end{equation}
with:
\begin{equation}
    \tilde{\mathcal{M}} = c_{k,l}\sigma_{\rm M,env}^{2}\bar{\rho}_{\rm env}\tilde{Y}_{k}^{m}(s_{\rm M, env})C_{l}^{m}(1/\nu_{\rm M,core})
\end{equation}
and
\begin{align}
    \tilde{\mathcal{N}} = c_{k,l}\left(\sigma_{\rm M,core}\bar{\rho}_{\rm core}\tilde{X}_{k}^{m}(s_{\rm M,env})P_{l}^{m}(1/\nu_{\rm M,core}) \right. \nonumber \\
    \left. + \Delta \bar{\rho}\frac{\mathcal{G}M_{\rm core}}{R_{\rm core}^{3}}\tilde{X}_{k}^{m}(s_{\rm M,env})C_{l}^{m}(1/\nu_{\rm M,core})\right)
\end{align}
Pursuing the same leading term analysis as before, the approximate coupling equation reads:
\begin{align}
    & \frac{\sigma_{\rm M,env}^{2}}{\sigma_{\rm M,core}^{2}}\left(1+\frac{\Delta\bar{\rho}}{\bar{\rho}_{\rm core}}\right)\cot \left(\frac{\pi^{2}s_{\rm M,env}}{\Omega_{\rm env}\Pi_{0,\rm M}}-\frac{\pi}{4}\right)\left[\frac{s_{\rm M,env}}{\Lambda_{k}^{m}(\nu_{\rm M,env})^{1/2}}\right]  \nonumber \\
    & \times F_{l}^{m}(\nu_{\rm M,core}) \simeq\tilde{\epsilon}\left(1-\frac{\Delta \bar{\rho}}{\bar\rho_{\rm core}}\frac{\mathcal{G}M_{\rm core}}{\sigma_{\rm M,core}^{2}R_{\rm core}^{3}}F_{l}^{m}(\nu_{\rm M,core})\right) \, .
\end{align}
We define:
\begin{align}
    \tilde{V}_{\rm M,\neq} = & -\left(1+\frac{\Delta \bar{\rho}}{\bar\rho_{\rm core}}\right) \nonumber \\
    & \times \left[\frac{\mathrm{d}F_{l}^{m}}{\mathrm{d}\nu_{\rm M,core}}\frac{\alpha_{\rm rot}^{2}(G_{\rm M}\circ u_{\rm core}^{-1})(\nu_{\rm M,core})}{2\Lambda_{k}^{m}(u_{\rm env}\circ G_{\rm M}\circ u_{\rm core}^{-1}(\nu_{\rm M,core}))^{1/2}}\right. \nonumber \\
    &\left. \times \left(\frac{u_{\rm core}^{-1}(\nu_{\rm M,core})}{G_{\rm M}\circ u_{\rm core}^{-1}(\nu_{\rm M,core})}\right)^{2}\right]_{\nu_{\rm M,core}^{*}} \, .
\end{align}
We expand $F_{l}^{m}$ from its zero $\nu_{\rm M,core}$ and use the expression of $\tilde{V}_{\rm M,\neq}$:
\begin{align}
    & \cot\left(\frac{\pi^{2}s_{\rm M,env}}{\Omega_{\rm env}\Pi_{0,\rm M}}-\frac{\pi}{4}\right)+\tilde{\epsilon}\left(\frac{\alpha_{\rm rot}G_{\rm M}^{-1}(s_{\rm M,env})}{2\Omega_{\rm env}}\right)^{2}\frac{\mathcal{G}M_{\rm core}}{R_{\rm core}^{3}}\frac{\Delta \bar{\rho}}{\bar{\rho}_{\rm env}} \nonumber \\
    & \times \left[\frac{2\Lambda_{k}^{m}(u_{\rm env}\circ G_{\rm M}\circ u_{\rm core}^{-1}(\nu_{\rm M, core}))^{1/2}}{\alpha_{\rm rot}^{2}(G_{\rm M}\circ u_{\rm core}^{-1})(\nu_{\rm M, core})}\right. \nonumber \\
    & \times \left. \left(\frac{G_{\rm M}\circ u_{\rm core}^{-1}(\nu_{\rm M, core})}{u_{\rm core}^{-1}(\nu_{\rm M, core})}\right)^{2} \right]_{\nu_{\rm M,core}^*} \simeq -\frac{\tilde{\epsilon}/\tilde{V}_{\rm M,\neq}}{\nu_{\rm M,core}-\nu_{\rm M,core}^{*}} \, .
\end{align}
This is the equivalent of Eq.(53) in \citetalias{Barrault2025ConstrainingDips} in the magnetic case.
The correction of the structural factor $\tilde{V}_{\rm M,\neq}$ from the solid-body, non magnetic case is:
\begin{align}
    \frac{\tilde{V}_{\rm M,\neq}}{\tilde{V}} = \left[\alpha_{\rm  rot}^{2}\left(\frac{G_{\rm M}\circ u_{\rm core}^{-1}(\nu_{\rm M,core})}{\nu_{\rm M, core}} \right)\left(\frac{u_{\rm core}^{-1}(\nu_{\rm M, core})}{G_{\rm M}\circ u_{\rm core}^{-1}(\nu_{\rm M, core})}\right)^{2} \right. \nonumber \\
    \left. \times \left(\frac{\Lambda_{k}^{m}(\nu_{\rm M, core})}{\Lambda_{k}^{m}(u_{\rm env}\circ G_{\rm M}\circ u_{\rm core}^{-1}(\nu_{\rm M, core}))}^{1/2} \right) \right]_{\nu_{\rm M,core}^*} \, .
\end{align}
After further manipulations of the cotangent, the dip profile is obtained with the magnetic variables:
\begin{align}
    \frac{1}{\Delta \rm P_{M}}-&\frac{1}{\Pi_{0,\rm M}}\simeq \nonumber \\
    &\frac{\dfrac{\tilde{\Gamma}_{\rm M,\neq}}{\pi}\dfrac{\mathrm{d} u_{\rm core}\circ G_{\rm M}^{-1}}{\mathrm d s_{\rm M, env}}\Bigg|_{\bar{s}_{\rm M}}}{\left(\left(\rm{P_{M}}-P_{M}^*\right)\dfrac{\mathrm{d} u_{\rm core}\circ G_{\rm M}^{-1}}{\mathrm d s_{\rm M, env}}\Bigg|_{\frac{\bar{s}_{\rm M,env}+s_{\rm M,env}^*}{2}}\right)^{2}+\tilde{\Gamma}_{\rm M,\neq}^{2}} \, ,
\end{align}
having defined $\tilde{\Gamma}_{\rm M,\neq} = \dfrac{\pi \tilde{\epsilon}}{\Omega_{\rm env}\tilde{V}_{\rm M,\neq}}$.

\section{Summary of the expressions derived in this work}\label{App:summary}
We provide in Table~\ref{tab:gen_expression} a summary of the expressions for the coupling equation, the correction of the structural factors and the dip profile for both a continuous and discontinuous treatment of $N$ at the core-to-envelope boundary.

\begin{sidewaystable*}[]
    \centering
    \begin{tabular}{|c|c|c|c|}
    \cline{3-4}
    \multicolumn{2}{c|}{} & 
          $\rm Le_{core} = Le_{env}$ &  General $\rm Le_{core} \neq \rm Le_{\rm env}$ \\
         \hline
         &&& \\
        & coupling &     $F_{l}^{m}(\nu_{\rm M})\dfrac{\sqrt{3}}{2} \dfrac{\alpha}{\Lambda_{k}^{m}(\nu_{\rm M})^{2/3}}s_{\rm M}^{2/3}$ & $F_{l}^{m}(\nu_{\rm M, core}) \dfrac{\sqrt{3}}{2} \dfrac{\alpha}{\Lambda_{k}^{m}(\nu_{\rm  M, env})^{2/3}}s_{\rm M,env}^{2/3}\dfrac{s_{\rm M, core}^{2}}{s_{\rm M, env}^{2}}\alpha_{\rm rot}^{2}$ \\ 
        &equation&$\times \left[\cot\left(\dfrac{\pi^2s_{\rm M}}{\Omega\Pi_{0,\rm M}}-\dfrac{\pi}{6}\right)+\dfrac{1}{\sqrt{3}}\right] \simeq \epsilon $&$\times \left[\cot\left(\dfrac{\pi^2s_{\rm M, env}}{\Omega_{\rm env}\Pi_{\rm 0,M}}-\dfrac{\pi}{6}\right)+\dfrac{1}{\sqrt{3}}\right] \simeq \epsilon$ \\
        &&& \\
        \cline{2-4}
        &&& \\
        continuous $N$ & $V_{\rm M}/V$ & $\left[\left(\dfrac{u^{-1}(\nu_{\rm M})}{\nu_{\rm M}}\right)^{2/3}\right]_{\nu_{\rm M}^{*}}$& $\left[\left(\dfrac{\Lambda_{k}^{m}(\nu_{\rm M, core})}{\Lambda_{k}^{m}(u_{\rm env}\circ G_{\rm M} \circ u^{-1}_{\rm core}(\nu_{\rm M, core}))}\right)^{2/3}\right.$\\
        &&& $\times \left. \left(\dfrac{G_{\rm M}\circ u_{\rm core}^{-1}(\nu_{\rm M, core})}{\nu_{\rm M, core}}\right)^{2/3} \left(\dfrac{u_{\rm core}^{-1}(\nu_{\rm M, core})}{{G_{\rm M}\circ u_{\rm core}^{-1}}(\nu_{\rm M, core})}\right)^{2}\alpha_{\rm rot}^{2} \right]_{\nu_{\rm M, core}^{*}}$ \\
        \cline{2-4}
        &&& \\
        & $\frac{1}{\Delta \mathrm{P_{M}}} - \frac{1}{\Pi_{0,\rm M}} \simeq $ & $\dfrac{\frac{\Gamma_{\rm M}}{\pi}\frac{\mathrm{d}u}{\mathrm{d}s_{\rm M}}\big|_{\bar{s}_{\rm M}}}{\left((\mathrm{P_{M}}-\mathrm{P}_{\rm M}^{*})\frac{\mathrm{d}u}{\mathrm{d}s_{\rm M}}\big|_{\frac{\bar{s}_{\rm M}+s^{*}_{\rm M}}{2}} + \frac{\Gamma_{\rm M}}{\sqrt{3}}\right)^{2}+\Gamma_{\rm M}^{2}}$ & 
    $\dfrac{\frac{\Gamma_{\rm M, \neq}}{\pi}\frac{\mathrm{d}u_{\rm core}\circ G_{\rm M}^{-1}}{\mathrm{d}s_{\rm M,env}}\big|_{\bar{s_{\rm M,env}}}}{\left((\mathrm{P_{M}}-\mathrm{P}_{\rm M}^{*})\frac{\mathrm{d}u_{\rm core}\circ G_{\rm M}^{-1}}{\mathrm{d}s_{\rm M, env}}\big|_{\frac{\bar{s}_{\rm M,env}+s^{*}_{\rm M,env}}{2}} + \frac{\Gamma_{\rm M,\neq}}{\sqrt{3}}\right)^{2}+\Gamma_{\rm M,\neq}^{2}}$ \\
        &&& \\
        \hline
        &&& \\
        & coupling & $F_{l}^{m}(\nu_{\rm M}) \dfrac{s_{\rm M}}{2\Lambda_{k}^{m}(\nu_{\rm M})^{1/2}}$& $F_{l}^{m}(\nu_{\rm M, core})\dfrac{s_{\rm M,env}}{2\Lambda_{k}^{m}(\nu_{\rm  M, env})^{1/2}}\dfrac{s_{\rm M, core}^{2}}{s_{\rm M, env}^{2}}\alpha_{\rm rot}^{2}$ \\
        & equation & $\times \cot\left(\dfrac{\pi^2s_{\rm M}}{\Omega\Pi_{0,\rm M}}-\dfrac{\pi}{4}\right) \simeq \tilde{\epsilon}$& $\times \cot\left(\dfrac{\pi^2s_{\rm M, env}}{\Omega_{\rm env}\Pi_{\rm 0,M}}-\dfrac{\pi}{4}\right) \simeq \tilde{\epsilon}$\\
        &&& \\
        \cline{2-4}
        &&& \\
        discontinuous $N$, &$\tilde{V}_{\rm M}/\tilde{V}$ & $\left[\dfrac{u^{-1}(\nu_{\rm M})}{\nu_{\rm M}}\right]_{\nu_{\rm M}^{*}}$& $\left[\left(\dfrac{\Lambda_{k}^{m}(\nu_{\rm M, core})}{\Lambda_{k}^{m}(u_{\rm env}\circ G_{\rm M} \circ u^{-1}_{\rm core}(\nu_{\rm M, core}))}\right)^{1/2}\left(\dfrac{G_{\rm M}\circ u_{\rm core}^{-1}(\nu_{\rm M, core})}{\nu_{\rm M,core}}\right)\right.$\\
        $\Delta \rho = 0$ &&& $\times \left.  \left(\dfrac{u_{\rm core}^{-1}(\nu_{\rm M, core})}{G_{\rm M}\circ u_{\rm core}^{-1}(\nu_{\rm M, core})}\right)^{2} \alpha_{\rm rot}^{2}\right]_{\nu_{\rm M,core}^{*}}$ \\
        \cline{2-4}
        &&& \\
        &$\frac{1}{\Delta \mathrm{P_{M}}} - \frac{1}{\Pi_{0,\rm M}} \simeq $ & $\dfrac{\frac{\tilde{\Gamma}_{\rm M}}{\pi}\frac{\mathrm{d}u}{\mathrm{d}s_{\rm M}}\big|_{\bar{s}_{\rm M}}}{\left((\mathrm{P_{M}}-\mathrm{P}_{\rm M}^{*})\frac{\mathrm{d}u}{\mathrm{d}s_{\rm M}}\big|_{\frac{\bar{s}_{\rm M}+s^{*}_{\rm M}}{2}}\right)^{2}+\tilde{\Gamma}_{\rm M}^{2}}$ & 
    $\dfrac{\frac{\tilde{\Gamma}_{\rm M, \neq}}{\pi}\frac{\mathrm{d}u_{\rm core}\circ G_{\rm M}^{-1}}{\mathrm{d}s_{\rm M,env}}\big|_{\bar{s}_{\rm M,env}}}{\left((\mathrm{P_{M}}-\mathrm{P}_{\rm M}^{*})\frac{\mathrm{d}u_{\rm core}\circ G_{\rm M}^{-1}}{\mathrm{d}s_{\rm M, env}}\big|_{\frac{\bar{s}_{\rm M, env}+s^{*}_{\rm M,env}}{2}}\right)^{2}+\tilde{\Gamma}_{\rm M,\neq}^{2}}$ \\
        &&& \\
        \hline
        \multicolumn{4}{c}{} \\
        \end{tabular}
    \caption{Expressions for a continuous treatment of $N$ (top panels) and a discontinuous one (bottom panels), considering case 1 (left, uniform rotation rate and Alfvén frequency) or the more general case 2 (right, bi-layer rotation rate and Alfvén frequency).}
    \label{tab:gen_expression}
\end{sidewaystable*}

\end{appendix}

\end{document}